\documentclass[a4paper,12pt]{report}
\usepackage{amsfonts}
\usepackage{amssymb}
\usepackage{amsmath}
\usepackage{graphicx}
\usepackage{annulus}
\usepackage[a4paper,marginratio={1:1}]{geometry}

\allowdisplaybreaks[3]

\begin{document}

\vspace*{-1.5cm}
\begin{flushright}
  {\small
  MPP-2009-161
  }
\end{flushright}

\vspace{1.5cm}
\begin{center}
  {\LARGE One-loop and D-instanton corrections to the effective action of open string models
  }
\end{center}

\vspace{0.25cm}
\begin{center}
  {\small
  Maximilian~Schmidt-Sommerfeld\\
  }
\end{center}

\vspace{0.1cm}
\begin{center}
  \emph{
  Max-Planck-Institut f\"ur Physik\\
  F\"ohringer Ring 6\\
  80805 M\"unchen\\
  Germany } \\
\end{center}

\vspace{-0.1cm}
\begin{center}
  \tt{
  Maximilian.Schmidt-Sommerfeld@mpp.mpg.de\\
  }
\end{center}

\vspace{1.5cm}
\begin{center}
  \bf{
  Abstract\\
  }
\end{center}
\noindent
One-loop corrections to the gauge coupling and the gauge kinetic function
in certain classes of four-dimensional D-brane models are computed. It is described how to
determine D-instanton corrections to the superpotential and
the gauge kinetic function in such models. The Affleck-Dine-Seiberg superpotential
is re\-de\-rived in string theory. The existence of a new class of multi-instantons, dubbed poly-instantons,
is conjectured. 


\thispagestyle{empty}
\clearpage
\tableofcontents
\chapter{Summary}
It is outlined how to determine certain corrections to effective actions of four-dimensional quantum field theories capturing the low energy physics of string compactifications with open strings. To set the stage, the general form of such actions is described and some examples of open string compactifications are introduced. Orientifolds of type IIA string theory on Calabi-Yau manifolds with intersecting D6-branes are described. D6-brane models on toroidal orbifolds, for which a CFT description exists, are discussed and the partition functions for two orbifolds are presented. A particular orbifold model of the type I string, for which a dual heterotic description is known, is introduced. Finally, some aspects of models based on abstract CFTs are outlined and the general forms of open string partition functions and vertex operators for supersymmetric string compactifications with D-branes are given.

Corrections to the gauge coupling constant and the holomorphic gauge kinetic function are discussed. After showing how to determine one-loop gauge threshold corrections in four-dimensional D-brane models they are computed for intersecting D6-brane models on two different toroidal orbifolds as well as the aforementioned type I model and its heterotic dual. It turns out that gauge threshold corrections do generically depend non-holomorphically on the moduli of the compactification space. It is shown that this is not in contradiction with the holomorphy of the gauge kinetic function and how the one-loop corrections to the latter can be extracted from the aforementioned results. A complete cancellation of non-holomorphic terms only takes place if some of the closed string moduli are redefined at one loop. This redefinition can also be extracted from the gauge threshold corrections.

Next, D-brane instantons and their effects on the low energy effective action are considered in great detail. After describing the relevant instantons, their zero modes including the vertex operators are discussed at length. It is shown how zero mode counting and global abelian symmetries can be exploited in order to find out which instantons can contribute to which quantities. A formula for the computation of spacetime correlators of charged matter fields in a D-instanton background is given. Although it is expected that these correlators can be encoded in a superpotential in the effective action, it is, given how they are computed, not clear a priori how this should work. The reason is that the resulting expressions seem to be at variance with the holomorphy of the superpotential. It is shown that non-holomorphic terms partly cancel and partly rearrange such that a result in agreement with the holomorphy of the superpotential comes out.

The D-instanton calculus is then used to rederive the ADS superpotential known from field theory in a string theory model of SQCD. After engineering SQCD in a local intersecting D-brane model, the D-instanton responsible for the generation of the superpotential is identified and its zero mode structure is analysed. The relevant CFT disc diagrams are computed and the integration over zero modes is performed. The expected result is found. The analysis is redone for models with other gauge groups. The fact that one is able to rederive results known from field theory should be interpreted as a successful test of the D-instanton calculus.

The latter is then extended to corrections to the gauge kinetic function. S-duality between the heterotic and the type I string is used to infer what the zero mode structure of the relevant instantons looks like. It is explained how the fermionic zero modes are absorbed and how the instanton calculus yields a holomorphic gauge kinetic function. The calculus is then applied to the aforementioned type I model. The relevant instantons are described and the one-loop diagram through which the zero modes are absorbed is determined. The expected result, namely the corrections to the gauge kinetic function due to worldsheet instantons in the dual heterotic description, can be reproduced. The D-instanton calculus for corrections to the gauge kinetic function has thus passed an important test.

Finally, the existence of a new class of D-instanton corrections to holomorphic quantities is conjectured. The equality of the D-instanton action and the gauge kinetic function on a stack of (fictitious) D-branes suggests that the D-instanton action should receive instanton corrections, because the gauge kinetic function does. Instanton corrections to instanton actions are rephrased in terms of new so-called poly-instanton corrections to holomorphic quantities. It is outlined how to determine them, and some poly-instanton amplitudes are computed in the aforementioned type I orbifold model. Their contribution to the gauge kinetic function has no counterpart in the dual heterotic model. It is not clear what this discrepancy means. It is possible that there are new corrections also in the heterotic string which arise as the effect of several mutually interacting worldsheet instantons.
\chapter{Four-dimensional effective actions}
\label{fourdimensionaleffectiveactions}
The low energy physics of a four-dimensional string compactification can be described by an effective field theory. This is true as long as the energies characteristic of the processes one is interested in are small compared to the string scale and to the scale set by the size of the compactification manifold. In order to determine the effective action, one needs to identify all fields whose masses are smaller than the scale up to which one wants the field theory to be valid, write down the most general Lagrangian for these fields which respects the relevant symmetries and determine its parameters by equating S-matrix elements computed in string theory and in a quantum field theory based on this Lagrangian.

There are two different objects which are frequently referred to as effective actions \cite{Shifman:1986zi,Kaplunovsky:1994fg,Louis:1996ya}, namely the one-particle-irreducible effective action $\Gamma(\mu)$, where $\mu$ is the renormalisation scale, and the Wilsonian effective action $S_W(\mu)$ \cite{Wilson:1973jj}, where $\mu$ is the cutoff scale, below which the effective theory is defined. The Wilsonian action is obtained by integrating out all fluctuations whose momenta $p$ are bigger than $\mu$. This means in particular that all particles with masses greater than $\mu$ are integrated out. The Wilsonian action is local. When one uses it as the starting point to determine correlation functions one has to compute Feynman diagrams including loops. The loop integrals have to be cut off at $\mu$. By contrast, correlation functions are obtained from the one-particle-irreducible effective action just by functional derivation. All quantum effects, including virtual particles with low momenta, have already been integrated out. The couplings in the one-particle-irreducible effective action are thus physical quantities that can be measured, e.g. in scattering experiments. Due to infrared divergences the one-particle-irreducible effective action is non-local.

It can be obtained from the Wilsonian action by computing correlation functions, e.g. in a perturbative expansion using Feynman diagrams. Schematically, one can write \cite{Shifman:1986zi}
\begin{eqnarray}
 \exp\left( i \Gamma(\mu) \right) = \langle \exp \left( i S_W(\mu) \right) \rangle .
\end{eqnarray}
Ultimately, one is interested in the one-particle-irreducible effective action because it directly contains the information needed to compare the predictions of the theory with experiment. Especially when dealing with supersymmetric theories, it is however often the Wilsonian action that is determined, in particular when deriving the low energy field theory of a string compactification. This is because it is usually easier to compute as there are non-renormalisation theorems implying that some couplings in the Wilsonian action of a supersymmetric theory receive only certain corrections. This will be elaborated on in the following. Note that in passing from $S_W(\mu)$ to $\Gamma(\mu)$ one only has to take low momentum modes into account, the details of a possible high energy theory, e.g. a string theory, are unimportant. This means that no information is lost in making the intermediate step of computing the Wilsonian action rather than the one-particle-irreducible one directly.

The general form of the effective action capturing the low energy physics of a four-dimensional string compactification will now be described. String theory comprises gravitational interactions {\`a} la general relativity, so the Lagrangian contains an Einstein-Hilbert term and is generally covariant. If the compactification preserves supersymmetry, or slightly breaks it dynamically, the low energy effective theory is a supergravity theory with vector, chiral and linear multiplets in addition to the gravity multiplet. The linear multiplets can usually be dualised into chiral multiplets. Thus the focus will here be on a locally supersymmetric field theory with a bunch of vector and chiral superfields. The two-derivative Wilsonian action of such a theory \cite{Cremmer:1982en} is characterised by the Kaehler potential $K(\Phi,\Phi^*)$, the superpotential $W(\Phi)$, the gauge kinetic function $f_{ab}(\Phi)$ and, if there are abelian factors in the gauge group, Fayet-Iliopoulos constants $\xi_a$. The Kaehler potential is a real gauge invariant function of the chiral multiplets $\Phi$ and their complex conjugates $\Phi^*$, whereas the superpotential and the gauge kinetic function depend holomorphically on the chiral superfields $\Phi$.

The Lagrangian will now be written down in the limit of global supersymmetry, i.e. gravity effects are neglected. The reason for this is that the full Lagrangian is terribly lengthy and will not be needed in the following. The vector superfields and their field strengths are denoted by $V^a$ and $W^{\alpha a}$. In superspace notation the Lagrangian reads
\begin{eqnarray}
 L &=& \left(\int d^2\theta f_{ab}(\Phi) W^{\alpha a} W^b_\alpha + c.c.\right)
     + \int d^4 \theta \xi_a V^a
     \nonumber \\ &&
     + \int d^4\theta K(\Phi^* e^{2gV},\Phi)
     + \left( \int d^2\theta W(\Phi) + c.c. \right) .
 \label{susylagrangiansuperspace}
\end{eqnarray}
The holomorphy of the gauge kinetic function and the superpotential has its reason in the fact that the relevant terms in the Lagrangian are integrated only over chiral superspace, as can be seen in \eqref{susylagrangiansuperspace}. The bosonic part of \eqref{susylagrangiansuperspace} contains the kinetic and the topological term
\begin{eqnarray}
 - \frac{1}{4} {\rm Im}(f_{ab}(\phi)) F^a_{\mu \nu} F^{b\mu\nu} - \frac{1}{8} {\rm Re}(f_{ab}(\phi))
 \epsilon^{\mu\nu\sigma\rho} F^a_{\mu\nu} F^b_{\sigma\rho}
 \label{kinetictermsgaugebosons}
\end{eqnarray}
for the gauge field, whose field strength is denoted by $F^a_{\mu \nu}$. The prefactors ${\rm Im} (f_{ab} (\phi) ) $ and ${\rm Re}(f_{ab}(\phi))$ depend on the scalars $\phi$ of the chiral supermultiplets $\Phi$. The kinetic term
\begin{eqnarray}
 - K_{ij}(\phi,\phi^*) D_\mu \phi^{*i} D^\mu \phi^j
\end{eqnarray}
for these scalars is expressed in terms of the gauge covariant derivative $D^\mu \phi^j$ and the Kaehler metric
\begin{eqnarray}
 K_{ij}(\phi,\phi^*) = \frac{\partial^2 K(\phi,\phi^*)}{\partial \phi^{*i} \partial \phi^j} .
\end{eqnarray}
By comparing \eqref{kinetictermsgaugebosons} with the standard kinetic term $g^{-2} F^a_{\mu\nu} F^{a\mu\nu}$ of a renormalisable gauge theory, one sees that the (inverse squares of the) gauge couplings $g_a$, where $a$ labels the different gauge group factors, are given by the imaginary parts of holomorphic functions. Note that this is true only for the gauge couplings in the Wilsonian effective action. The one-particle-irreducible effective action is in general not of the form \eqref{susylagrangiansuperspace} and the running, loop-corrected, physical gauge couplings $g_a(\mu^2)$ appearing in it are not imaginary parts of holomorphic functions. The non-holomorphic parts of $g_a(\mu^2)$ come from infrared effects and therefore only from massless modes. This means that they can be computed entirely in the low energy theory.

One finds that the running, physical gauge couplings $g_a(\mu^2)$ do not only depend on the the gauge kinetic function(s), but also on the Kaehler potential $K$ and the Kaehler metrics $K^{ab}_r(\mu^2)$ of the charged matter fields transforming in some representation $r$ of the gauge group. Focusing for simplicity on diagonal gauge kinetic functions, i.e. $f_{ab}=f_a \delta_{ab}$, the formula relating $g_a(\mu^2)$ and $f_a$ is \cite{Shifman:1986zi,LopesCardoso:1991zt,Ibanez:1992hc,Kaplunovsky:1994fg,Kaplunovsky:1995jw}
\begin{eqnarray}
 16\pi^2 g_a^{-2}(\mu^2) &=& 16\pi^2 {\rm Im}(f_a) + b_a \ln\frac{\Lambda^2}{\mu^2} + c_a K +
             2 T_a(adj) \ln  g_a^{-2}(\mu^2) 
             \nonumber \\ &&
             - 2 \sum_r T_a(r) \ln \det K^{ab}_r(\mu^2) .
 \label{physicalgaugecouplingINTRO}
\end{eqnarray}
The beta function coefficient $b_a$ of the gauge group factor $G_a$ is given by $b_a=\sum_r n_r T_a(r)-3T_a(adj)$ and $c_a$ is defined as $c_a=\sum_r n_r T_a(r)-T_a(adj)$. The sums over $r$ run over the representations of $G_a$, $n_r$ is the number of chiral multiplets transforming in the representation $r$ and $T_a(r)=Tr_r(T^2_{(a)})$, where $T_{(a)}$ are the generators of $G_a$. The scale at which the gauge coupling is defined is denoted by $\Lambda$.

The holomorphy of $W$ and $f_{ab}$ puts strong constraints on which fields these quantities can depend on. The reason is that there are often symmetries in string theory models under which the real parts of certain complex fields shift. Due to holomorphy, $W$ and $f_{ab}$ cannot depend only on the imaginary parts of those fields and due to the shift symmetry they cannot depend on the full complex fields. This means that there are fields of which $W$ and $f_{ab}$ are independent as long as the symmetries remain intact. The latter are usually only broken by instantons which means that the only terms in $W$ and $f_{ab}$ which depend on the aforementioned fields are generated non-perturbatively. As will be explicated in section \ref{holomorphygaugekineticfunction}, it is possible to formulate so called non-renormalisation theorems. These theorems state which kind of perturbative and non-perturbative corrections certain quantities can receive and what these corrections look like.

It was already mentioned that the parameters of the effective field theory capturing the low energy physics of a string compactification are determined by equating S-matrix elements computed in string and field theory. For the effective supergravity action just described this amounts to determining the gauge kinetic function, the superpotential and the Kaehler potential. In the next chapter, some examples of string compactifications will be introduced. It will be described which fields are contained in the low energy effective theories of these compactifications and what the tree-level expressions for the gauge kinetic functions look like. In the following chapter it will be discussed how to compute one-loop corrections to gauge coupling constants and gauge kinetic functions. The subsequent chapters are concerned with non-perturbative contributions to superpotentials and gauge kinetic functions. In order to be able to compute them the holomorphy of $W$ and $f_{ab}$ will be crucial.

\chapter{Overview of some examples of four-dimensional open string compactifications}
\section{Intersecting D6-brane models on Calabi-Yau manifolds}
\label{D6branemodelsCY}

The starting point for models with intersecting D6-branes  \cite{Blumenhagen:2000wh,Aldazabal:2000dg,Aldazabal:2000cn,Uranga:2003pz,MarchesanoBuznego:2003hp,Lust:2004ks,Blumenhagen:2005mu,Blumenhagen:2006ci}on Calabi-Yau
manifolds \cite{Blumenhagen:2002wn} is the ten-dimensional type IIA theory. In this theory, massless bosonic fields
arise in the NSNS as well as in the RR sector. In the former there are
the graviton, the Kalb-Ramond two-form $B_2$ with three-form field strength $H_3$
and the (ten-dimensional) dilaton $\phi_{10}$, whereas in the latter one finds p-form potentials $C_p$ with
odd p. These have (p+1)-form field strengths which are subject to
duality relations. To get a four-dimensional model preserving
eight supercharges, one compactifies the theory on a three complex dimensional
Calabi-Yau manifold $CY_3$,
i.e. one makes the following ansatz for the ten-dimensional spacetime:
\begin{eqnarray}
 M_{10} = \mathbb{R}^4 \times CY_3
\end{eqnarray}
The Calabi-Yau manifold comes equipped with a Kaehler form $J$ and a holomorphic three-form $\Omega_3$. Its volume will be denoted by $V_{CY_3}$.

In order to have non-abelian gauge interactions in the low energy effective theory and to break another half of
the supersymmetries one can orientifold this theory and introduce stacks of D6-branes filling
out the external four-dimensional space as well as a three-dimensional
submanifold of the internal Calabi-Yau space. Orientifolding means that one divides the
theory by the symmetry
\begin{eqnarray}
 \Omega (-1)^{F_L} \bar{\sigma}.
 \label{orientifoldprojection}
\end{eqnarray}
$\Omega$ is the world-sheet parity operator, i.e. it inverts the orientation
of the string. $F_L$ is the spacetime fermion number in the left-moving sector and
$\bar{\sigma}$ is an antiholomorphic involution of the internal manifold. The fixed point set
of this involution is a three-cycle, whose homology class will be denoted by $\pi_{O6}$. The
product of this three-cycle with the four-dimensional external space is referred to
as the orientifold plane.

The topological data of the Calabi-Yau manifold allows one to determine the massless (closed string) spectrum of the four-dimensional low energy effective theory. The fields relevant in the following are the complex structure moduli $U^{(i)}$, $i\in \{0,...,h_{21}\}$, and the Kaehler moduli $T^{(i)}$, $i\in\{1,...,h_{11}^-\}$. $h_{21}$ is the number of harmonic $(2,1)$-forms on the Calabi-Yau manifold and $h_{11}^-$ is the number of two-cycles that are anti-invariant under the antiholomorphic involution $\bar{\sigma}$. In order to be able to properly define these moduli, one first introduces a basis $(A_i,B_i)$, $i\in \{0,...,h_{21}\}$, of three-cycles
satisfying $A_i \circ A_j = 0$, $B_i \circ B_j = 0$ and $A_i \circ B_j = \delta_{ij}$ as well as a basis
$(a_i,b_i)$, $i\in \{0,...,h_{21}\}$, of harmonic three-forms obeying
$\int_{CY_3} a_i \wedge a_j = 0 = \int_{CY_3} b_i \wedge b_j$ and $\int_{CY_3} a_i \wedge b_j = \delta_{ij}$. It
is convenient to choose theses bases dual to each other, i.e. $\int_{A_i} a_j = \delta_{ij}$ and
$\int_{B_i} b_j = \delta_{ij}$, and such that the cycles $A_i$ are invariant under the involution, and the $B_i$ anti-invariant. The complex structure moduli are then given by
\begin{eqnarray}
 U^{(i)} = \int_{A_i} C_3 + i e^{-\phi_4} \int_{A_i} {\rm Re}(\Omega_3) \ \ .
 \label{complexstructuremoduli}
\end{eqnarray}
The four-dimensional dilaton $\phi_4$ and the ten-dimensional one $\phi_{10}$ are related by $\phi_4=\phi_{10}-\ln(V_{CY_3})/2$. Similarly, denoting a basis of anti-invariant two-cycles by $C_i$, $i\in\{1,...,h_{11}^-\}$, the Kaehler moduli are
\begin{eqnarray}
 T^{(i)} = \int_{C_i} B_2 + i  \int_{C_i} J \ \ .
 \label{kaehlermoduli}
\end{eqnarray}

The next step is to introduce stacks of D6-branes wrapping three-dimen\-sional subspaces of the internal
manifold and filling out the external four-dimensional space. In order to be able to perform the orientifold projection consistently, the model has to contain the orientifold images of the branes, too. A brane stack labelled $a$ consists of $N_a$ branes and wraps a three-cycle in
the homology class $\pi_a$. Both the orientifold plane and the D-branes are charged
under the RR seven-form potential $C_7$. These couplings are described by the
terms
\begin{eqnarray}
 S_{O6}^{CS} = -4\mu_6\int_{\mathbb{R}^4\times\pi_{O6}} C_7 + ...
 \label{ChernSimonsactionO6}
\end{eqnarray}
and
\begin{eqnarray}
 S_{D6_a}^{CS} = \mu_6 N_a \int_{\mathbb{R}^4\times\pi_a} C_7 + ...
 \label{ChernSimonsactionD6}
\end{eqnarray}
in the Chern-Simons actions.
Using \eqref{ChernSimonsactionO6}, \eqref{ChernSimonsactionD6} and the kinetic term
\begin{eqnarray}
 S_{kin} \varpropto \int_{M_{10}} d C_7 \wedge \star d C_7
\end{eqnarray}
for the seven-form, which is part of the ten-dimensional action of the Type IIA theory, one derives
the equation of motion for the seven-form. Summing over all stacks of branes and taking their
orientifold images into account, too, it becomes
\begin{eqnarray}
 d\star d C_7 \varpropto \sum_a N_a \left( \delta(\pi_a) + \delta(\pi'_a) \right) - 4 \delta(\pi_{O6}).
 \label{eomc7}
\end{eqnarray}
$\delta(\pi_a)$ denotes the Poincare dual three-form of $\pi_a$ and $\pi'_a$ is the orientifold image
of $\pi_a$. Equation \eqref{eomc7} determines the tadpole cancellation condition to be
\begin{eqnarray}
 \sum_a N_a ( \pi_a + \pi'_a ) = 4 \pi_{O6} .
 \label{tadpolecancellationcondition}
\end{eqnarray}
If a D6-brane is to preserve some supersymmetry, it has to wrap a special Lagrangian cycle \cite{Becker:1995kb,Kachru:1999vj}. The latter is
a submanifold fulfilling the following conditions:
\begin{eqnarray}
 J |_{\pi_a} &=& 0 \\
 {\rm Im} \left( \exp(i\phi_a) \Omega_3 \right) |_{\pi_a} &=& 0
\end{eqnarray}
$\phi_a$ is a calibration phase. If the whole model is to be supersymmetric, all branes have to
be calibrated with the same phase. This phase is determined by the antiholomorphic involution $\bar{\sigma}$
that is part of the orientifold projection \eqref{orientifoldprojection} via
\begin{eqnarray}
 \bar{\sigma} (\Omega_3) = \exp ( i \phi ) \bar{\Omega}_3.
\end{eqnarray}
The supersymmetry condition is thus:
\begin{eqnarray}
 \phi = \phi_a \qquad \underset{a}{\forall}
 \label{susyconditionCY}
\end{eqnarray}

The full gauge group of the model is a product of unitary, orthogonal and unitary symplectic groups. A stack
$a$ of branes not invariant under the orientifold projection yields a factor $U(N_a)$, whereas a stack
that is mapped to itself by the orientifold projection yields a factor $SO(N_a)$ or $USp(N_a)$. The
chiral spectrum is determined by the topological intersection numbers of the three-cycles the D-branes wrap
and is given in table \ref{chiralspectrum}.
\begin{table}
 \begin{center}
 \begin{tabular}{|c|c|}
  \hline
  Sym. rep. of $U(N_a)$ & 1/2 ($\pi'_a \circ \pi_a - \pi_{O6} \circ \pi_a$) \\
  Antisym. rep. of $U(N_a)$ & 1/2 ($\pi'_a \circ \pi_a + \pi_{O6} \circ \pi_a$) \\
  Antifund.$\times$Fund. rep. of $U(N_a)\times U(N_b)$ & $\pi_a \circ \pi_b$ \\
  Fund.$\times$Fund. rep. of $U(N_a)\times U(N_b)$ & $\pi'_a \circ \pi_b$ \\
  \hline
 \end{tabular}
 \caption{Intersecting D6-branes: Chiral spectrum\label{chiralspectrum}}
 \end{center}
\end{table}

The tadpole cancellation condition \eqref{tadpolecancellationcondition} ensures that all purely non-abelian
anomalies cancel, i.e. all triangle graphs with three non-abelian gauge bosons are zero. However, this is not
true for anomalies involving abelian gauge bosons. More precisely, the mixed abelian/non-abelian anomaly, related to
a triangle graph with two non-abelian and one abelian gauge bosons, the mixed abelian/gravitational anomaly,
which comes from a triangle with two gravitons and an abelian gauge boson, and the purely abelian anomaly,
arising from a graph with three abelian gauge bosons, only vanish upon taking the Green-Schwarz mechanism \cite{Green:1984sg,Aldazabal:1998mr,Ibanez:1998qp,Ibanez:1999pw,Aldazabal:2000dg} into
account.

To see how this happens, one first writes the homology classes of the sub\-ma\-ni\-folds which brane stack $a$, its orientifold image and the orientifold plane wrap in terms of the basis of three-cycles introduced earlier:
\begin{eqnarray}
 \pi_a = m_a^i A_i + n_a^i B_i, \quad
 \pi'_a = m_a^i A_i - n_a^i B_i, \quad
 \pi_{O6} = m_O^i A_i
\end{eqnarray}
The RR three- and five-forms are expanded in the basis of three-forms as follows:
\begin{eqnarray}
 C_3 = {\rm Re}(U^{(i)}) a_i , \qquad
 C_5 = x^{(2)}_i b_i
\end{eqnarray}
As $C_3$ and $C_5$ are Hodge dual in ten dimensions, ${\rm Re}(U^{(i)})$ and $x^{(2)}_i$ are dual in four dimensions.

The Chern-Simons actions for the D-branes and the orientifold plane contain the terms
\begin{eqnarray}
 S_{D6_a}^{CS} &=& \int_{\mathbb{R}^4\times\pi_a} C_5 \wedge tr(F) +
      C_3 \wedge \left( tr(F\wedge F) - \frac{tr(1)}{48} tr(R\wedge R) \right) \label{CSD6} + ... \qquad \\
 S_{O6}^{CS} &=& \int_{\mathbb{R}^4\times\pi_{O6}}
      C_3 \wedge tr(R\wedge R) \label{CSO6} + ... \ \ ,
\end{eqnarray}
where $R$ is the spacetime curvature two-form and $F$ the field strength of the $U(N)$ gauge field on the D-brane.
Upon dimensional reduction \eqref{CSD6} and \eqref{CSO6} lead to the following terms in the four-dimensional Lagrangian:
\begin{eqnarray}
 L &=& \frac{1}{24} tr(R\wedge R) \left( m_{O}^i {\rm Re}(U^{(i)}) \right) \nonumber \\
   && + \left( tr(F\wedge F) - \frac{tr(1)}{48} tr(R\wedge R) \right) \left( 2 m_a^i {\rm Re}(U^{(i)})
      \right) \nonumber \\
   && + tr(F) \wedge \left( 2 n_a^i x^{(2)}_i \right) + ...
 \label{axionmassterms}
\end{eqnarray}
It is possible to show that if these terms are taken into account all the aforementioned anomalies cancel
and the gauge bosons of the seemingly anomalous symmetries become massive, the longitudinal degree of
freedom being a linear combination of RR sector fields. It is also possible that abelian gauge
bosons associated with non-anomalous symmetries become massive. Thus, in order to determine the massless
spectrum correctly, one has to take the couplings \eqref{axionmassterms} into account. Furthermore,
the Green-Schwarz mechanism implies that under a $U(1)_a$, where $U(1)_a$ is
the diagonal $U(1)$ subgroup of the $U(N_a)$ gauge group on brane stack $a$, gauge transformation
\begin{eqnarray}
 A^a_\mu \rightarrow A^a_\mu + \partial_\mu \alpha^a
\end{eqnarray}
the real parts of the complex structure moduli \eqref{complexstructuremoduli} transform by shifts:
\begin{eqnarray}
 U^{(i)} \rightarrow U^{(i)} - 2 N_a n_a^i \alpha^a
 \label{gaugetransformationcomplexstructuremoduli}
\end{eqnarray}
The abelian symmetries, whose associated gauge bosons are massive, do not appear as local gauge symmetries in
the low energy effective theory, but are global symmetries to all orders in the string perturbation series.
They are in general broken by instantons as will be explained in chapter \ref{Dinstantons}.

The gauge coupling on a stack $a$ of D6-branes can be computed by dimensionally reducing the
Dirac-Born-Infeld action
\begin{eqnarray}
 S_{D6_a}^{DBI} \varpropto \int_{\mathbb{R}^4\times\pi_a} d^7 x \exp(-\phi_{10}) \sqrt{-det(g_{\mu\nu}+F_{\mu\nu})}
\end{eqnarray}
of the D6-branes. Here, $g_{\mu\nu}$ is the pullback of the spacetime metric onto the
worldvolume of the brane and $F_{\mu\nu}$ once more denotes the field strength of the brane's gauge field.
Upon taking the Chern-Simons terms in \eqref{CSD6} into account, too, one finds
that the gauge kinetic function becomes
\begin{eqnarray}
 f_a = \int_{\pi_a} C_3 + i e^{-\phi_{4}} \int_{\pi_a} {\rm Re} ( \exp (i \phi_a ) \Omega_3 )
       = \sum_{i=0}^{h_{21}} m_a^i U^{(i)} .
 \label{gaugekineticfunction}
\end{eqnarray}
\section{Intersecting D6-brane models on toroidal or\-bi\-folds}
\label{D6branemodelsorbifolds}
Models on orbifolds \cite{Dixon:1985jw,Dixon:1986jc} can be defined as certain two-dimensional conformal field theories \cite{Belavin:1984vu,Ginsparg:1988ui,Ketov:1995yd,DiFrancesco:1997nk}, which can be constructed explicitly. In such models, D-branes can be described by boundary states. Equivalently, one considers open strings whose endpoints
are confined to a certain subspace of the full space, i.e. open strings with appropriate boundary conditions. In both descriptions, one is dealing with boundary CFT.

Toroidal orbifolds are tori divided by a discrete group. To get a four-dimensional model one considers
a compactification on an orbifold of a six-torus $T^6$ with orbifold group $\mathbb{Z}_N \times \mathbb{Z}_M$, where $N$ and $M$ are integers.
As these backgrounds are limiting cases of Calabi-Yau manifolds, much of what was said in the previous section carries over
to these models.

To start with, one has to compute the torus amplitude, or, in other words, the modular invariant partition function
of closed strings on this background. It is given by a trace over all states in all
(twisted and untwisted; (NS$\oplus$R)$\otimes$(NS$\oplus$R) ) sectors of the CFT
\begin{eqnarray}
 T = \int_\mathcal{F} \frac{d^2\tau}{\tau_2} tr^{T,U}_{NS,R}
      \left(P_{GSO} P_{orb} \exp(2\pi i \tau (L_0-\frac{c}{24})) \exp(-2\pi i \tau^* (\widetilde{L}_0-\frac{\widetilde{c}}{24})) \right) , \quad
\end{eqnarray}
where
\begin{eqnarray}
 P_{GSO} = \frac{\Big(1+(-1)^F\Big)\Big(1+(-1)^{\widetilde{F}+\widetilde{\alpha}}\Big)}{4}
\end{eqnarray}
is an operator implementing the Gliozzi-Scherk-Olive projection. $F$ and $\widetilde{F}$ are the left- and right-moving worldsheet fermion numbers and $\widetilde{\alpha}$ is 1 in the right-moving R sector and 0 in the right-moving NS sector. The orbifold projector is given by
\begin{eqnarray}
 P_{orb} = \frac{1}{N M} \sum_{h_1\in\mathbb{Z}_N} \sum_{h_2\in\mathbb{Z}_M} h_1 h_2 \ \ .
\end{eqnarray}

The next step is to define an orientifold projection, which allows one to compute the Klein bottle amplitude
\begin{eqnarray}
K &=& \int \frac{dt}{t} tr^{T,U}_{NS,R} \Big( \Omega (-1)^{F_L} R P_{GSO} P_{orb}
         \nonumber \\ &&
         \exp(2\pi i \tau (L_0-\frac{c}{24})) \exp(-2\pi i \tau^* (\widetilde{L}_0-\frac{\widetilde{c}}{24})) \Big) ,
\end{eqnarray}
where $R$ is an operator that inverts three coordinates of the six-torus. Alternatively, the orientifolding of the theory can be accomplished by introducing so-called crosscap states $|C_6\rangle$. The Klein bottle amplitude can also be computed as an overlap of these crosscap states
\begin{eqnarray}
 K = \int dl \langle C_6 | \exp (-2 \pi l H) | C_6 \rangle ,
\end{eqnarray}
with $H$ the closed string worldsheet Hamiltonian.
The orientifolding leads to tadpole divergences which need to be cancelled by the introduction of D6-branes. The latter are described by boundary states $|B_6\rangle$ and the annulus diagram can be computed as the overlap of these boundary states
\begin{eqnarray}
 A = \int dl \langle B_6 | \exp (-2 \pi l H) | B_6 \rangle .
\end{eqnarray}
Again, there is an alternative way to determine the annulus amplitude which is by computing a trace in the appropriate open string Hilbert space
\begin{eqnarray}
 A = \int \frac{dt}{t} tr^{T,U}_{NS,R} \left(P_{GSO} P_{orb} \exp(- 2 \pi t (L_0-\frac{c}{24})) \right) .
\end{eqnarray}
Finally, the Moebius strip amplitude is either given by an overlap of a boundary and a crosscap state or by the trace in the open string Hilbert space with the orientifold projection operator inserted:
\begin{eqnarray}
 M &=& \int dl \langle C_6 |  \exp (-2 \pi l H) | B_6 \rangle \nonumber \\
      &=&  \int \frac{dt}{t} tr^{T,U}_{NS,R} \left(\Omega (-1)^{F_L} R P_{GSO} P_{orb} \exp(- 2 \pi t (L_0-\frac{c}{24})) \right)
\end{eqnarray}
In these models, the gauge group is a product of unitary, orthogonal and unitary symplectic groups, too. Its precise form as well as the spectrum can be obtained from the open and closed string partition functions, i.e. the torus, the Klein bottle, the annulus and the Moebius strip. As in the case of D6-brane models on Calabi-Yau manifolds, the Green-Schwarz mechanism has to be taken into account in order to obtain the exact spectrum. The supersymmetry condition in the orbifold CFT models is just the vanishing of the partition functions. The gauge kinetic functions are most easily obtained from \eqref{gaugekineticfunction} applied to the orbifold case.
\subsection{An example with bulk branes}
\label{examplesbulk}
In this and the next subsections some more details about the orbifolds with orbifold group $\mathbb{Z}_2\times\mathbb{Z}_2$ are given \cite{Blumenhagen:2005mu,Blumenhagen:2005tn}. There are two variants of this orbifold, differing in how the generators of the $\mathbb{Z}_2$ factors act on the fixed points of the other factor, and therefore in their Hodge numbers. This subsection is concerned with the example with Hodge number $h_{21}=3$, the case with $h_{21}=51$ is discussed in the next subsection.

\begin{figure}[ht]
\begin{center}
 \includegraphics[width=0.7\textwidth]{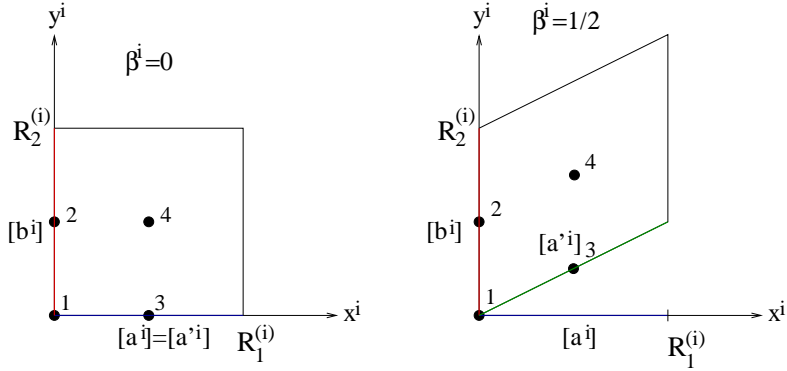}
\end{center}
\caption{\small Geometry of the two-tori, orbifold fixed points and one-cycles}
\label{twotorigeometry}
\end{figure}

Due to the orbifolding the six-torus splits into a direct product of three two-tori. The non-trivial orbifold group elements each invert two of the two-tori and leave the third invariant. The fundamental one-cycles of these tori will be denoted $[a^i]$ and $[b^i]$, $i\in\{1,2,3\}$, and their sizes are measured by the radii $R_1^{(i)}$ and $R_2^{(i)}$. There are also discrete degrees of freedom ($\beta^i\in\{0,1/2\}$) given by possible tilts of the tori. The complex structure and Kaehler moduli are again denoted by $U^{(i)}$ and $T^{(i)}$ and their imaginary parts can be expressed in terms of the dilaton and the radii.
Each of the three two-tori contains four points that are fixed points of the orbifold action. These properties of the tori are illustrated in figure \ref{twotorigeometry}.
In the following, only D6-branes wrapping a three-cycle that is a product of a one-cycle on each of the three two-tori will be considered. These three one-cycles are written $n^i [{a'}^i] + m^i [b^i]$, where $[{a'}^i]=[a^i]+\beta^i [b^i]$. $(n^i,m^i)$ are called the wrapping numbers and encode the homological charge of the branes.  Upon defining $\widetilde{m}^i=m^i+\beta^i n^i$, the one-cycles can also be written as $n^i [a^i] + \widetilde{m}^i [b^i]$. The length of the one-cycle wrapped by the brane on the $i$'th torus is given by
\begin{eqnarray}
 L^i = \sqrt{(n^i R_1^{(i)})^2+(\widetilde{m}^i R_2^{(i)})^2}
 \label{onecyclelength}
\end{eqnarray}
and the (tree level) gauge coupling on the brane becomes \cite{Lust:2004cx}
\begin{eqnarray}
 g_{tree}^{-2} &=& e^{-\phi_{10}} \prod_i L^i = e^{-\phi_4} \prod_i ({\rm Im}(T^{(i)}))^{-1/2} L^i \nonumber \\
                        &=& ({\rm Im}(U^{(0)}))^{1/4} \prod_i ({\rm Im}(U^{(i)}))^{1/4} ({\rm Im}(T^{(i)}))^{-1/2} L^i .
 \label{treelevelgaugecoupling}
\end{eqnarray}
The supersymmetry condition amounts to $\sum_{i=1}^3 \theta^i = 0\ mod\ 2\pi$ with $\tan \theta^i = \widetilde{m}^i R_2^{(i)}/n^i R_1^{(i)}$. If it is fulfilled, the gauge kinetic function on the brane is given by \cite{Lust:2004cx}
\begin{eqnarray}
 f_{tree} = U^{(0)} n^1 n^2 n^3 - \sum_{i\neq j\neq k=1}^3 U^{(i)} n^i \widetilde{m}^j \widetilde{m}^k .
 \label{treelevelgaugekineticfunction}
\end{eqnarray}
The antiholomorphic involution, whose fixed point set defines the orientifold \linebreak[4] plane(s), inverts the $y$-coordinate (see figure \ref{twotorigeometry}) on each of the three tori. The homology class of the orientifold plane is
\begin{eqnarray}
 \pi_O = 8 [a^1] [a^2] [a^3] - \sum_{i\neq j\neq k\neq i} 2^{3-2\beta^j-2\beta^k} [a^i][b^j][b^k]
\end{eqnarray}
and can be encoded in the following set of wrapping numbers:
\begin{eqnarray}
 (n^1_{O_0},\widetilde{m}^1_{O_0},n^2_{O_0},\widetilde{m}^2_{O_0},n^3_{O_0},\widetilde{m}^3_{O_0}) &=& (2,0,2,0,2,0) \\
 (n^1_{O_1},\widetilde{m}^1_{O_1},n^2_{O_1},\widetilde{m}^2_{O_1},n^3_{O_1},\widetilde{m}^3_{O_1}) &=& (2,0,0,2^{1-2\beta^2},0,-2^{1-2\beta^3}) \\
 (n^1_{O_2},\widetilde{m}^1_{O_2},n^2_{O_2},\widetilde{m}^2_{O_2},n^3_{O_2},\widetilde{m}^3_{O_2}) &=& (0,-2^{1-2\beta^1},2,0,0,2^{1-2\beta^3}) \\
 (n^1_{O_3},\widetilde{m}^1_{O_3},n^2_{O_3},\widetilde{m}^2_{O_3},n^3_{O_3},\widetilde{m}^3_{O_3}) &=& (0,2^{1-2\beta^1},0,-2^{1-2\beta^2},2,0)
 \label{orientifoldplanewrappingnumbers}
\end{eqnarray}
The wrapping numbers of the orientifold image of a brane are given by $(n^i,-\widetilde{m}^i)$.
With this information one can either determine the tadpole cancellation conditions from \eqref{tadpolecancellationcondition} or from the partition functions that will be given later on. They read
\begin{eqnarray}
 \sum_a N_a n_a^1 n_a^2 n_a^3 &=& 16 \nonumber \\
 \sum_a N_a n_a^1 \widetilde{m}_a^2 \widetilde{m}_a^3 &=& - 2^{4-2\beta^2-2\beta^3}\nonumber \\
 \sum_a N_a n_a^2 \widetilde{m}_a^3 \widetilde{m}_a^1 &=& - 2^{4-2\beta^3-2\beta^1}\nonumber \\
 \sum_a N_a n_a^3 \widetilde{m}_a^1 \widetilde{m}_a^2 &=& - 2^{4-2\beta^1-2\beta^2} ,
 \label{tadpolecancellationconditionexamplebulk}
\end{eqnarray}
where the sums run over the different stacks of branes labelled $a$ and $N_a$ is the number of branes on stack $a$.
Here and in the following, several quantities carry a label (e.g. $a$ in the above formulas) denoting the brane stack to which they refer.

Using the notation introduced above, it is now possible to write down the open string partition functions for this background. They can be determined by constructing the boundary and crosscap states and computing overlaps \cite{Gaberdiel:1999ch,Gaberdiel:2000jr,Stefanski:2000fp,Gaberdiel:2000fe,Quiroz:2001xz,Craps:2001xw,Maiden:2006qe}. For the annulus diagrams, three cases will be distinguished.

{\bf Case 1:} Both boundaries of the annulus are on the same stack of branes. The amplitude can be written as \cite{Blumenhagen:2000wh}
\begin{eqnarray}
 A_a = N_a^2 \int_0^\infty dl \frac{\vartheta_3^4 - \vartheta_4^4 - \vartheta_2^4 + \vartheta_1^4}{\eta^{12}}  \prod_i 
         \frac{(L^i_a)^2 Z^i_a(l)}{R_1^{(i)} R_2^{(i)}} ,
 \label{partitionfunctionbulk1}
\end{eqnarray}
where $\vartheta = \vartheta(0,2il)$, $\eta=\eta(2il)$ and the following lattice sum has been defined \cite{Blumenhagen:1999md}:
\begin{eqnarray}
 Z^i_a(l) = \sum_{p,q} \exp \left( - \frac{\pi l (L^i_a)^2}{({\rm Im}(T^{(i)}))^2} \left|p+T^{(i)}q\right|^2 \right)
\end{eqnarray}

{\bf Case 2:} The boundaries of the annulus are on branes that are parallel on one torus and intersect at non-trivial angles on the other two. In the following it will be assumed without loss of generality that they are parallel on the first torus($i=1$). The amplitude is \cite{Blumenhagen:2000wh}:
\begin{eqnarray}
 A_{ab} = N_a N_b \int_0^\infty dl \sum_{\alpha ,\beta} (-1)^{2(\alpha+\beta)}
                \frac{\vartheta\genfrac[]{0pt}{}{\alpha}{\beta}(0)^2}{\eta^6} \frac{(L^1_a)^2 Z^1_a(l)}{R_1^{(1)} R_2^{(1)}}
                \prod_{j=2}^3 I^j_{ab}
                \frac{\vartheta\genfrac[]{0pt}{}{\alpha}{\beta}(\theta^j_{ab})}{\vartheta\genfrac[]{0pt}{}{1/2}{1/2}(\theta^j_{ab})}
 \label{partitionfunctionbulk2}
\end{eqnarray}
The branes intersect at an angle $\theta^j_{ab}=\theta^j_a-\theta^j_b$ on the $j$'th torus. The intersection number is given by $I^j_{ab} = (\widetilde{m}^j_a n^j_b - \widetilde{m}^j_b n^j_a)$.

{\bf Case 3:} The annulus is stretched between two branes intersecting non-trivially on all three tori \cite{Blumenhagen:2000wh}:
\begin{eqnarray}
 A_{ab} = N_a N_b \int_0^\infty dl \sum_{\alpha,\beta} (-1)^{2(\alpha+\beta)} \frac{\vartheta\genfrac[]{0pt}{}{\alpha}{\beta}(0)}{\eta^3} \prod_{i=1}^3 I^i_{ab} \frac{\vartheta\genfrac[]{0pt}{}{\alpha}{\beta}(\theta^i_{ab})}{\vartheta\genfrac[]{0pt}{}{1/2}{1/2}(\theta^i_{ab})}
 \label{partitionfunctionbulk3}
\end{eqnarray}
From the partition functions (transformed into loop channel) one can read off the open string spectrum. One finds the following massless states: In case 1 there are a vector multiplet and three chiral multiplets transforming in the adjoint representation of $U(N_a)$, in case 2 one finds $I_{ab}=\prod_{j=2}^3 I^j_{ab}$ hypermultiplets in the bifundamental representation of $U(N_a)\times U(N_b)$ and case 3 yields $I_{ab}=\prod_{i=1}^3 I^i_{ab}$ chiral multiplets in the bifundamental representation of $U(N_a)\times U(N_b)$. In order to determine the full open string spectrum one needs to take all annuli stretching between the different stacks of branes (as well as the Moebius strip diagrams to be given in the sequel) into account.

When writing down the Moebius strip diagrams, it is also useful to distinguish three cases.

{\bf Case 1:} The brane and the orientifold plane are parallel on all three tori. The amplitude can be written as
\begin{eqnarray}
 M_a = N_a \int_0^\infty dl \frac{\vartheta_3^4 - \vartheta_4^4 - \vartheta_2^4 + \vartheta_1^4}{\eta^{12}}  \prod_i 
         \frac{(L^i_a)^2 Z^i_a(4l)}{R_1^{(i)} R_2^{(i)}} ,
 \label{partitionfunctionmoebiusbulk1}
\end{eqnarray}
where $\vartheta = \vartheta(0,2il+1/2)$ and $\eta=\eta(2il+1/2)$.

{\bf Case 2:} The brane and the orientifold plane are parallel on one torus and intersect at non-trivial angles on the other two. In the following it will be assumed without loss of generality that they are parallel on the first torus($i=1$). The amplitude is:
\begin{eqnarray}
 M_{aO_k} = N_a \int_0^\infty dl \sum_{\alpha ,\beta} (-1)^{2(\alpha+\beta)} \frac{\vartheta\genfrac[]{0pt}{}{\alpha}{\beta}(0)^2}{\eta^6} \frac{(L^1_a)^2 Z^1_a(4l)}{R_1^{(1)} R_2^{(1)}} \prod_{j=2}^3 I^j_{aO_k} \frac{\vartheta\genfrac[]{0pt}{}{\alpha}{\beta}(\theta^j_{aO_k})}{\vartheta\genfrac[]{0pt}{}{1/2}{1/2}(\theta^j_{aO_k})}
 \nonumber \\ 
 \label{partitionfunctionmoebiusbulk2}
\end{eqnarray}
The intersection numbers $I^j_{aO_k}$ and angles $\theta^j_{aO_k}$ involving one stack of branes and the orientifold planes are defined in analogy to the quantities involving two stacks of branes using the wrapping numbers of the orientifold plane \eqref{orientifoldplanewrappingnumbers}.

{\bf Case 3:} The brane and the orientifold plane intersect non-trivially on all three tori:
\begin{eqnarray}
 M_{aO_k} = N_a \int_0^\infty dl \sum_{\alpha,\beta} (-1)^{2(\alpha+\beta)} \frac{\vartheta\genfrac[]{0pt}{}{\alpha}{\beta}(0)}{\eta^3} \prod_{i=1}^3 I^i_{aO_k} \frac{\vartheta\genfrac[]{0pt}{}{\alpha}{\beta}(\theta^i_{O_k})}{\vartheta\genfrac[]{0pt}{}{1/2}{1/2}(\theta^i_{aO_k})}
 \label{partitionfunctionmoebiusbulk3}
\end{eqnarray}
As was already mentioned, the Moebius strip diagrams need to be taken into account when determining the open string spectrum. By doing so one finds that the gauge symmetry is reduced from $U(N_a)$ to $USp(N_a)$ or $O(N_a)$ if the brane is mapped to itself by the orientifold projection. The Moebius strip diagrams, together with the annulus diagrams stretching between a brane and its orientifold image, are also important in order to determine the number of multiplets in the symmetric and the antisymmetric representation of $U(N_a)$.

\subsection{An example with fractionally charged branes}
\label{examplesfractional}
This subsection is concerned with a toroidal orbifold which is similar to that described in the previous one. The orbifold group is again $\mathbb{Z}_2\times\mathbb{Z}_2$, but in this case $h_{21}=51$ which means that there are many more three-cycles the D6-branes can wrap around. However, in the orbifold limit considered here, most of them are collapsed to zero size. The moduli whose imaginary parts are the complex structure moduli describing the sizes of these collapsed three-cycles are twisted sector fields denoted by $W_{ikl}$ and $\widetilde{W}_{ikl}$, $i\in\{1,2,3\}$, $k,l\in\{1,2,3,4\}$. They arise at the fixed point denoted $kl$ in the twisted sector which comes from the orbifold group element leaving the $i$'th torus invariant. The real parts of these fields are twisted RR sector fields.

The boundary states describing the D-branes on the background under discussion are a sum of two parts. One is identical to the boundary states of the previous section (up to normalisation) and the other one consists of states in the twisted sectors of the orbifold CFT \cite{Green:1996um,Gaberdiel:1999ch,Gaberdiel:2000jr,Stefanski:2000fp,Gaberdiel:2000fe,Quiroz:2001xz,Craps:2001xw,Maiden:2006qe}. Some more data is therefore needed to fully characterise a D-brane on this background \cite{Blumenhagen:2005tn}. More precisely, in addition to the wrapping numbers one needs to specify the twisted RR charges $\epsilon^i\in\{-1,1\}$, subject to $\epsilon^1=\epsilon^2 \epsilon^3$, the positions $\delta^i \in \{0,1\}$ and the discrete Wilson lines $\lambda^i \in \{0,1\}$. If the brane is charged under fixed point 1 (see figure \ref{twotorigeometry}) on the $i$'th torus, $\delta^i$=0, otherwise $\delta^i=1$. The values of $\epsilon^i$, $\delta^i$ and $\lambda^i$ can be used to determine the charges $\epsilon^i_{kl} \in \{-1,0,1\}$, $i\in\{1,2,3\}$, $k,l\in\{1,2,3,4\}$, of the brane under the fixed point labelled $kl$ in the $i$'th twisted sector. All the symbols introduced above can carry a further index denoting the brane stack to which they refer. The quantities $\sigma^i_{ab} = \sum_{k,l=1}^4 \epsilon^i_{a,kl} \epsilon^i_{b,kl}/4$ and $\sigma_{ab}=\sum_{i=1}^3 \sigma_{ab}^i$ will be useful in the following.

Most formulas of the last subsection, notably \eqref{onecyclelength} and \eqref{treelevelgaugecoupling}, are still valid, \eqref{treelevelgaugekineticfunction} however is replaced by
\begin{eqnarray}
 \label{treelevelgaugekineticfunction2}
  f_{tree} &=& U^{(0)} n^1 n^2 n^3 - \sum_{i\neq j\neq k} U^{(i)} n^i \widetilde{m}^j \widetilde{m}^k \\
                    && + \sum_i \sum_{k,l}
                    n^i ( \epsilon^i_{kl} + \epsilon^i_{R(k)R(l)} ) W_{ikl} + \widetilde{m}^i
                    ( \epsilon^i_{kl} - \epsilon^i_{R(k)R(l)} ) \widetilde{W}_{ikl} \nonumber .
\end{eqnarray}
The function $R$ (which should really carry an index $i$) is given by $R(k)=k$ for $\beta^i=0$ and $R(\{1,2,3,4\})=\{1,2,4,3\}$ for $\beta^i=1/2$. The coupling to the twisted sector fields can be determined by an anomaly analysis \cite{Blumenhagen:2007ip}. The tadpole cancellation conditions are modified by some signs and completed by those arising in the twisted sectors.
\begin{eqnarray}
 \sum_a N_a n_a^1 n_a^2 n_a^3 &=& -16 \nonumber \\
 \sum_a N_a n_a^1 \widetilde{m}_a^2 \widetilde{m}_a^3 &=& - 2^{4-2\beta^2-2\beta^3}\nonumber \\
 \sum_a N_a n_a^2 \widetilde{m}_a^3 \widetilde{m}_a^1 &=& - 2^{4-2\beta^3-2\beta^1}\nonumber \\
 \sum_a N_a n_a^3 \widetilde{m}_a^1 \widetilde{m}_a^2 &=& - 2^{4-2\beta^1-2\beta^2} \\
 \sum_a N_a n_a^i (\epsilon^i_{a,kl} + \epsilon^i_{a,R(k)R(l)} ) &=& 0 \qquad \forall_{i,k,l} \nonumber \\
 \sum_a N_a \widetilde{m}_a^i (\epsilon^i_{a,kl} - \epsilon^i_{a,R(k)R(l)} ) &=& 0 \qquad \forall_{i,k,l}
 \label{tadpolecancellationconditionsfractionaltwisted}
\end{eqnarray}
The open string partition functions for the background considered here will be given in the following \cite{Dudas:2005jx,Blumenhagen:2007ip}. Four cases will be distinguished.

{\bf Case 1:}
Both boundaries of the annulus are on the same stack of branes. The amplitude is
\begin{eqnarray}
 A_a &=& N_a^2 \int_0^\infty dl \Bigg[ \frac{\vartheta_3^4 - \vartheta_4^4 - \vartheta_2^4 + \vartheta_1^4}{\eta^{12}}  \prod_{i=1}^3 
         \frac{(L^i_a)^2 Z^i_{aa}}{R_1^{(i)} R_2^{(i)}} \nonumber \\ && + 16 \sum_{i=1}^3 \sigma^i_{aa}
         \frac{\vartheta_3^2\vartheta_2^2 - \vartheta_4^2 \vartheta_1^2 - \vartheta_2^2 \vartheta_3^2 + \vartheta_1^2 \vartheta_4^2}
              {\eta^6 \vartheta_4^2} \frac{(L^i_a)^2 Z^i_{aa}}{R_1^{(i)} R_2^{(i)}} \Bigg] ,
\end{eqnarray}
where the lattice sum \cite{Blumenhagen:1999md}
\begin{eqnarray}
 Z^i_{ab} = \sum_{p,q} \exp \left( - \frac{\pi l (L^i_a)^2}{({\rm Im}(T^{(i)}))^2} \left| p + T^{(i)} q \right|
               + i \pi p (\delta^i_a-\delta^i_b) + i \pi q (\lambda^i_a-\lambda^i_b) \right) \nonumber
\end{eqnarray}
has been used.

{\bf Case 2:}
The two branes lie on top of each other on the torus before orbifolding and carry the same Wilson lines, i.e. $\theta^i_a=\theta^i_b$, $L^i_a=L^i_b$, $\delta^i_a=\delta^i_b$ and $\lambda^i_a=\lambda^i_b$, but differ in (some of) their twisted charges $\epsilon^i$ such that they define different brane stacks. The equalities above imply $\sigma^i_{ab}=\pm1$. The amplitude becomes
\begin{eqnarray}
  A_{ab} &=& N_a N_b \int_0^\infty dl \Bigg[ \frac{\vartheta_3^4 - \vartheta_4^4 - \vartheta_2^4 + \vartheta_1^4}{\eta^{12}}  \prod_{i=1}^3 
         \frac{(L^i_a)^2 Z^i_{ab}}{R_1^{(i)} R_2^{(i)}} \nonumber \\ && + 16 \sum_{i=1}^3 \sigma^i_{ab}
         \frac{\vartheta_3^2\vartheta_2^2 - \vartheta_4^2 \vartheta_1^2 - \vartheta_2^2 \vartheta_3^2 + \vartheta_1^2 \vartheta_4^2}
              {\eta^6 \vartheta_4^2} \frac{(L^i_a)^2 Z^i_{ab}}{R_1^{(i)} R_2^{(i)}} \Bigg] .
\end{eqnarray}

{\bf Case 3:}
The annulus has its boundaries on branes that are in the same bulk homology class of the torus, which implies $\theta^i_a=\theta^i_b$ and $L^i_a=L^i_b$, but differ in (some of) their positions and Wilson lines, i.e. $\delta^i_a \neq \delta^i_b$ and $\lambda^i_a\neq\lambda^i_b$. The amplitude takes the same form as that of case 2.

{\bf Case 4:}
The branes intersect at non-trivial angles on all three two-tori. The amplitude can be written
\begin{eqnarray}
  A_{ab} &=& N_a N_b \int_0^\infty dl \Bigg[ 8 \sum_{\alpha,\beta} (-1)^{2(\alpha+\beta)} \frac{\vartheta\genfrac[]{0pt}{}{\alpha}{\beta}(0)}{\eta^3} \prod_{i=1}^3 I^i_{ab} \frac{\vartheta\genfrac[]{0pt}{}{\alpha}{\beta}(\theta^i_{ab})}{\vartheta\genfrac[]{0pt}{}{1/2}{1/2}(\theta^i_{ab})} \\
  && + 32 \sum_{i=1}^3 I^i_{ab} \sigma^i_{ab} \sum_{\alpha,\beta} (-1)^{\alpha+\beta} \frac{\vartheta\genfrac[]{0pt}{}{\alpha}{\beta}(0)}{\eta^3} \frac{\vartheta\genfrac[]{0pt}{}{\alpha}{\beta}(\theta^i_{ab})}{\vartheta\genfrac[]{0pt}{}{1/2}{1/2}(\theta^i_{ab})} \prod_{j=1,\neq i}^3 \frac{\vartheta\genfrac[]{0pt}{}{|\alpha-1/2|}{\beta}(\theta^j_{ab})}{\vartheta\genfrac[]{0pt}{}{0}{1/2}(\theta^j_{ab})} \Bigg] \nonumber .
\end{eqnarray}

The massless open string spectrum can be read off from the partition functions above after transforming them into loop channel. In case 1 one finds a vector multiplet in the adjoint representation of $U(N_a)$. Case 2 yields a hypermultiplet in the bifundamental representation of $U(N_a)\times U(N_b)$, whereas there is no massless matter in case 3. Finally, in case 4, there are $|\Upsilon_{ab}|$, $\Upsilon_{ab}=(\prod_{i=1}^3 I^i_{ab})/4+\sum_{i=1}^3 I^i_{ab} \sigma^i_{ab}$, chiral multiplets in the bifundamental representation of $U(N_a)\times U(N_b)$.
\section{An orbifold compactification of the type I string}
\label{examplestypeImodel}
Constructions of orbifold models based on the type I string are rather similar to those based on orientifolds of type IIA strings, which were discussed in the previous section. The underlying closed string theory is however the type IIB theory and the orientifold projection acts only on the worldsheet and not in spacetime. The projector is the worldsheet parity operator $\Omega$. In the case of orbifold compactifications one nevertheless has to deal with operators acting both on the worldsheet and in spacetime when combining the orientifold projector with elements of the orbifold group.

The orbifold discussed in this section \cite{Camara:2007dy} is again an orbifold of a six-torus that splits into a direct product of three two-tori. The orbifold group is once more $\mathbb{Z}_2\times\mathbb{Z}_2$, but in this case its elements not only invert two tori but also shift some coordinates by half a lattice vector. More precisely, the three non-trivial elements $\Theta$, $\Theta'$ and $\Theta''$ of $\mathbb{Z}_2\times\mathbb{Z}_2$ act on the coordinates $x^a$, $a\in\{1,...,6\}$, of the torus as follows:
\begin{eqnarray}
 (x^1,x^2,x^3,x^4,x^5,x^6) &\overset{\Theta}{\rightarrow}& (x^1+1/2,x^2,-x^3,-x^4,-x^5+1/2,-x^6) \quad \nonumber \\
 (x^1,x^2,x^3,x^4,x^5,x^6) &\overset{\Theta'}{\rightarrow}& (-x^1+1/2,-x^2,x^3+1/2,x^4,-x^5,-x^6) \quad \nonumber \\
 (x^1,x^2,x^3,x^4,x^5,x^6) &\overset{\Theta''}{\rightarrow}& (-x^1,-x^2,-x^3+1/2,-x^4,x^5+1/2,x^6)
\end{eqnarray}
The massless spectrum of this model contains Kaehler and complex structure moduli parameterising the size and shape of the tori, too. In this case, the tilts of the two-tori are continuous parameters such that both the real and the imaginary part of the complex structure moduli $U^{(i)}$ are NSNS sector fields and describe the shape of the torus. The sizes of the tori are again given by the imaginary parts of the Kaehler moduli $T^{(i)}$. Their real parts are the RR two-form, which is part of the massless spectrum of the ten-dimensional Type I theory, integrated over the two-tori. There is another modulus, denoted $S$, that will be important in the following. Its imaginary and real parts are the dilaton and the universal axion, the latter being the Hodge dual of the four-dimensional RR 2-form.

The torus partition function can be determined to be \cite{Angelantonj:2002ct,Camara:2007dy}
\begin{eqnarray}
 T&=& \frac{1}{4} \int_\mathcal{F} \frac{d^2\tau}{\tau_2^3} \Bigg[
      \frac{|\vartheta_3^4-\vartheta_4^4-\vartheta_2^4-\vartheta_1^4|^2}{4|\eta|^{24}} \prod_{i=1}^3
      \Lambda_i \genfrac[]{0pt}{}{0}{0} 
      \nonumber \\ &&
      +\frac{4|\vartheta_3^2\vartheta_4^2-\vartheta_4^2\vartheta_3^2+\vartheta_2^2\vartheta_1^2
      +\vartheta_1^2\vartheta_2^2|^2}{|\eta|^{12}|\vartheta_2|^4}
      \left(\sum_{i=1}^3 \Lambda_i \genfrac[]{0pt}{}{0}{1/2} \right)
      \nonumber \\ &&
      +\frac{4|\vartheta_3^2\vartheta_2^2-\vartheta_4^2\vartheta_1^2-\vartheta_2^2\vartheta_3^2
     -\vartheta_1^2\vartheta_4^2|^2}{|\eta|^{12}|\vartheta_4|^4} \left(\sum_{i=1}^3
     \Lambda_i \genfrac[]{0pt}{}{1/2}{0} \right)
      \nonumber \\ &&
      +\frac{4|\vartheta_3^2\vartheta_1^2-\vartheta_4^2\vartheta_2^2+\vartheta_2^2\vartheta_4^2
     +\vartheta_1^2\vartheta_3^2|^2}{|\eta|^{12}|\vartheta_3|^4}
     \left(\sum_{i=1}^3 \Lambda_i  \genfrac[]{0pt}{}{1/2}{1/2} \right)
     \Bigg]
\end{eqnarray}
where the lattice sums are given by
\begin{eqnarray}
 \Lambda_i \genfrac[]{0pt}{}{\alpha}{\beta}
 &=&\frac{{\rm Im} (T^{(i)})}{\tau_2} \sum_{w_1,w_2,l_1,l_2}
     \exp \Bigg[ 2\pi i T^{(i)} det(A) \nonumber \\  && -\frac{\pi {\rm Im}(T^{(i)})}{\tau_2 {\rm Im}(U^{(i)})}
     \left|\left(1,U^{(i)}\right) A
     \left( \begin{matrix} \tau \\ -1 \end{matrix} \right) \right|^2
     \Bigg] ,
\end{eqnarray}
with the matrix of winding and (Poisson resummed) momentum modes
\begin{eqnarray}
 A = \left( \begin{matrix} w_1+\alpha && l_1 +\beta \\ w_2 && l_2 \end{matrix} \right) .
\end{eqnarray}
The Klein bottle amplitude is
\begin{eqnarray}
 K &=& \int_0^\infty \frac{dt}{8 t^3} \frac{\vartheta_3^4-\vartheta_4^4-\vartheta_2^4-\vartheta_1^4}{\eta^{12}}
 \times \nonumber \\ &&
 \left( \prod_{i=1}^3 \Lambda^M_i [0,0,0] +
  \sum_{i=1}^3 \Lambda^M_i [1/2,0,0] \prod_{j=1,\neq i}^3 \Lambda^W_j [0] \right)
\end{eqnarray}
with the momentum respectively winding sums given by
\begin{eqnarray}
 \Lambda^M_i [\alpha,\beta,\gamma] &=& \sum_{m_1,m_2} \exp \Bigg[ 2\pi i \alpha (m_1+\beta) \nonumber \\ &&
 - \frac{\pi t}{{\rm Im}(U^{(i)}) {\rm Im}(T^{(i)})} | (m_2+\gamma) + U^{(i)} (m_1+\beta)|^2 \Bigg] \label{typeImomentumsum} \\
 \Lambda^W_i [\alpha] &=& \sum_{w_1,w_2} \exp \left[ 
  - \frac{\pi t {\rm Im}(T^{(i)})}{{\rm Im}(U^{(i)})} | (w_1 + \alpha)+ U^{(i)} w_2|^2 \right] . \label{typeIwindingsum}
\end{eqnarray}
Finally, the annulus and Moebius strip diagrams are:
\begin{eqnarray}
 A &=& \frac{(32)^2}{8} \int_0^\infty \frac{dt}{t^3} \Bigg[
        \frac{\vartheta_3^4-\vartheta_4^4-\vartheta_2^4-\vartheta_1^4}{\eta^{12}} 
        \prod_{i=1}^3 \Lambda^M_i [0,0,0] \nonumber \\ && + 4
        \frac{\vartheta_3^2\vartheta_4^2-\vartheta_4^2\vartheta_3^2+\vartheta_2^2\vartheta_1^2
      +\vartheta_1^2\vartheta_2^2}{\eta^6\vartheta_2^2}
      \left( \sum_{i=1}^3 \Lambda^M_i[1/2,0,0] \right) \Bigg] \\
 M &=& -\frac{32}{8} \int_0^\infty \frac{dt}{t^3} \Bigg[
        \frac{\vartheta_3^4-\vartheta_4^4-\vartheta_2^4-\vartheta_1^4}{\eta^{12}} 
        \prod_{i=1}^3 \Lambda^M_i [0,0,0] \nonumber \\ && + 4
        \frac{\vartheta_3^2\vartheta_4^2-\vartheta_4^2\vartheta_3^2+\vartheta_2^2\vartheta_1^2
      +\vartheta_1^2\vartheta_2^2}{\eta^6\vartheta_2^2}
      \left( \sum_{i=1}^3 \Lambda^M_i[1/2,0,0] \right) \Bigg]
\end{eqnarray}
In the Klein bottle amplitude the argument of the $\vartheta$/$\eta$-functions is $2it$, in the annulus amplitude it is $it$ and in the Moebius strip amplitude $it+1/2$.

The partition functions given above allow one to extract the string spectrum. At the massless level one finds the gravity multiplet, vector multiplets transforming in the adjoint representation of $SO(32)$ and the aforementioned moduli chiral multiplets. The low energy effective theory is therefore a pure $SO(32)$ gauge theory coupled to supergravity and contains in addition seven neutral chiral multiplets.

As the model described in this section is based on a freely acting orbifold, one expects, according to the adiabatic argument \cite{Vafa:1995gm}, that it should have an S-dual heterotic description. It is indeed possible to find this dual model, but there is a subtlety \cite{Camara:2007dy}. In order to preserve the full $SO(32)$ gauge group, the orbifold generators must act trivially on the left-moving fermions of the heterotic string. Choosing the same action of the orbifold group on the six torus coordinates as in the Type I case would not lead to a modular invariant partition function. The way out is to replace the purely geometric orbifold action of the Type I case with a non-geometric one \cite{Camara:2007dy}. More precisely, instead of the shift
\begin{eqnarray}
 X \rightarrow X + \pi R
\end{eqnarray}
one takes the asymmetric shift \cite{Camara:2007dy}
\begin{eqnarray}
 X_L \rightarrow X_L + \frac{\pi R}{2} + \frac{\pi \alpha'}{2R} \qquad
 X_R \rightarrow X_R + \frac{\pi R}{2} - \frac{\pi \alpha'}{2R} .
\end{eqnarray}
The modular invariant torus partition function of the dual heterotic model can then be determined to be \cite{Camara:2007dy}
\begin{eqnarray}
 \label{heteroticpartitionfunction}
 T &=& \int_\mathcal{F} \frac{d^2\tau}{4\tau_2^3} \Bigg[
      \frac{{\vartheta^*_3}^4-{\vartheta^*_4}^4-{\vartheta^*_2}^4-{\vartheta^*_1}^4}{|\eta|^{16} {\eta^*}^4}
      \prod_{i=1}^3 \Lambda^a_i \genfrac[]{0pt}{}{0}{0} \\ &&
      + 16 \frac{{\vartheta^*_3}^2{\vartheta^*_4}^2-{\vartheta^*_4}^2{\vartheta^*_3}^2
      +{\vartheta^*_2}^2{\vartheta^*_1}^2
      +{\vartheta^*_1}^2{\vartheta^*_2}^2}{|\eta\vartheta_2|^4{\eta^*}^4}
      \sum_{i=1}^3 \Lambda^a_i \genfrac[]{0pt}{}{0}{1/2} \nonumber \\ &&
      + 16 \frac{{\vartheta^*_3}^2{\vartheta^*_2}^2-{\vartheta^*_4}^2{\vartheta^*_1}^2
      -{\vartheta^*_2}^2{\vartheta^*_3}^2
      -{\vartheta^*_1}^2{\vartheta^*_4}^2}{|\eta\vartheta_4|^4{\eta^*}^4}
      \sum_{i=1}^3 \Lambda^a_i \genfrac[]{0pt}{}{1/2}{0}
      \nonumber \\ &&
      - 16 \frac{{\vartheta^*_3}^2{\vartheta^*_1}^2-{\vartheta^*_4}^2{\vartheta^*_2}^2
      +{\vartheta^*_2}^2{\vartheta^*_4}^2
      +{\vartheta^*_1}^2{\vartheta^*_3}^2}{|\eta\vartheta_3|^4{\eta^*}^4}
      \sum_{i=1}^3 \Lambda^a_i \genfrac[]{0pt}{}{1/2}{1/2} \Bigg]
      \times \sum_{a=1}^4 \frac{\vartheta_a^{16}}{2\eta^{16}} \nonumber \ ,
\end{eqnarray}
where the lattice sum for the asymmetric shift orbifold is
\begin{eqnarray}
 \Lambda^a_i \genfrac[]{0pt}{}{\alpha}{\beta}
 &=&\frac{{\rm Im} (T^{(i)})}{\tau_2} \sum_{w_1,w_2,l_1,l_2}
     \exp \Bigg[ 2 \pi i (\alpha l_1+\beta w_1)
     + 2\pi i T^{(i)} det(A) \nonumber \\  && -\frac{\pi {\rm Im}(T^{(i)})}{\tau_2 {\rm Im}(U^{(i)})}
     \left|\left(1,U^{(i)}\right) A
     \left( \begin{matrix} \tau \\ -1 \end{matrix} \right) \right|^2
     \Bigg] ,
 \label{latticesumasymmetric}
\end{eqnarray}
with the matrix of winding and (Poisson resummed) momentum modes
\begin{eqnarray}
 A = \left( \begin{matrix} w_1+\alpha && l_1 +\beta \\ w_2 && l_2 \end{matrix} \right) .
 \label{matrixwindingmomentummodes}
\end{eqnarray}
The massless spectrum of the heterotic model can be extracted from the partition function \eqref{heteroticpartitionfunction}. As it must be, it is identical to the one found in the Type I description.
\section{On models based on abstract CFTs}
\label{abstractCFTs}
A string compactification to four dimensions can be defined by a tensor product of three CFTs. One of them is a CFT describing the propagation of a superstring in four-dimensional Minkowski spacetime (It is clearly also possible to consider other backgrounds, but will not be done here.), i.e. four free bosons $X^\mu$ plus four free fermions $\psi^\mu$, $\mu\in\{0,1,2,3\}$. Another one is the CFT of the reparameterisation ghosts and superghosts. Finally, one needs a CFT of appropriate central charge to cancel the Weyl anomaly in the Polyakov path integral. This latter CFT will henceforth be called "internal".

In the following, only models preserving spacetime supersymmetry will be considered. This amounts to requiring the internal CFT to have extended (worldsheet) supersymmetry. The extended superconformal algebra in two dimensions has two Cartan generators. This implies that the states in the CFT are not only labelled by their conformal weights $h$, but also by their $U(1)_R$ charge $q$. The corresponding operators will be denoted $\mathcal{O}^h_q$. Selection rules follow from $U(1)_R$ charge conservation. The most prominent examples of such models are the Gepner models \cite{Gepner:1987vz}, which are based on the discrete series of minimal models of the minimally extended superconformal algebra.

In order to have open strings, one has to introduce boundary states in these models. If one wants a globally consistent one, one also has to perform an orientifold projection to achieve tadpole cancellation. There are a number of further consistency conditions to be satisfied.

The annulus partition functions are a sum over the spin structures of the worldsheet fermions. For each spin structure one has to multiply the amplitudes of the different CFTs. The free boson/fermion and ghost CFTs together yield the universal factor $\vartheta\genfrac[]{0pt}{}{\alpha}{\beta}(0)/\eta^3$. The amplitude in the internal CFT depends on the form of the boundary states representing the brane stacks $a$ and $b$ and will be denoted $A^{int}_{ab}\genfrac[]{0pt}{}{\alpha}{\beta}$ such that the full amplitude is
\begin{eqnarray}
 A_{ab} = \int_0^\infty dl \sum_{\alpha,\beta} (-1)^{2(\alpha+\beta)}
 \frac{\vartheta\genfrac[]{0pt}{}{\alpha}{\beta}(0,2il)}{\eta^3(2il)}
 A^{int}_{ab}\genfrac[]{0pt}{}{\alpha}{\beta}(2il) .
 \label{generalannuluspartitionfunction}
\end{eqnarray}
Similarly, the Moebius strip amplitude can be written as follows:
\begin{eqnarray}
 M_{a} = \int_0^\infty dl \sum_{\alpha,\beta} (-1)^{2(\alpha+\beta)}
 \frac{\vartheta\genfrac[]{0pt}{}{\alpha}{\beta}(0,2il+\frac{1}{2})}{\eta^3(2il+\frac{1}{2})}
 M^{int}_{a}\genfrac[]{0pt}{}{\alpha}{\beta}(2il+\frac{1}{2})
 \label{generalmoebiuspartitionfunction}
\end{eqnarray}

The vertex operators for a number of massless open string states relevant in the following will now be described. Firstly, there are gauge bosons. They arise universally with every boundary state, or, in other words, D-brane. This universality is reflected in the fact that the vertex operator
\begin{eqnarray}
 V_{gauge\ boson} = e^{-\phi(z)} \psi^\mu(z) e^{i k_\mu X^\mu(z)}
\end{eqnarray}
acts trivially in the internal CFT. $\phi(z)$ is a field arising upon bosonisation of the superghosts and $k_\mu$ is the four-dimensional momentum of the gauge boson. In a supersymmetric theory, the gauge boson will have a gaugino as its superpartner, whose vertex operators (one for each helicity) are
\begin{eqnarray}
 V_{gaugino^+} &=&  e^{-\phi(z)/2} S^{\dot{\alpha}}(z) e^{i k_\mu X^\mu(z)} \mathcal{O}^{3/8}_{3/2}(z)
 \nonumber \\
 V_{gaugino^-} &=&  e^{-\phi(z)/2} S^{\alpha}(z) e^{i k_\mu X^\mu(z)} \mathcal{O}^{3/8}_{-3/2}(z) ,
 \label{gauginovertexoperator}
\end{eqnarray}
where $S^{\dot{\alpha}}$ and $S^{\alpha}$ are spin fields in the free fermion CFT of the $\psi^\mu$'s. The operators $\mathcal{O}^{3/8}_{\pm3/2}(z)$ are the spectral flow operators of the internal CFT, which must exist if some supersymmetry in four dimensions is to be preserved. The gauge bosons and gauginos transform in the adjoint representation of the gauge group. Depending on the internal CFT, there can also be chiral superfields transforming in the adjoint representation of the gauge group. In terms of D-branes these fields are moduli related to the brane position or Wilson line moduli. The vertex operators for these fields and their fermionic superpartners, called modulini, are
\begin{eqnarray}
 V_{modulus} &=& e^{-\phi(z)}  e^{i k_\mu X^\mu(z)} \mathcal{O}^{1/2}_{\pm1}(z) \nonumber \\
 V_{modulino^+} &=& e^{-\phi(z)/2} S^{\dot{\alpha}}(z) e^{i k_\mu X^\mu(z)} \mathcal{O}^{3/8}_{1/2}(z)
 \nonumber \\
 V_{modulino^-} &=& e^{-\phi(z)/2} S^{\alpha}(z) e^{i k_\mu X^\mu(z)} \mathcal{O}^{3/8}_{-1/2}(z) .
 \label{chiralsuperfieldvertexoperators}
\end{eqnarray}
In order to determine the number of massless chiral supermultiplets transforming in the bifundamental representation of the gauge group $U(N_a)\times U(N_b)$, one has to compute the overlap of two distinct boundary states and modularly transform it such that the resulting expression can be interpreted as an open string partition function. The vertex operators for these fields are boundary changing operators. They "change" the boundary conditions for the world-sheet fields from those describing one brane to those describing the other brane and take the form \eqref{chiralsuperfieldvertexoperators}. Finally, there can be massless states transforming in the symmetric or antisymmetric representation of $U(N_a)$. Their number can be obtained from the overlaps of a boundary state and its orientifold image, or the crosscap state, respectively. They are chiral multiplets and their vertex operators are boundary changing operators and look like \eqref{chiralsuperfieldvertexoperators}.

Of course, the form of the annulus partition functions and vertex operators given here is correct for all the models described in this chapter. For the toroidal models, the operators $\mathcal{O}^h_q$ can be written down explicitly \cite{Cvetic:2003ch,Abel:2003yx,Abel:2003vv,Lust:2004cx,Bertolini:2005qh}.
\chapter{The gauge coupling at one loop}
\label{gaugecouplingoneloop}
The formulas for the gauge kinetic function/gauge coupling given in the previous chapter are tree-level expressions. This chapter is concerned with one-loop corrections to these quantities \cite{Kaplunovsky:1987rp,Dixon:1990pc,Kaplunovsky:1992vs,Mayr:1993mq,Berg:2004ek,Bianchi:2005sa,Anastasopoulos:2006hn}.
\section{Computing gauge threshold corrections}
There are (at least) two ways to compute one-loop corrections to the gauge coupling on a stack of D-branes.  One method consists in computing correlation functions of two gauge boson vertex operators on annulus and Moebius strip diagrams. The other one, which is used here, is the background field method \cite{Bachas:1996zt,Antoniadis:1999ge,Lust:2003ky}. It amounts to determining the one-loop partition function in the background of a magnetic field $B$ in the four-dimensional spacetime, expanding it in a series in $B$ and extracting the quadratic term. In order to compute corrections to the gauge coupling, which is associated with a $CP$-even term in the Lagrangian, one has to take only the even spin structures into account. Clearly, when computing the corrections to the gauge coupling on some brane stack $a$, one has to sum over all annulus diagrams with one boundary on stack $a$ and the other on any brane and take the Moebius strip diagram with the boundary on stack $a$ into account, too.

The one-loop partition functions in the background of a magnetic field can be determined from the (usual) partition functions. To do so, one has to replace the universal factors in \eqref{generalannuluspartitionfunction} and \eqref{generalmoebiuspartitionfunction} as follows \cite{Lust:2003ky}:
\begin{eqnarray}
 \frac{\vartheta\genfrac[]{0pt}{}{\alpha}{\beta}(0,2il)}{\eta^3(2il)} &\rightarrow& 2 i \pi B q_a
 \frac{\vartheta\genfrac[]{0pt}{}{\alpha}{\beta}(-\epsilon_a,2il)}{\vartheta\genfrac[]{0pt}{}{1/2}{1/2}(-\epsilon_a,2il)}
 \\
  \frac{\vartheta\genfrac[]{0pt}{}{\alpha}{\beta}(0,2il+\frac{1}{2})}{\eta^3(2il+\frac{1}{2})} &\rightarrow& 2 i \pi B q_a
 \frac{\vartheta\genfrac[]{0pt}{}{\alpha}{\beta}(-\epsilon_a/2,2il+\frac{1}{2})}{\vartheta\genfrac[]{0pt}{}{1/2}{1/2}(-\epsilon_a/2,2il+\frac{1}{2})}
\end{eqnarray}
Here, $q_a$ is the charge of the open string ending on brane $a$ and $\pi \epsilon_a=\arctan(\pi q_a B)$. Expanding the above expressions in powers of $B$, one finds that the quadratic terms are multiplied by
\begin{eqnarray}
 &&
 \frac{-i \pi q_a^2}{\vartheta'_1(0,2il)} \left( \vartheta''\genfrac[]{0pt}{}{\alpha}{\beta}(0,2il) + 
 \vartheta\genfrac[]{0pt}{}{\alpha}{\beta}(0,2il) (\frac{2\pi^2}{3}-\frac{\vartheta'''_1}{3\vartheta'_1}) \right) ,
 \\ \nonumber &&
  \frac{-i \pi q_a^2}{2\vartheta'_1(0,2il+\frac{1}{2})}
  \left( \vartheta''\genfrac[]{0pt}{}{\alpha}{\beta}(0,2il+\frac{1}{2}) + 
 \vartheta\genfrac[]{0pt}{}{\alpha}{\beta}(0,2il+\frac{1}{2}) (\frac{4\pi^2}{3}-\frac{\vartheta'''_1}{6\vartheta'_1}) \right) .
\end{eqnarray}
One now has to put this together with the rest of the partition functions \eqref{generalannuluspartitionfunction}, \eqref{generalmoebiuspartitionfunction} and use that the (usual) partition functions vanish in the supersymmetric case. (Other cases will not be discussed here.) One is then left with the following rather general formula for the one-loop correction to the gauge coupling on brane stack $a$ induced by brane stack $b$:
\begin{eqnarray}
 \left(g_{ab}^{1-loop}\right)^{-2} = \int_0^\infty dl \sum_{\alpha,\beta} (-1)^{2(\alpha+\beta)}
 \frac{\vartheta''\genfrac[]{0pt}{}{\alpha}{\beta}(0,2il)}{\eta^3(2il)}
 A^{int}_{ab}\genfrac[]{0pt}{}{\alpha}{\beta}(2il)
 \label{generalformulathresholdcorrections}
\end{eqnarray}
Analogously, the general form of the Moebius strip diagram is
\begin{eqnarray}
 \left(g_{aO}^{1-loop}\right)^{-2} = \int_0^\infty dl \sum_{\alpha,\beta} (-1)^{2(\alpha+\beta)}
 \frac{\vartheta''\genfrac[]{0pt}{}{\alpha}{\beta}(0,2il+\frac{1}{2})}{\eta^3(2il+\frac{1}{2})}
 M^{int}_{a}\genfrac[]{0pt}{}{\alpha}{\beta}(2il+\frac{1}{2})
 \label{generalformulathresholdcorrections2} .
\end{eqnarray}
In these expressions it is understood that the sums run only over the even spin structures, i.e. $\alpha,\beta\in\{0,1/2\}$, $(\alpha,\beta)\neq(1/2,1/2)$. The integrals in \eqref{generalformulathresholdcorrections} and \eqref{generalformulathresholdcorrections2} are in general divergent both for small and large $l$. The divergence at large $l$ cancels in a globally consistent model when summing over all branes $b$, taking the Moebius strip diagram into account and using the tadpole cancellation condition. The divergence for small $l$ is due to massless open string modes. As 
the latter are dynamical degrees of freedom in the low-energy effective field theory, their effects should not be included in the gauge threshold corrections and should therefore be removed from \eqref{generalformulathresholdcorrections} and \eqref{generalformulathresholdcorrections2}. These massless modes lead to the running of the gauge coupling $g_a$, i.e. its dependence on the renormalisation scale $\mu$, in the field theory. One might therefore just as well replace the divergence for small $l$ in the formula for the one-loop corrected running gauge coupling $g_a(\mu)$ by the term $b_a \ln M_s^2/\mu^2$, which determines the scale dependence. Here, $b_a$ is the beta-function coefficient of the gauge theory on brane stack $a$ and $M_s$ is the string scale, the scale below which the low energy theory is defined. It is however important to stress that the divergence for small $l$ is an infrared divergence that also appears in the low energy field theory and that the ultraviolet divergence, which leads to the running of the gauge coupling, is absent in string theory. Note that the concept of a gauge coupling only exists in the low energy field theory.

In a more careful treatment, one would compute a correlation function of two, three or four gauge bosons at one loop both in string theory and in the low energy field theory. The field theory correlator would be both infrared and ultraviolet divergent, the string theory correlator only infrared divergent. One would then absorb the ultraviolet divergence in the field theory expression into a renormalised, scale-dependent gauge coupling. Finally, one would equate the field and string theory results and drop the infrared divergence, that must be the same on both sides. The resulting equation would define the renormalised gauge coupling at the string scale, which must be used in the low energy effective field theory that is to reproduce the full string theory at low energies.
\subsection{An orbifold model with bulk D6-branes}
\label{gaugethresholdsbulk}
The aforementioned formulas will now be applied to intersecting D6-brane models on the $\mathbb{Z}_2\times\mathbb{Z}_2$ toroidal orbifold with $h_{21}=3$, which was discussed in section \ref{examplesbulk}. By comparing the general formula \eqref{generalannuluspartitionfunction} with the partition functions \eqref{partitionfunctionbulk1}, \eqref{partitionfunctionbulk2} and \eqref{partitionfunctionbulk3} one can extract the internal partition function $A^{int}_{ab}\genfrac[]{0pt}{}{\alpha}{\beta}$, which can then be used in \eqref{generalformulathresholdcorrections} to find the following expressions for the gauge threshold corrections \cite{Lust:2003ky,Akerblom:2007np}.

{\bf Case 1:}
\begin{eqnarray}
 \left(g_{aa}^{(1)}\right)^{-2} &=& N_a^2 \int_0^\infty dl \sum_{\alpha,\beta} (-1)^{2(\alpha+\beta)}
        \frac{\vartheta''\genfrac[]{0pt}{}{\alpha}{\beta}(0,2il)}{\eta^3(2il)}
        \frac{\vartheta^3\genfrac[]{0pt}{}{\alpha}{\beta}(0,2il)}{\eta^9(2il)}
        \prod_i \frac{(L^i_a)^2 Z^i_a}{R_1^{(i)} R_2^{(i)}}
        \nonumber \\ &=& 0
 \label{thresholdcorrectionsbulk1}
\end{eqnarray}
A theta function identity implies that the gauge threshold corrections in such a sector vanish. This was to be expected as the sector preserves sixteen supercharges.

{\bf Case 2:}
\begin{eqnarray}
 \left(g_{ab}^{(2)}\right)^{-2} &=& N_a N_b I_{ab} \int_0^\infty dl \sum_{\alpha,\beta} (-1)^{2(\alpha+\beta)}
 \frac{\vartheta''\genfrac[]{0pt}{}{\alpha}{\beta}(0,2il)}{\eta^3(2il)}
 \frac{\vartheta\genfrac[]{0pt}{}{\alpha}{\beta}(0)}{\eta^3} \frac{(L^1_a)^2 Z^1}{R_1^{(1)} R_2^{(1)}} 
 \nonumber \\ &&
 \prod_{j=2}^3
 \frac{\vartheta\genfrac[]{0pt}{}{\alpha}{\beta}(\theta^j_{ab})}{\vartheta\genfrac[]{0pt}{}{1/2}{1/2}(\theta^j_{ab})}
 = (2\pi)^2 N_a N_b I_{ab} \int_0^\infty dl
 \frac{(L^1_a)^2 Z^1}{R_1^{(1)} R_2^{(1)}}
 \nonumber \\ &=&
 (2\pi)^2 N_a N_b I_{ab} \Bigg[
      \int_0^\infty dl \frac{(L^1_a)^2}{R_1^{(1)} R_2^{(1)}} + \ln \frac{M_s^2}{\mu^2} - \ln \left( (L^1_a)^2 \right)
       \nonumber \\ && - 4 {\rm Im}( i \ln \eta ( T^{(1)} ) )- \ln(4\pi)
 \Bigg]
 \label{thresholdcorrectionsbulk2}
\end{eqnarray}
In the first step a theta function identity has been used and in the second one the divergence for $l\rightarrow0$ has been replaced by $\ln M_s^2/\mu^2$, as explained previously. This replacement will be made in various expressions in the following.

{\bf Case 3:}
\begin{eqnarray}
 \left(g_{ab}^{(3)}\right)^{-2} &=& N_a N_b I_{ab} \int_0^\infty dl \sum_{\alpha,\beta} (-1)^{2(\alpha+\beta)} 
 \frac{\vartheta''\genfrac[]{0pt}{}{\alpha}{\beta}(0)}{\eta^3} \prod_{i=1}^3
 \frac{\vartheta\genfrac[]{0pt}{}{\alpha}{\beta}(\theta^i_{ab})}{\vartheta\genfrac[]{0pt}{}{1/2}{1/2}(\theta^i_{ab})}
 \nonumber \\ &=& 4\pi^2 N_a N_b I_{ab} \int_0^\infty dl
 \sum_{i=1}^3 \frac{\vartheta'_1(\theta^i_{ab})}{\vartheta_1(\theta^i_{ab})}
 \nonumber \\ &=& 4\pi^3 N_a N_b I_{ab}
 \sum_{i=1}^3 \cot (\pi \theta^i_{ab}) \int_0^\infty dl 
 \label{thresholdcorrectionsbulk3} \\ &&
 - 2\pi^3 N_a N_b I_{ab} \Bigg[
 \ln \frac{M_s^2}{\mu^2} \sum_{i=1}^3 sign(\theta^i_{ab}) \nonumber \\  &&
 - \ln \prod_{i=1}^3 \left( \frac{\Gamma(|\theta^i_{ab}|)}{\Gamma(1-|\theta^i_{ab}|)} \right)^{sign(\theta^i_{ab})}
  - \sum_{i=1}^3 sign(\theta^i_{ab}) (\ln2-\gamma) \Bigg] \nonumber
\end{eqnarray}
Again, a theta function identity has been used in the first step.

Computing the Moebius strip diagrams yields results rather similar to those just obtained from the annulus diagrams. For case 1 of the Moebius strip diagrams of section \ref{examplesbulk} one finds that the gauge threshold corrections vanish, for case 2 one finds a result similar to \eqref{thresholdcorrectionsbulk2}. Finally, case 3 yields
\begin{eqnarray}
 \left(g_{aO_k}^{(3)}\right)^{-2} &=& N_a I_{aO_k} \int_0^\infty dl \sum_{\alpha,\beta} (-1)^{2(\alpha+\beta)} 
 \frac{\vartheta''\genfrac[]{0pt}{}{\alpha}{\beta}(0)}{\eta^3} \prod_{i=1}^3
 \frac{\vartheta\genfrac[]{0pt}{}{\alpha}{\beta}(\theta^i_{aO_k})}{\vartheta\genfrac[]{0pt}{}{1/2}{1/2}(\theta^i_{aO_k})}
 \nonumber \\ &=& 4\pi^2 N_a I_{aO_k} \int_0^\infty dl
 \sum_{i=1}^3 \frac{\vartheta'_1(\theta^i_{aO_k})}{\vartheta_1(\theta^i_{aO_k})}
 \nonumber \\ &=& 4\pi^3 N_a I_{aO_k}
 \sum_{i=1}^3 \cot (\pi \theta^i_{aO_k}) \int_0^\infty dl 
 - \pi^3 N_a I_{aO_k} \nonumber \\ && \Bigg[
 \ln \frac{M_s^2}{\mu^2} \sum_{i=1}^3 sign(\theta^i_{aO_k}) 
 - \ln \prod_{i=1}^3 \left( \frac{\Gamma(2|\theta^i_{aO_k}|)}{\Gamma(1-2|\theta^i_{aO_k}|)} \right)^{sign(\theta^i_{aO_k})} \nonumber \\  &&
  - \sum_{i=1}^3 sign(\theta^i_{aO_k}) (3\ln2-\gamma) \Bigg]
 \label{thresholdcorrectionsbulkmoebius} .
\end{eqnarray}
Note that here the argument of the theta and eta functions is $2il+1/2$ and that the final expression is only valid for a restricted range of the angles. The expression for other values of the angles \cite{Akerblom:2007np} will not be needed in the following.

It remains to show that the prefactors of the divergent terms, i.e. those multiplied by $\int_0^\infty dl$, in the final expressions of \eqref{thresholdcorrectionsbulk2}, \eqref{thresholdcorrectionsbulk3} and \eqref{thresholdcorrectionsbulkmoebius} sum to zero \cite{Lust:2003ky}.
Using the formulas for the one-cycle volumes $L^i_a$, the intersection number $I_{ab}$ and the intersection angles $\theta^i_{ab}$ given in section \ref{D6branemodelsorbifolds} as well as trigonometric identities, one can rewrite
\begin{eqnarray}
 N_a N_b I_{ab}(L_a^1)^2/(R_1^{(1)}R_2^{(1)}) \qquad \textrm{and} \qquad N_a N_b I_{ab}\sum_i \cot(\pi\theta^i_{ab}) ,
\end{eqnarray}
which appear in \eqref{thresholdcorrectionsbulk2}, and \eqref{thresholdcorrectionsbulk3}, respectively, as
\begin{eqnarray}
 N_a N_b \sum_{i\neq j\neq k\neq i} && \frac{R_1^{(i)}}{R_2^{(i)}} \Big( n_a^i \widetilde{m}_a^j \widetilde{m}_a^k n_b^i n_b^j n_b^k
 - n_a^i n_a^j \widetilde{m}_a^k n_b^i \widetilde{m}_b^j n_b^k \nonumber \\ &&
 - n_a^i \widetilde{m}_a^j n_a^k n_b^i n_b^j \widetilde{m}_b^k
 + n_a^i n_a^j n_a^k n_b^i \widetilde{m}_b^j \widetilde{m}_b^k \Big)
 \nonumber \\ &&
 + \frac{R_2^{(i)}}{R_1^{(i)}} \Big( \widetilde{m}_a^i \widetilde{m}_a^j \widetilde{m}_a^k \widetilde{m}_b^i n_b^j n_b^k
 - \widetilde{m}_a^i n_a^j \widetilde{m}_a^k \widetilde{m}_b^i \widetilde{m}_b^j n_b^k \nonumber \\ &&
 - \widetilde{m}_a^i \widetilde{m}_a^j n_a^k \widetilde{m}_b^i n_b^j \widetilde{m}_b^k
 + \widetilde{m}_a^i n_a^j n_a^k \widetilde{m}_b^i \widetilde{m}_b^j \widetilde{m}_b^k \Big) .
 \label{annulusdivergence}
\end{eqnarray}
Similarly, the prefactor of the divergent term in \eqref{thresholdcorrectionsbulkmoebius} can be cast into
\begin{eqnarray}
  N_a \sum_{i\neq j\neq k\neq i} && \frac{R_1^{(i)}}{R_2^{(i)}} \Big( 8 n_a^i \widetilde{m}_a^j \widetilde{m}_a^k
 + 2^{3-2\beta^j-2\beta^k} n_a^i n_a^j n_a^k \Big)
 \nonumber \\ &&
 - \frac{R_2^{(i)}}{R_1^{(i)}} \Big( 2^{3-2\beta^i-2\beta^j} \widetilde{m}_a^i n_a^j \widetilde{m}_a^k
 + 2^{3-2\beta^i-2\beta^k} \widetilde{m}_a^i \widetilde{m}_a^j n_a^k \Big) ,
 \label{moebiusdivergence}
\end{eqnarray}
where a sum over all four orientifold planes has already been performed. By adding the contribution \eqref{annulusdivergence} of all brane stacks $b$, their orientifold images as well as the orientifold image of stack $a$, one finds that, using the tadpole cancellation conditions \eqref{tadpolecancellationconditionexamplebulk}, the divergences from the annulus diagrams \eqref{thresholdcorrectionsbulk2}, \eqref{thresholdcorrectionsbulk3} cancel those from the Moebius strip diagrams \eqref{thresholdcorrectionsbulkmoebius}.

In order to determine the full one-loop threshold corrections to the gauge coupling on some brane stack $a$ one has to sum the finite parts of \eqref{thresholdcorrectionsbulk2}, or \eqref{thresholdcorrectionsbulk3}, respectively, over all brane stacks and take the contribution \eqref{thresholdcorrectionsbulkmoebius} from the Moebius strip diagrams into account as well.
\subsection{An orbifold model with fractionally charged D6-branes}
\label{gaugethresholdsfractional}
The gauge threshold corrections in models on the orbifold described in section \ref{examplesfractional} can be computed rather similarly to those discussed in the previous section \cite{Blumenhagen:2007ip}. The results for the annulus diagrams are given in appendix \ref{appendixgaugethresholdcorrectionsfractional}. The Moebius strip diagrams are equal (up to some signs) to those of the previous section. In contradistinction to the annulus diagrams there are no new contributions, as the orientifold planes do not carry fractional charges.

One important difference to the case discussed in the previous section is that there are tadpole cancellation conditions in the twisted sectors in addition to those in the untwisted sectors. The expressions for the divergences arising in the untwisted sectors are nearly identical to those of the previous section, some signs are different. The prefactor of the divergent integral in \eqref{appendixAuntwtadpole} is proportional to that in the final expression of \eqref{thresholdcorrectionsbulk3} and can be rewritten as in \eqref{annulusdivergence}. Therefore, the divergences arising in the untwisted sectors can be shown to cancel analogously to those discussed in the previous section. It remains to be shown that the divergences  \eqref{appendixAtwtadpole1}, \eqref{appendixAtwtadpole2}, \eqref{appendixAtwtadpole3} and \eqref{appendixAtwtadpole4} cancel when summing over all branes \cite{Blumenhagen:2007ip}. With the help of the formulas for the one-cycle volumes $L^i_a$ and the intersection angles $\theta^i_{ab}$ these four terms can all be cast into
\begin{eqnarray}
 8 \pi N_a N_b \int_0^\infty dl \sum_{i=1}^3 \sum_{k,l=1}^4 \epsilon^i_{a,kl} \epsilon^i_{b,kl}
      \frac{n_a^i n_b^i (R_1^{(i)})^2 + \widetilde{m}_a^i \widetilde{m}_b^i
 (R_2^{(i)})^2}{R_1^{(i)} R_2^{(i)}} .
 \label{twistedtadpolerewritten}
\end{eqnarray}
Summing \eqref{twistedtadpolerewritten} over all branes $b$, their orientifold images and the orientifold image of brane $a$ yields an expression that vanishes when taking the twisted sector tadpole cancellation conditions \eqref{tadpolecancellationconditionsfractionaltwisted} into account.

As before, the full one-loop correction to the gauge coupling is given by summing the finite parts of the expressions given in appendix \ref{appendixgaugethresholdcorrectionsfractional} over all branes.
\subsection{A type I model and its heterotic dual}
\label{gaugethresholdstypeImodel}
This section is concerned with the one-loop (in the string perturbation expansion) corrections to the gauge coupling in the type I model and its heterotic dual which were discussed in section \ref{examplestypeImodel}. The computation in the type I model is rather similar to the computations performed in the preceding sections. It therefore suffices to just state the result \cite{Camara:2007dy}:
\begin{eqnarray}
 \label{gaugethresholdstypeI}
 \left(g_I^{1-loop}\right)^{-2} &\varpropto& \int_0^\infty dl \sum_{i=1}^3 \widetilde{\Lambda}^M_i [1/2] \\ &\varpropto&
          \sum_{i=1}^3 \ln \frac{M_s^2}{\mu^2} - \ln \left({\rm Im}(U^{(i)}){\rm Im}(T^{(i)})\right)
          + 2 \ln \left| \frac{\vartheta_4}{\eta^3}(2U^{(i)}) \right| , \nonumber
\end{eqnarray}
where
\begin{eqnarray}
 \widetilde{\Lambda}^M_i [\alpha] &=& {\rm Im}(T^{(i)}) \sum_{m_1,m_2} \exp \left[
 - \frac{2 \pi l {\rm Im}(T^{(i)})}{{\rm Im}(U^{(i)})} | m_1 + \alpha + U^{(i)} m_2|^2 \right]
\end{eqnarray}
is the Poisson resummed form of the momentum sum $\Lambda^M_i$ given in \eqref{typeImomentumsum}.

The computation of the gauge threshold corrections in the heterotic model requires some new techniques \cite{Kaplunovsky:1987rp,Kaplunovsky:1992vs} that will not be explained here. Similar to the case of open string models, one can write down a rather general formula for the threshold corrections to the gauge coupling associated with some gauge group factor $G$. It requires one to compute a trace in the Hilbert state of the internal CFT describing the compactification space of a heterotic string model. With $H$ and $\widetilde{H}$ the left- and right-moving worldsheet Hamiltonians, $\widetilde{F}$ the right-moving worldsheet fermion number and $Q$ the charge of a string state under the group $G$ the formula is \cite{Kaplunovsky:1987rp,Kaplunovsky:1992vs}
\begin{eqnarray}
 \left( g_{G}^{1-loop} \right)^{-2} &=& \frac{i}{4\pi} \int_\mathcal{F} \frac{d^2\tau}{\tau_2 |\eta|^2} \sum_{\alpha,\beta}
       \partial_{\tau^*} \left(\frac{\vartheta^*\genfrac[]{0pt}{}{\alpha}{\beta}}{\eta^*}\right) \times \nonumber \\ &&
       Tr_\alpha \left[\left( Q^2-\frac{1}{4\pi\tau_2}\right)(-1)^{\beta \widetilde{F}} q^H {q^*}^{\widetilde{H}} \right] ,
\end{eqnarray}
where, as before, the sum only runs over the even spin structures.

This formula now has to be applied to the heterotic string model described in section \ref{examplestypeImodel}, whose partition function is given in	 \eqref{heteroticpartitionfunction}. One first notes that applying the charge operator $Q^2$ to the partition function $\sum_a \vartheta_a^{16}/\eta^{16}$ of the left-moving current algebra yields $\sum_a \vartheta_a^{16}/\eta^{16} \times \vartheta''_a/\vartheta_a$. Using some theta function identities the gauge threshold corrections for the model under consideration can then be written as \cite{Camara:2007dy,Blumenhagen:2008ji}
\begin{eqnarray}
 \left( g_{h}^{1-loop} \right)^{-2} &=& \int_\mathcal{F} \frac{d^2\tau}{\tau_2}
        \sum_{i=1}^3 \Bigg( \frac{1}{\eta^2\vartheta_2^2} \Lambda_i^a \genfrac[]{0pt}{}{0}{1/2}
                          - \frac{1}{\eta^2\vartheta_4^2} \Lambda_i^a \genfrac[]{0pt}{}{1/2}{0} \nonumber \\ &&
                          - \frac{i}{\eta^2\vartheta_3^2} \Lambda_i^a \genfrac[]{0pt}{}{1/2}{1/2} \Bigg) \times
        \sum_{a=1}^4 \frac{\vartheta_a^{16}}{\eta^{16}} \left( \frac{\vartheta''_a}{\vartheta_a} + \frac{\pi}{\tau_2} \right) .
 \label{gaugethresholdsheteroticraw}
\end{eqnarray}
The next step is to evaluate \eqref{gaugethresholdsheteroticraw}, which can be done as follows. One first notices that when all three summands in the first bracket of \eqref{gaugethresholdsheteroticraw} are taken into account, one effectively sums over all matrices \eqref{matrixwindingmomentummodes} with half integer or integer, but not both integer at the same time, entries in the first row and integer entries in the second row. The prefactors (theta/eta functions) in front of the lattice sums \eqref{latticesumasymmetric} in \eqref{gaugethresholdsheteroticraw} are different, but are transformed into one another by modular transformations. The idea \cite{Dixon:1990pc} is to split the two by two matrices $A$ into $A=BM$, $M\in SL(2,\mathbb{Z})$, and to sum only over a restricted set of matrices $B$, but to therefor integrate over the image of the fundamental domain $\mathcal{F}$ under the action of $M$ on the modular parameter $\tau$. The set of matrices $B$ has to be chosen such that every matrix $A$ is taken into account precisely once when unfolding the integral by the action of $M$ on $\tau$ as described. It turns out that two cases differing in whether the determinant of $A$ (or $B$; They are equal.) is zero or not have to be distinguished.

For the matrices with $\det A=\det B=0$ one can choose the matrices $B$ to take the form \cite{Camara:2007dy}
\begin{eqnarray}
 B = \left( \begin{matrix} 0 && j+1/2 \\ 0 && p \end{matrix} \right) \qquad j,p\in\mathbb{Z} \ \ ,
 \label{matrixrepresentativesdegenerate}
\end{eqnarray}
with the identification $(j,p)\sim(-j-1,p)$. In order to determine the domain of integration one notes that matrices of the form $M_0=\left( \begin{matrix} 1 && m \\ 0 && 1 \end{matrix} \right)$, which are contained in $SL(2,\mathbb{Z})$, do not change the form of the matrix $B$ in \eqref{matrixrepresentativesdegenerate}, i.e. $BM_0=B$. Taking this into account, it turns out that one has to integrate over the double cover of the strip $\{\tau\in\mathbb{C};\tau_2>0,|\tau_1|<1/2\}$. To take care of the double covering and the aforementioned identification one can sum over all $j$ and $p$ and just integrate once over the strip. The integral to be evaluated becomes
\begin{eqnarray}
 && {\rm Im}(T) \int_{-1/2}^{1/2} d\tau_1\int_0^\infty\frac{d\tau_2}{\tau_2^2} \sum_{j,p} \exp
           \left( \frac{-\pi {\rm Im}(T)}{\tau_2 {\rm Im}(U)} |-j-\frac{1}{2}+pU|^2\right) \times \nonumber \\ &&
           \qquad \frac{1}{\eta^2\vartheta_2^2} \sum_a \frac{\vartheta_a^{16}}{\eta^{16}}
           \left( \frac{\vartheta''_a}{\vartheta_a} + \frac{\pi}{\tau_2} \right) .
 \label{gaugethresholdsheteroticdegenerateindermediate}
\end{eqnarray}
The combination of theta/eta functions in the last line of \eqref{gaugethresholdsheteroticdegenerateindermediate} can be written as a double series expansion in powers of $q=\exp(2\pi i \tau)$ and inverse powers of $\tau_2$. The integral over the strip is only non-vanishing for the terms of order $q^0$ \cite{Bachas:1997mc}.

Next, the contributions from terms involving matrices of non-vanishing determinant have to be evaluated. The matrices $B$ can be chosen to be of the form \cite{Blumenhagen:2008ji}
\begin{eqnarray}
 B = \left( \begin{matrix} k && j \\ 0 && p \end{matrix} \right)
 \label{matrixrepresentativesnondegenerate}
\end{eqnarray}
with $2j,2k,p\in\mathbb{Z}$, $0\leq j<k$, but not both $j$ and $k$ integer. In this case there are no matrices $M_0\in SL(2,\mathbb{Z})$ that leave the matrices \eqref{matrixrepresentativesnondegenerate} invariant. The domain of integration therefore has to be the image of the fundamental domain $\mathcal{F}$ under the full group $SL(2,\mathbb{Z})$, which is the double cover of the upper half complex plane. The integrals to be evaluated are
\begin{eqnarray}
 \int_{-\infty}^{\infty} d\tau_1\int_0^\infty\frac{d\tau_2}{\tau_2} \Lambda_i^a \genfrac[]{0pt}{}{\alpha}{\beta} X(q) 
 \label{gaugethresholdsheteroticnondegenerateindermediate}
\end{eqnarray}
with
\begin{eqnarray}
 X(q) =
           \frac{1}{\eta^2\vartheta_\gamma^2} \sum_a \frac{\vartheta_a^{16}}{\eta^{16}}
           \left( \frac{\vartheta''_a}{\vartheta_a} + \frac{\pi}{\tau_2} \right) ,
\end{eqnarray}
where the values of $\alpha$, $\beta$ and $\gamma$ depend on whether $j$ and $k$ are integer or half integer. As before, the integral in \eqref{gaugethresholdsheteroticnondegenerateindermediate} can be performed after writing $X(q)$ as a series in powers of $q$ and inverse powers of $\tau_2$.

After evaluating the integrals \cite{Bachas:1997mc,Blumenhagen:2008ji} and putting everything together, the one-loop gauge threshold corrections for the heterotic orbifold model with gauge group $SO(32)$ become
\begin{eqnarray}
 \left(g_h^{1-loop}\right)^{-2} &\varpropto& \sum_{i=1}^3 \ln \frac{M_s^2}{\mu^2} - \ln \left({\rm Im}(U^{(i)}){\rm Im}(T^{(i)})\right)
          + 2 \ln \left| \frac{\vartheta_4}{\eta^3}(2U^{(i)}) \right| \nonumber \\ &&
          \label{gaugethresholdsheterotic}
          + \sum_{i=1}^3 \frac{c_1}{{\rm Im}(T^{(i)})} E(2,-\frac{1}{2},0,U^{(i)}) \\ \nonumber &&
          + \sum_{i=1}^3 \sum_{p,k,j} \frac{c_2 e^{2\pi i kp T^{(i)}}}{kp}
          \frac{f(j,k)}{\eta^2\vartheta_{g(j,k)}^2} \sum_a \frac{\vartheta_a^{16}}{\eta^{16}}
          \frac{\vartheta''_a}{\vartheta_a} \left(\frac{j+p U^{(i)}}{k}\right) ,
\end{eqnarray}
where $c_1$ and $c_2$ are numerical constants and the sum runs over the ranges of $p,k,j$ given above. The functions $\big(f(j,k),g(j,k)\big)$ take the values $\left((-1)^k,2\right)$ for $k$ integer, $j$ half integer, $\left(-(-1)^j,4\right)$ for $k$ half integer, $j$ integer and $\left(i(-1)^{k+j},3\right)$ for $k$, $j$ both half integer. The double series
\begin{eqnarray}
 E(s,a,b,\tau) = \sum_{m,n} \frac{\tau_2^s}{|m+a+(n+b)\tau|^{2s}}
\end{eqnarray}
can be considered as a generalisation of the non-holomorphic Eisenstein series. In \eqref{gaugethresholdsheterotic} only the terms holomorphic in $U^{(i)}$ and $T^{(i)}$ of the contributions coming from the summands in \eqref{gaugethresholdsheteroticraw} with matrices of non-zero determinant are displayed.

The terms in the first line of \eqref{gaugethresholdsheterotic} precisely match the one-loop gauge threshold corrections in the dual type I model \eqref{gaugethresholdstypeI} and those in the second line correspond to contributions of higher order in the perturbative expansion in the type I model. The terms in the third line are contributions of world-sheet instantons of area $kp T^{(i)}$, hence the factor $e^{2\pi i kp T^{(i)}}$, and correspond to D-instanton corrections in the type I model, which will be discussed in chapter \ref{Dinstantoncorrectionsgaugekineticfunction}.
\section{Holomorphy of the gauge kinetic function}
\label{holomorphygaugekineticfunction}
It was discussed in chapter \ref{fourdimensionaleffectiveactions} that the holomorphy of the superpotential and the gauge kinetic function puts strong constraints on which fields these quantities can depend on and that it is possible to formulate non-renormalisation theorems. Such theorems will now be explicated for the D6-brane models described in sections \ref{D6branemodelsCY} and \ref{D6branemodelsorbifolds} \cite{Akerblom:2007uc}. (Similar theorems hold for orientifolds of the type IIB theory featuring Dp-branes with odd p.) The gauge symmetries associated with the two- and three-form fields $B_2$ and $C_3$ in ten dimensions lead to symmetries of the low energy effective theory under which the real parts of the complex structure moduli $U^{(i)}$ \eqref{complexstructuremoduli} and Kaehler moduli $T^{(i)}$ \eqref{kaehlermoduli} transform by shifts.
These symmetries are only broken by instantons. More precisely, worldsheet instantons, whose action can be written as a linear combination of the Kaehler moduli, break the symmetry under which the latter shift. The action of (the relevant) spacetime instantons scales as the inverse of the string coupling and thus depends linearly on the complex structure moduli, in whose definition the dilaton, and thus the string coupling, enters.

The string perturbation expansion is a double series in powers of the string coupling and the inverse of the string tension. For the present case, this translates into an expansion in inverse powers of the Kaehler and complex structure moduli. Given that the tree-level superpotential is non-zero and independent of the moduli, it cannot acquire perturbative corrections, which would be terms with negative powers of the moduli. The latter are forbidden by the combination of holomorphy and the shift symmetry. Including instanton corrections, which always contain the factor $\exp(-S_{inst})$, where $S_{inst}$ is the instanton action, the full superpotential takes the form
\begin{eqnarray}
 W = W^{tree} + W^{np}\left(\exp(2\pi i U^{(i)}),\exp(2\pi i T^{(i)})\right) .
\end{eqnarray}
The gauge kinetic functions \eqref{gaugekineticfunction} are linear in the complex structure moduli. One-loop (in the string coupling) corrections, which contain an inverse power of $U^{(i)}$ compared to the tree level contribution, are therefore allowed, but, in analogy to the case of the superpotential, further perturbative corrections are forbidden. Considering for simplicity only diagonal gauge kinetic functions, i.e. $f_{ab}=f_a \delta_{ab}$, they thus look like
\begin{eqnarray}
 f_a = \sum_{i=0}^{h_{21}} m_a^i U^{(i)} + f_a^{1-loop}\left(\exp(2\pi i T^{(i)})\right) +
            f_a^{np}\left(\exp(2\pi i U^{(i)}),\exp(2\pi i T^{(i)})\right) .\nonumber \\ \label{generalformgaugekineticfunction}
\end{eqnarray}
The shift symmetry would allow the superpotential and gauge kinetic functions to depend on the imaginary parts of the moduli without depending on the real parts, but this is not allowed due to the holomorphy of $W$ and $f_a$. The tree level expression for the gauge kinetic functions does break the shift symmetry, but the real parts of the gauge kinetic functions only couple to the topological term in the Yang-Mills action and therefore to instantons which do indeed break the shift symmetry.

Recall from chapter \ref{fourdimensionaleffectiveactions} the relation \eqref{physicalgaugecouplingINTRO} between the running, loop-corrected, physical gauge couplings $g_a(\mu^2)$ depending on the renormalisation scale $\mu$ and the holomorphic Wilsonian gauge kinetic functions $f_a$, in which the Kaehler potential $K$ and the charged matter Kaehler metrics $K^{ab}_r(\mu^2)$ enter.
\begin{eqnarray}
 16\pi^2 g_a^{-2}(\mu^2) &=& 16\pi^2 {\rm Im}(f_a) + b_a \ln\frac{\Lambda^2}{\mu^2} + c_a K +
             2 T_a(adj) \ln  g_a^{-2}(\mu^2) 
             \nonumber \\ &&
             - 2 \sum_r T_a(r) \ln \det K^{ab}_r(\mu^2)
 \label{physicalgaugecoupling}
 \\
 b_a&=&\sum_r n_r T_a(r)-3T_a(adj)
 \\
 c_a&=&\sum_r n_r T_a(r)-T_a(adj)
\end{eqnarray}
The sums over $r$ run over the representations of the gauge group factor under consideration, $n_r$ counts the number of chiral multiplets transforming in the representation $r$ and $T_a(r)=Tr_r(T^2_{(a)})$, where $T_{(a)}$ are the group generators. The natural cutoff scale for a field theory supposed to capture the infrared physics of a string compactification is the Planck scale, i.e. $\Lambda^2=M_{Pl}^2$. The formula \eqref{physicalgaugecoupling} is to be understood recursively, so if one is interested in the $n$-loop corrected gauge coupling and/or gauge kinetic function, one has to use the $(n-1)$-loop corrected values for $K$, $K^{ab}_r(\mu^2)$ and the gauge coupling itself on the RHS of \eqref{physicalgaugecoupling}. As will be detailed later on, there can also be corrections to the RHS of \eqref{physicalgaugecoupling} arising through a redefinition at loop level of the complex structure moduli that enter the tree level expression of $f_a$.

In string theory, one usually computes physical, on-shell quantities. The gauge threshold corrections computed in the previous sections are one-loop corrections to such physical quantities and should be viewed as corrections to the LHS of \eqref{physicalgaugecoupling}. A non-trivial consistency check arises through the requirement that the non-holomorphic terms in these expressions must equal the non-holomorphic terms involving the Kaehler potential and the Kaehler metrics on the RHS of \eqref{physicalgaugecoupling}. If one knows $K$ and $K^{ab}_r(\mu^2)$ (in addition to the gauge threshold corrections) one can determine the one-loop corrections to the holomorphic gauge kinetic function. On the other hand, having computed the gauge threshold corrections, one can use \eqref{physicalgaugecoupling} to strongly restrict the form of the Kaehler metrics.

In the following the gauge threshold corrections computed in sections \ref{gaugethresholdsbulk} and \ref{gaugethresholdsfractional} will be analysed with the help of \eqref{physicalgaugecoupling} \cite{Akerblom:2007uc,Blumenhagen:2008ji}. The threshold corrections were computed at one-loop, so one has to use the tree-level values for $K$, $K^{ab}_r(\mu^2)$ and $g_a^{-2}(\mu^2)$ on the RHS of \eqref{physicalgaugecoupling}. It will be shown that the non-holomorphic terms are indeed equal on both sides and the one-loop corrections to the gauge kinetic functions will be determined.
\subsection{An orbifold model with bulk D6-branes}
The relevant formulas for the gauge threshold corrections on the $\mathbb{Z}_2\times\mathbb{Z}_2$ orbifold with $h_{21}=3$ are \eqref{thresholdcorrectionsbulk2}, \eqref{thresholdcorrectionsbulk3} and \eqref{thresholdcorrectionsbulkmoebius}. The first thing to notice is that the cutoff scale appearing in these expressions is the string scale, whereas the Planck scale appears in \eqref{physicalgaugecoupling}. These two scales are related by
\begin{eqnarray}
 \frac{M_s^2}{M_{Pl}^2} \varpropto \exp(\phi_4) \varpropto \left(\prod_{i=0}^3 {\rm Im}(U^{(i)})\right)^{-\frac{1}{2}} .
\end{eqnarray}
Next, one observes that all terms but the first on the RHS of \eqref{physicalgaugecoupling} are sums over the representations of the gauge group factor. It is therefore useful to consider the terms according to which representation they are related to.

Noting that on the orbifold under consideration there are three chiral multiplets in the adjoint representation of each gauge group factor, one finds that the terms in \eqref{physicalgaugecoupling} multiplied by $2T(adj)$ are
\begin{eqnarray}
 K + \ln (g_a^{tree})^{-2} - \ln \det K^i_{adj} .
 \label{termsadjoint}
\end{eqnarray}
Using
\begin{eqnarray}
 K = - \sum_{i=0}^3 \ln {\rm Im}(U^{(i)}) - \sum_{i=1}^3 \ln {\rm Im}(T^{(i)}) ,
 \label{kaehlerpotentialorbifold}
\end{eqnarray}
the Kaehler metrics for the three ($i\in\{1,2,3\}$) chiral multiplets
\begin{eqnarray}
 K^i_{adj} = \frac{1}{{\rm Im}(T^{(i)}){\rm Im}(U^{(i)})}
   \left| \frac{(n_a^j+iu^j\widetilde{m}_a^j)(n_a^k+iu^k\widetilde{m}_a^k)}{(n_a^i+iu^i\widetilde{m}_a^i)} \right|
 \quad i \neq j \neq k \neq i , \nonumber
\end{eqnarray}
where
\begin{eqnarray}
 (u^i)^2=\frac{{\rm Im}(U^{(j)}){\rm Im}(U^{(k)})}{{\rm Im}(U^{(i)}){\rm Im}(U^{(0)})} ,
\end{eqnarray}
the expression \eqref{treelevelgaugekineticfunction} for the tree level gauge kinetic function, and the supersymmetry condition, one finds that the expression \eqref{termsadjoint} vanishes. This was to be expected as states transforming in the adjoint representation are strings with both ends on the same stack of branes. Terms proportional to $T(adj)$ on the LHS of \eqref{physicalgaugecoupling} should therefore come from an annulus diagram with both boundaries on the same stack of branes. But such diagrams were shown not to contribute to the gauge threshold corrections \eqref{thresholdcorrectionsbulk1}.

The next case to be discussed are contributions from states transforming in the fundamental representation of $G_a$. Such states arise at the intersection of brane stack $a$ with another stack $b$. The intersection is characterised by the intersection angles $\theta^i_{ab}$ and the number of such states is counted by the product of the intersection number $I_{ab}$ and the number of branes on stack $b$, $N_b$. As for the computation of the gauge threshold corrections in \ref{gaugethresholdsbulk}, two cases have to be distinguished.

The first one is characterised by all three intersection angles being non-trivial. Certain scattering amplitudes can be used to determine the Kaehler metric of the chiral fields transforming in the fundamental representation of $G_a$ to be \cite{Kors:2003wf,Lust:2004cx,Bertolini:2005qh,Blumenhagen:2006ci,DiVecchia:2008tm,DiVecchia:2009yw}
\begin{eqnarray}
 K^{ab}_f &=& \left({\rm Im}(U^{(0)})\right)^{-\alpha} \prod_{i=1}^3
      \left({\rm Im}(U^{(i)})\right)^{-(\beta+\xi\theta^i_{ab})} \left({\rm Im}(T^{(i)})\right)^{-(\gamma+\zeta\theta^i_{ab})} \times
 \nonumber \\ && \left[ \prod_{i=1}^3 
 \left(\frac{\Gamma(1-|\theta^i_{ab}|)}{\Gamma(|\theta^i_{ab}|)}\right)^{sign(\theta^i_{ab})} \right]^{-1/[2\sum_j sign(\theta^j_{ab})]} ,
 \label{kaehlermetricfundamental1}
\end{eqnarray}
where $\alpha$, $\beta$, $\gamma$, $\xi$ and $\zeta$ are undetermined constants. Using $n_f=|I_{ab}|N_b$ and some of the formulas given above, the terms proportional to $T_a(f)$ on the RHS of \eqref{physicalgaugecoupling} can be seen to reproduce the second and third term\footnote{The first term cancels upon summing over all branes and using the tadpole cancellation condition. The last term is a moduli-independent constant and can be absorbed into $M_s$ or be viewed as a correction to the gauge kinetic function.} of the last expression in \eqref{thresholdcorrectionsbulk3} if
\begin{eqnarray}
 \alpha=\beta=\frac{1}{4} \quad , \quad \gamma=\frac{1}{2}
 \label{kaehlermetricfundamental1constants}
\end{eqnarray}
and $\xi=\zeta=0$. To get all signs right one has to distinguish several cases according to the signs of the intersection angles and intersection numbers. It will be shown in section \ref{Redefinitionmoduli} that $\xi$ and $\zeta$ can actually be non-zero. This is related to the aforementioned redefinition of the moduli at loop level.

The second case differs from the first in that one of the intersection angles is zero. One hypermultiplet or, equivalently, two chiral multiplets arise at each intersection point of the brane stacks. The relevant formula for the gauge threshold corrections is \eqref{thresholdcorrectionsbulk2}. It contains the term
${\rm Im}(i\ln\eta(T^{(1)}))$, which manifestly is the imaginary part of a holomorphic function. One therefore concludes that the gauge kinetic function receives the one-loop correction
\begin{eqnarray}
 f_a^{1-loop} \varpropto iI_{ab}N_b\ln\eta(T^{(1)}) \ \ ,
\end{eqnarray}
whose dependence on the moduli is in agreement with the form \eqref{generalformgaugekineticfunction} predicted by the non-renormalisation theorem. Using the Kaehler metric \cite{Lust:2004cx,Blumenhagen:2006ci}
\begin{eqnarray}
 K^{ab,1}_f = \frac{|n_a^1+i u^1 \widetilde{m}_a^1|}
              {\Big({\rm Im}(U^{(2)}){\rm Im}(U^{(3)}){\rm Im}(T^{(2)}){\rm Im}(T^{(3)})\Big)^{\frac{1}{2}}}
\end{eqnarray}
for the fundamental matter in the sector under discussion and proceeding as before, one finds that the terms arising on the RHS of \eqref{physicalgaugecoupling} from this sector reproduce the second and third term in the last expression of \eqref{thresholdcorrectionsbulk2}.

The conclusion is that also in the sectors that give rise to fields in the fundamental representation of the gauge group, the non-holomorphic terms on both sides of equation \eqref{physicalgaugecoupling} are equal, as required by consistency.

Finally, if the gauge group factor is $G_a=SU(N_a)$, there can be fields transforming in the symmetric and/or the antisymmetric representation. These are strings stretching between brane stack $a$ and its orientifold image $a'$. Therefore, one has to take the annulus diagram with boundaries on brane $a$ and its orientifold image as well as the Moebius strip diagram into account. They are given by \eqref{thresholdcorrectionsbulk2} and \eqref{thresholdcorrectionsbulk3} with $\theta^i_{ab}$ and $I_{ab}N_b$ replaced by $\theta^i_{aa'}=2\theta^i_a$ and $I_{aa'}$ as well as \eqref{thresholdcorrectionsbulkmoebius}. The Kaehler metric for the relevant chiral multiplets is given by \eqref{kaehlermetricfundamental1} with the same replacements. One again finds that the non-holomorphic terms on both sides of \eqref{physicalgaugecoupling} are equal.	
\subsection{An orbifold model with fractionally charged D6-branes}
The analysis of the gauge threshold corrections in models on the $\mathbb{Z}_2\times\mathbb{Z}_2$ orbifold \cite{Blumenhagen:2007ip} with $h_{21}=51$ is similar to that of the previous section. When writing down the partition functions in section \ref{examplesfractional} and when computing the gauge threshold corrections in section \ref{gaugethresholdsfractional} and appendix \ref{appendixgaugethresholdcorrectionsfractional}, four cases were distinguished. This distinction will be made here, too.

{\bf Case 1:} Both boundaries of the annulus are on the same stack of branes. Therefore, the open string modes in this sector transform in the adjoint representation of the gauge group. In contradistinction to the orbifold with $h_{21}=3$, there are no chiral multiplets in this representation. So the terms proportional to $T(adj)$ on the RHS of \eqref{physicalgaugecoupling} do not cancel amongst each other, but yield a contribution that matches \eqref{appendixAannurunning1annukaehlermet1}. There is a further term, \eqref{appendixAholocorr1}, which is the imaginary part of a holomorphic function, so one concludes that the gauge kinetic function receives a one-loop correction
\begin{eqnarray}
 \delta_a f_a^{1-loop} \varpropto i N_a \sum_{i=1}^3 \ln \eta(T^{(i)}) .
\end{eqnarray}

{\bf Case 2:} This sector yields $n_f=2N_b$ chiral multiplets in the fundamental representation of $G_a$. The vertex operators for these fields are identical to those for the chiral multiplets in the adjoint representation on the orbifold with $h_{21}=3$. One concludes that the Kaehler metrics for these fields are identical. Upon changing variables, they can be written as
\begin{eqnarray}
 K_f^{ab,2} = \left(\prod_{i=0}^3{\rm Im}(U^{(i)})\right)^{-\frac{1}{4}} \left(\prod_{i=1}^3{\rm Im}(T^{(i)})\right)^{-\frac{1}{2}}
              \left(\prod_{i=1}^3 (L_a^i)^{\sigma^i_{ab}} \right)^{\frac{1}{\sigma_{ab}}} .
\end{eqnarray}
Proceeding as in the previous section, one finds that the terms on the RHS of \eqref{physicalgaugecoupling} reproduce the terms \eqref{appendixAannurunning2annukaehlermet2}. The term \eqref{appendixAholocorr2} yields a correction to the gauge kinetic function
\begin{eqnarray}
 \delta_{b^{(2)}} f_a^{1-loop} \varpropto i \sum_b N_b \sigma_{ab} \sum_{i=1}^3 \sigma^i_{ab} \ln \eta(T^{(i)}) .
\end{eqnarray}

{\bf Case 3:} There are no massless open string modes in this sector and therefore no contributions to the RHS of \eqref{physicalgaugecoupling}. As required by consistency, no non-holomorphic terms appear in the gauge threshold corrections. But there is a holomorphic term and therefore a correction to the gauge kinetic function
\begin{eqnarray}
 \delta_{b^{(3)}} f_a^{1-loop} \varpropto i \sum_b \frac{N_b}{\sigma_{ab}} \sum_{i=1}^3 \sigma^i_{ab}
      \ln\frac{\vartheta
      \genfrac[]{0pt}{}{(1-|\delta_a^i-\delta_b^i|)/2}{(1-|\lambda_a^i-\lambda_b^i|)/2}
      (0,T^{(i)})}{\eta(T^{(i)})} .
\end{eqnarray}
{\bf Case 4:} This is the sector yielding chiral bifundamentals. The Kaehler metrics are identical to those on the orbifold with $h_{21}=3$ and given in \eqref{kaehlermetricfundamental1} with \eqref{kaehlermetricfundamental1constants}. As before, up to terms related to a redefinition of the moduli at loop level, the non-holomorphic terms on both sides of \eqref{physicalgaugecoupling} are equal. Those appearing on the LHS given by \eqref{appendixAannurunning4annukaehlermet4}.
\section{Redefinition of the moduli}
\label{Redefinitionmoduli}
Loop corrections to the low energy effective action of a string compactification can modify the proper definition of the chiral superfields \cite{Derendinger:1991hq,Derendinger:1991kr,Kaplunovsky:1995jw}, on which the superpotential and gauge kinetic function depend holomorphically. For example, this is possible if the low energy fields should really be linear instead of chiral multiplets, because the duality transformation relating the two may be corrected at loop-level. The possibility/necessity of a one-loop redefinition of the closed string moduli in D6-brane models on the $\mathbb{Z}_2\times\mathbb{Z}_2$ orbifolds is the subject of this section \cite{Akerblom:2007uc,Blumenhagen:2007ip}.
\subsection{A model with bulk D6-branes}
It was mentioned after \eqref{kaehlermetricfundamental1constants} that a naive application of \eqref{physicalgaugecoupling} would imply that the constants $\xi$ and $\zeta$ appearing in \eqref{kaehlermetricfundamental1} must vanish. It will now be shown that this does not have to be the case if one takes the possibility of a redefinition of the complex structure moduli at one loop into account.

If $\xi$ and $\zeta$ are non-zero, one gets an extra contribution from the term depending on the Kaehler metric to the RHS of \eqref{physicalgaugecoupling}, which does not have a counterpart on the LHS. It is non-holomorphic, so it cannot be interpreted in terms of a correction to the gauge kinetic function. The contribution can however cancel against terms from the holomorphic tree-level gauge kinetic function if the complex structure moduli appearing there are redefined at one loop. In order for this to be true, the $\xi$- and $\zeta$-dependent terms in \eqref{kaehlermetricfundamental1} must actually be \cite{Akerblom:2007uc,Blumenhagen:2007ip}
\begin{eqnarray}
 f(I_{ab},\theta^i_{ab}) = \prod_{i=1}^3 \left({\rm Im}(U^{(i)})\right)^{-\xi\theta^i_{ab} sign(I_{ab})}
                                       \left({\rm Im}(T^{(i)})\right)^{-\zeta\theta^i_{ab} sign(I_{ab})} ,
 \label{kaehlermetricfundamentalextrafactor}
\end{eqnarray}
where $\xi$ and $\zeta$ have to be equal for all brane stacks/brane intersections in the model. As before, one has to replace $\theta_{ab}$ and $I_{ab}$ by $\theta_{aa'}$ and $I_{aa'}-I_{aO}$, or $I_{aa'}+I_{aO}$, respectively, for fields transforming in the symmetric or antisymmetric representation of $SU(N_a)$. Summing over all relevant representations one finds the following extra contribution due to the factor \eqref{kaehlermetricfundamentalextrafactor} to the RHS of \eqref{physicalgaugecoupling}
\begin{eqnarray}
 \sum_{r=f,a,s} T_a(r) \ln \det {K'}^r &=& \frac{|I_{ab}|N_b}{2} \ln f(I_{ab},\theta^i_{ab}) +
                                       \frac{|I_{ab'}|N_b}{2} \ln f(I_{ab'},\theta^i_{ab'}) + \nonumber \\ &&
                                       \frac{N_a+2}{2}\frac{|I_{aa'}-I_{aO}|}{2} \ln f(I_{aa'}-I_{aO},2\theta^i_{a}) + \nonumber \\ &&
                                       \frac{N_a-2}{2}\frac{|I_{aa'}+I_{aO}|}{2} \ln f(I_{aa'}+I_{aO},2\theta^i_{a}) .
\end{eqnarray}
Using the tadpole cancellation conditions, this can be written as
\begin{eqnarray}
 && -n_a^1 n_a^2 n_a^3 \left[ \sum_b N_b \widetilde{m}_b^1 \widetilde{m}_b^2 \widetilde{m}_b^3
   \sum_{i=1}^3 \theta_b^i \left(\xi\ln {\rm Im}(U^{(i)})+\zeta\ln {\rm Im}(T^{(i)})\right) \right] \\ &&
 - \sum_{j \neq k \neq l \neq j}^3 n_a^j \widetilde{m}_a^k \widetilde{m}_a^l \left[ \sum_b N_b \widetilde{m}_b^j n_b^k n_b^l
   \sum_{i=1}^3 \theta_b^i \left(\xi\ln {\rm Im}(U^{(i)})+\zeta\ln {\rm Im}(T^{(i)})\right) \right] \nonumber .
 \label{extratermsDTredefinition1}
\end{eqnarray}
By comparing with \eqref{treelevelgaugekineticfunction} it can be seen that this contribution to the RHS of \eqref{physicalgaugecoupling} is cancelled if the imaginary parts of the complex structure moduli are redefined as follows:
\begin{eqnarray}
 \label{moduliredefinitionbulk}
 {\rm Im}(U^{(0)}) &\rightarrow& {\rm Im}(U^{(0)}) - \frac{1}{8\pi^2} \sum_b N_b \widetilde{m}_b^1 \widetilde{m}_b^2 \widetilde{m}_b^3 \times
             \\ \nonumber &&
             \sum_{i=1}^3 \theta_b^i \left(\xi\ln {\rm Im}(U^{(i)})+\zeta\ln {\rm Im}(T^{(i)})\right) \\
 {\rm Im}(U^{(j)}) &\rightarrow& {\rm Im}(U^{(j)}) + \frac{1}{8\pi^2} \sum_b N_b \widetilde{m}_b^j n_b^k n_b^l \times
             \nonumber \\ \nonumber &&
             \sum_{i=1}^3 \theta_b^i \left(\xi\ln {\rm Im}(U^{(i)})+\zeta\ln {\rm Im}(T^{(i)})\right)
             \qquad j \neq k \neq l \neq j
\end{eqnarray}
In conclusion, knowledge of the gauge threshold corrections for the model under consideration is not enough to completely fix the Kaehler metrics for the chiral bifundamental matter fields using \eqref{physicalgaugecoupling}. Also, it is not possible to determine whether the complex structure moduli are redefined at one-loop. The two constants $\xi$ and $\zeta$ are still free parameters, which have to be determined by other means. If they are zero there is no one-loop redefinition of the complex structure moduli.
\subsection{A model with fractionally charged D6-branes}
The analysis for the case of the orbifold with $h_{21}=51$ is similar to that of the previous section, important differences arise as the gauge kinetic functions in this case depend on the complex structure moduli in the twisted sectors \eqref{treelevelgaugekineticfunction2}.

The form of the vertex operators for the chiral bifundamental fields arising at the intersection of two branes does not depend on whether the underlying orbifold is the one with $h_{21}=3$ or the one with $h_{21}=51$. The extra factor \eqref{kaehlermetricfundamentalextrafactor} in the Kaehler metric therefore has to be the same on both backgrounds. It turns out that it is more convenient for the following analysis to rewrite it as\footnote{Note that the quantity $\Upsilon_{ab}$ is only defined for models on the orbifold with $h_{21}=51$, so this rewriting can only be done for such models.}
\begin{eqnarray}
 g(\Upsilon_{ab},\theta^i_{ab}) = \prod_{i=1}^3 \left({\rm Im}(U^{(i)})\right)^{-\xi\theta^i_{ab} sign(\Upsilon_{ab})}
                                       \left({\rm Im}(T^{(i)})\right)^{-\zeta\theta^i_{ab} sign(\Upsilon_{ab})} .
 \label{kaehlermetricfundamentalextrafactor2}
\end{eqnarray}
The two expressions \eqref{kaehlermetricfundamentalextrafactor} and \eqref{kaehlermetricfundamentalextrafactor2} differ in whether $sign(I_{ab})$  or $sign(\Upsilon_{ab})$ appears. A physical argument shows that these two signs must be equal. The orbifold projection removes some string states, but cannot change their spacetime chirality. As the latter is determined by the aforementioned signs they must be equal.

Proceeding as in the previous section, one finds two contributions to the RHS of \eqref{physicalgaugecoupling} due to the extra factor \eqref{kaehlermetricfundamentalextrafactor2} in the Kaehler metric. One is identical to \eqref{extratermsDTredefinition1} and is cancelled by a redefinition of the imaginary parts of the $U^{(i)}$ as in \eqref{moduliredefinitionbulk}, but with $1/8\pi^2$ replaced by $1/32\pi^2$. The other can be written as
\begin{eqnarray}
 \sum_{i;k,l} n_a^i \epsilon^i_{a,kl} \sum_b N_b \widetilde{m}^i_b ( \epsilon^i_{b,kl}+\epsilon^i_{b,R(k)R(l)} )
       \sum_j \theta^j_b \left( \xi \ln {\rm Im}(U^{(j)}) + \zeta \ln {\rm Im}(T^{(j)}) \right) \nonumber \\
 - \sum_{i;k,l} \widetilde{m}_a^i \epsilon^i_{a,kl} \sum_b N_b n^i_b ( \epsilon^i_{b,kl}-\epsilon^i_{b,R(k)R(l)} )
       \sum_j \theta^j_b \left( \xi \ln {\rm Im}(U^{(j)}) + \zeta \ln {\rm Im}(T^{(j)}) \right) \nonumber \\
 \label{extratermsDTredefinition2}
\end{eqnarray}
and cancels if the twisted sector complex structure moduli $W_{ikl}$ and $\widetilde{W}_{ikl}$ are redefined. There is another term \eqref{appendixAannuuni4} that gives contributions to this redefinition. It is part of the gauge threshold corrections and therefore appears on the LHS of \eqref{physicalgaugecoupling}. Summing over all branes including the orientifold images and using the tadpole cancellation conditions \eqref{tadpolecancellationconditionsfractionaltwisted}, it can be cast into the form
\begin{eqnarray}
 \sum_{i;k,l} n^i_a \epsilon^i_{a,kl} \sum_b N_b \widetilde{m}^i_b \theta^i_b ( \epsilon^i_{b,kl} + \epsilon^i_{b,R(k)R(l)} )
 \nonumber \\
 - \sum_{i;k,l} \widetilde{m}^i_a \epsilon^i_{a,kl} \sum_b N_b n_b^i \theta_b^i ( \epsilon^i_{b,kl} - \epsilon^i_{b,R(k)R(l)} ) .
 \label{extratermsDTredefinition3}
\end{eqnarray}
Taking the contributions \eqref{extratermsDTredefinition2} and \eqref{extratermsDTredefinition3} to the RHS, respectively LHS, of \eqref{physicalgaugecoupling} into account and using the form \eqref{treelevelgaugekineticfunction2} of $f_a$ as well as the identity
\begin{eqnarray}
 \sum_{k,l} \epsilon^i_{a,kl}(\epsilon^i_{b,kl}\pm\epsilon^i_{b,R(k)R(l)}) =
 \sum_{k,l} \epsilon^i_{b,kl}(\epsilon^i_{a,kl}\pm\epsilon^i_{a,R(k)R(l)})
\end{eqnarray}
one finds that the imaginary parts of the twisted sector complex structure moduli should be redefined as
\begin{eqnarray}
 {\rm Im}(W_{ikl}) &\rightarrow& {\rm Im}(W_{ikl}) - \frac{1}{64\pi^2} \sum_b N_b \widetilde{m}^i_b \epsilon^i_{b,kl}
             \nonumber \\ &&
             \sum_j \theta^j_b \left(\xi \ln {\rm Im}(U^{(j)}) + \zeta \ln {\rm Im}(T^{(j)}) + \ln 4 \delta_{ij} \right) \\
 {\rm Im}(\widetilde{W}_{ikl}) &\rightarrow& {\rm Im}(\widetilde{W}_{ikl}) + \frac{1}{64\pi^2} \sum_b N_b n^i_b \epsilon^i_{b,kl}
             \nonumber \\ &&
             \sum_j \theta^j_b \left(\xi \ln {\rm Im}(U^{(j)}) + \zeta \ln {\rm Im}(T^{(j)}) + \ln 4 \delta_{ij} \right) .
\end{eqnarray}
Note that even if $\xi$ and $\zeta$ vanish, $W_{ikl}$ and $\widetilde{W}_{ikl}$ acquire a one-loop redefinition.

The conclusion of this section is that also on the $\mathbb{Z}_2\times\mathbb{Z}_2$ orbifold with $h_{21}=51$, the non-holomorphic terms on both sides of \eqref{physicalgaugecoupling} are equal, as required by consistency.
\chapter{D-instantons in four-dimensional brane models}
\label{Dinstantons}
The only definition of string theory that exists today is that of the perturbative expansion of scattering amplitudes. The latter is an expansion in powers of the string coupling $g_s$. At each order of the perturbation series the contribution to a scattering amplitude is given by an integral over an amplitude of a conformal field theory defined on a Riemann surface. The Euler characteristic of the surface determines with which power of $g_s$ the term contributes to the full amplitude.

Various arguments can be given \cite{Polchinski:1998rq} in support of the claim that this perturbation series does not give the full result, but that there should exist non-perturbative contributions whose dependence on the string coupling is given e.g. by $\exp(-g_s^{-1})$. Given that the action of a D-brane depends on the string coupling as $g_s^{-1}$ and that instanton amplitudes always contain a factor $\exp(-S_{inst})$, where $S_{inst}$ is the instanton action, it is natural to suppose that there are corrections to various observables in string theory induced by instantons that are D-branes localised in time. Such instantons are called D-instantons \cite{Green:1997tv,Gutperle:1997iy,Witten:1999eg} and the subject of this chapter.
\section{D2-instantons in D6-brane models}
Given that, as just mentioned, only a definition of the string perturbation series as an expansion in $g_s$ exists, D-instantons yielding effects non-perturbative in $g_s$ at first sight seem to be very hard to describe. The situation luckily is somewhat better, as D-instantons are D-branes and as such described by open string theories \cite{Polchinski:1995mt}. This description will be the guiding principle in the following discussion of how to compute D-instanton effects \cite{Blumenhagen:2006xt,Haack:2006cy,Florea:2006si,Ibanez:2006da,Buican:2006sn,Blumenhagen:2009qh}. For concreteness, D-instantons in orientifold models of type IIA string theory will be considered in this section, but things are very similar in type I models or other orientifolds of the type IIB theory. The focus will be on type IIA orientifolds of Calabi-Yau manifolds, which were described in section \ref{D6branemodelsCY}. In order that their action is finite, the instantons have to be localised not only in time, but also in the three non-compact space directions. Type IIA string theory contains Dp-branes with even p, such that the candidates for D-instantons are D0-, D2- and D4-branes wrapping one-, three- and five-cycles of the internal manifold. As Calabi-Yau manifolds do not contain topologically non-trivial one- and five-cycles, only D2-instantons are relevant. The instantons must be (local) minima of the action, which for D-instantons in compactifications preserving supersymmetry implies that they are BPS-states. D2-instantons thus wrap special Lagrangian three-cycles and are calibrated with the same phase as the orientifold plane, just as the spacefilling D-branes \eqref{susyconditionCY}. The BPS property implies that the instantons contribute to F-terms, i.e. the superpotential and the gauge kinetic function, rather than D-terms in the low energy effective action.

The D2-instanton action $S_{D2}$ is the sum of the Dirac-Born-Infeld and Chern-Simons actions integrated over the three-cycle
\begin{eqnarray}
 \pi_{D2} = m_{D2}^i A_i + n_{D2}^i B_i
\end{eqnarray}
the instanton wraps. The characteristic exponential factor $\exp(-S_{inst})$ becomes
\begin{eqnarray}
 \exp(-S_{D2}) &=& \exp \left( - 2\pi e^{-\phi_4} \int_{\pi_{D2}} {\rm Re}(\Omega_3) + 2\pi i \int_{\pi_{D2}} C_3 \right) 
               \nonumber \\
               &=& \exp \left( 2\pi i \sum_{i=0}^{h_{21}} m^i_{D2} U^{(i)} \right) ,
 \label{exponentialinstantonfactor}
\end{eqnarray}
where the orientifold image has already been taken into account.
\section{Zero modes}
\label{zeromodes}
As in every instanton computation \cite{'tHooft:1976fv,Coleman:1978ae,Dorey:2002ik}, the zero modes, which in the case of D-instantons are massless open strings with at least one end on the instantonic brane, are of crucial importance. In the case at hand it is useful to distinguish two types of zero modes \cite{Blumenhagen:2006xt}. Neutral zero modes are given by open strings with both ends on the instanton or strings stretched between the instanton and its orientifold image. Strings with one end on the instanton and the other one on some spacefilling D-brane \cite{Ganor:1996pe} can also give rise to zero modes. These zero modes are called charged zero modes because they transform non-trivially under the gauge group of the four-dimensional theory. All the zero modes have analogues in terms of ordinary particle states which would arise from strings ending on a fictitious space-filling D-brane (a space-filling D-brane is a brane which fills out the non-compact four-dimensional space) wrapping the cycle of the internal space the instanton wraps, or, in more abstract CFT terms, that is described by the same boundary state in the internal CFT. The vertex operators of the instanton zero modes are similar to those of these fictitious particle states. For neutral zero modes they differ in that the zero modes have no momentum in the four non-compact directions as they are confined to the instanton worldvolume and cannot move in the external space. For charged zero modes only that part of the vertex operator which acts in the internal CFT is identical.

The different neutral zero modes shall be discussed first. The instanton breaks four-dimensional translational invariance. There will thus be four universal bosonic zero modes $x^\mu$, $\mu\in\{0,1,2,3\}$, whose vertex operators
\begin{eqnarray}
 V_{x^\mu} = e^{-\phi(z)} \psi^\mu(z)
 \label{vertexoperatorsx}
\end{eqnarray}
are quite similar to the gauge boson vertex operators given in section \ref{abstractCFTs}.

The instanton also breaks some supersymmetries. The number of broken supercharges and associated zero modes depends on whether the cycle wrapped by the instanton is mapped to itself by the antiholomorphic involution that is part of the orientifold projection or not \cite{Argurio:2007qk,Argurio:2007vq,Bianchi:2007wy,Ibanez:2007tu}. If it is not mapped to itself, the instanton breaks four of the eight supersymmetries which are preserved by a compactification of type IIA string theory on a Calabi-Yau manifold and which are therefore present in the bulk of the compactification space, away from the orientifold plane. There are thus four fermionic zero modes $\theta^\alpha$ and $\bar{\tau}^{\dot{\alpha}}$, $\alpha,\dot{\alpha}\in\{1,2\}$, with vertex operators \cite{Blumenhagen:2006xt}
\begin{eqnarray}
 V_{\bar{\tau}^{\dot{\alpha}}} &=&  e^{-\phi(z)/2} S^{\dot{\alpha}}(z) \mathcal{O}^{3/8}_{3/2}(z)
 \nonumber \\
 V_{\theta^\alpha} &=&  e^{-\phi(z)/2} S^{\alpha}(z) \mathcal{O}^{3/8}_{-3/2}(z) ,
 \label{vertexoperatorsthetatau}
\end{eqnarray}
similar to the gaugino vertex operators \eqref{gauginovertexoperator} of section \ref{abstractCFTs}. Indeed, $\mathcal{O}^{3/8}_{\pm3/2}(z)$ are once more the spectral flow operators of the internal CFT. The instanton, being a D-brane, carries a gauge theory on its worldvolume. The gauge group in the case under discussion, i.e. where the instanton is not invariant under the orientifold projection, is $U(1)$. There can also be multi-instantons. The gauge group is $U(k)$ for a $k$-instanton realised by a stack of $k$ D-brane instantons wrapping the same cycle of the compactification manifold. In this case, a $k\times k$ Chan-Paton matrix must be included in the vertex operators and the aforementioned zero modes transform in the adjoint representation of $U(k)$.

If the cycle wrapped by the instanton is invariant under the involution, some of the zero modes are removed by the orientifold projection. Two cases have to be distinguished. If the vertex operators of the zero modes $x^\mu$ and $\theta^\alpha$ are left invariant by the projection and those of $\bar{\tau}^{\dot{\alpha}}$ are anti-invariant, the Chan-Paton matrices of the former are symmetric $k\times k$ matrices, those of the latter anti-symmetric ones and the gauge group is $O(k)$. If $x^\mu$ and $\theta^\alpha$ are anti-invariant and $\bar{\tau}^{\dot{\alpha}}$ invariant, they have anti-symmetric respectively symmetric Chan-Paton matrices. This is only possible if $k$ is even and the gauge group is $USp(k)$. The most relevant case are instantons with gauge group $O(1)$ as they have four bosonic ($x^\mu$) and two fermionic ($\theta^\alpha$) zero modes, which make up precisely the integration over chiral superspace that is required for an F-term contribution to the effective action. 

Further neutral zero modes arise if the three-cycle $\pi_{D2}$ wrapped by the D2-instanton is not rigid. The number of additional zero modes is counted by the first Betti number $b_1(\pi_{D2})$ of the three-cycle. More precisely, there are $4b_1(\pi_{D2})$ fermionic and $2b_1(\pi_{D2})$ bosonic zero modes in the case of a unitary gauge group on the instanton worldvolume. The analogue of such zero modes for space-filling D6-branes are chiral multiplets transforming in the adjoint representation of the gauge group. In the case of orthogonal or symplectic gauge group on the instanton worldvolume there are $2b_1(\pi_{D2})$ fermionic zero modes in the symmetric as well as $2b_1(\pi_{D2})$ fermionic and $2b_1(\pi_{D2})$ bosonic zero modes in the antisymmetric representation or vice versa. The precise structure depends on how the antiholomorphic involution acts on the cycle wrapped by the instanton. The vertex operators of these zero modes are, in analogy to \eqref{chiralsuperfieldvertexoperators}, given by
\begin{eqnarray}
 V_{instanton\ modulus} &=& e^{-\phi(z)} \mathcal{O}^{1/2}_{\pm1}(z) \nonumber \\
 V_{instanton\ modulino^+} &=& e^{-\phi(z)/2} S^{\dot{\alpha}}(z) \mathcal{O}^{3/8}_{1/2}(z)
 \nonumber \\
 V_{instanton\ modulino^-} &=& e^{-\phi(z)/2} S^{\alpha}(z) \mathcal{O}^{3/8}_{-1/2}(z) .
 \label{instantonmodulivertexoperators}
\end{eqnarray}

In the case of instantons with gauge group $U(k)$, there can also be zero modes arising at the intersection of the instanton with its orientifold image. The analogue for D6-branes are states transforming in the symmetric or antisymmetric representation of a unitary gauge group (on a space-filling D-brane in this case). The vertex operators look as those in \eqref{instantonmodulivertexoperators}, but they are operators changing the boundary conditions from those describing the instanton to those describing its orientifold image. The number of such zero modes is counted by the physical (not topological) intersection numbers of the cycles wrapped by the instanton, its orientifold image and the orientifold plane.

Some of the non-universal neutral zero modes can under certain circumstances be lifted in flux backgrounds \cite{Tripathy:2005hv,Bergshoeff:2005yp,Park:2005hj,Lust:2005cu,Blumenhagen:2007bn,Billo':2008sp,Billo':2008pg,Uranga:2008nh}.

A class of neutral zero modes that will not be relevant in the following are modes arising at the intersection of two branes in a multi-instanton configuration, where the different branes wrap different cycles \cite{GarciaEtxebarria:2007zv,Cvetic:2008ws,GarciaEtxebarria:2008pi}.

Finally, there can be so-called charged zero modes. These zero modes are strings with one end on the instanton and the other one on one of the space-filling D-branes so that they are charged under the gauge group of the four-dimensional theory. They transform in the fundamental or anti-fundamental representation of the gauge group on the brane and in the fundamental or anti-fundamental representation of the gauge group on the instanton. As one of their ends, the one which can move along the space-filling brane, is subject to Neumann boundary conditions in the four external dimensions and the other one, which is confined to the instanton, to Dirichlet boundary conditions (in the external space), their vertex operators contain the operators $T_{X^\mu}$, $\mu\in\{0,1,2,3\}$, which change the boundary conditions from Neumann to Dirichlet or vice versa in the four free boson CFTs.

The number of charged zero modes in a given sector, i.e. massless open strings between the instanton and a (fixed) brane, depends on whether or not they wrap the same cycle in the internal manifold. If they do, the instanton can be viewed as the string theory realisation of a gauge instanton. In such a configuration, there will always be four bosonic and two fermionic zero modes with vertex operators \cite{Billo:2002hm}
\begin{eqnarray}
 V_{w^{\dot{\alpha}}} &=& e^{-\phi(z)} S^{\dot{\alpha}}(z) \prod_{\mu=0}^3 T_{X^\mu}(z) \qquad
 V_{\bar{w}^{\dot{\alpha}}} = e^{-\phi(z)} S^{\dot{\alpha}}(z) \prod_{\mu=0}^3 T_{X^\mu}(z) 
 \label{vertexoperatorsw} \\
 V_\mu &=& e^{-\phi(z)/2} \prod_{\mu=0}^3 T_{X^\mu}(z) \mathcal{O}^{3/8}_{3/2}(z) \quad
 V_{\bar{\mu}} = e^{-\phi(z)/2} \prod_{\mu=0}^3 T_{X^\mu}(z) \mathcal{O}^{3/8}_{-3/2}(z) . \qquad
 \label{vertexoperatorsmu}
\end{eqnarray}
The operators $\mathcal{O}^{3/8}_{\pm3/2}(z)$ are once more the spectral flow operators. The zero modes $w^{\dot{\alpha}}$ and $\bar{w}^{\dot{\alpha}}$ as well as $\mu$ and $\bar{\mu}$ have opposite orientation. If the first Betti number of the cycle the brane and the instanton wrap is not zero, there are $2b_1(\pi_{D2})$ further fermionic zero modes with vertex operators
\begin{eqnarray}
 V_{\lambda_{adj}} &=& e^{-\phi(z)/2} \prod_{\mu=0}^3 T_{X^\mu}(z) \mathcal{O}^{3/8}_{1/2}(z)
 \\
 V_{\bar{\lambda}_{adj}} &=& e^{-\phi(z)/2} \prod_{\mu=0}^3 T_{X^\mu}(z) \mathcal{O}^{3/8}_{-1/2}(z) .
\end{eqnarray}
$\lambda_{adj}$ and $\bar{\lambda}_{adj}$ have opposite orientation. The label $adj$ has its origin in the fact that these modes can be viewed as the zero modes of fermions in chiral multiplets transforming in the adjoint representation of the gauge theory instanton that this D-instanton realises. There can be further zero modes if the instanton intersects the orientifold image of the brane. Such modes could be viewed as gauge instanton zero modes of fields transforming in the symmetric or antisymmetric representation of the gauge group.

The zero mode structure is different if the cycle $\pi_{D2}$ wrapped by $k$ instantonic branes is different from the cycle $\pi_a$ wrapped by a stack of $N_a$ spacefilling D6-branes. Let $\pi_a \cap \pi_{D2}$ and $\pi_{D2} \cap \pi_a$ be the number of positive and negative intersections of the two cycles and denote the orientifold image of $\pi_a$ by $\pi'_a$. There are only fermionic zero modes in such sectors and their number is given in table \ref{chargedzeromodes} together with the representation of the product $G_a\times G_{D2}$ of the D-brane and instanton gauge groups in which they transform. The vertex operators for these states are \cite{Blumenhagen:2006xt}
\begin{eqnarray}
 V_{charged\ zero\ mode} = e^{-\phi(z)/2} \prod_{\mu=0}^3 T_{X^\mu}(z) \mathcal{O}^{3/8}_{\pm1/2}(z) ,
\end{eqnarray}
where the sign in $\mathcal{O}^{3/8}_{\pm1/2}(z)$ depends on the precise structure of the internal CFT.
\begin{table}
 \begin{center}
 \begin{tabular}{|c|c|c|}
  \hline
   zero mode & number & representation of $G_a\times G_{D2}$ \\
  \hline
   $\lambda_a$ & $\pi_a \cap \pi_{D2}$ & fundamental$\times$anti-fundamental \\
   $\bar{\lambda}_a$ & $\pi_{D2} \cap \pi_a$ & anti-fundamental$\times$fundamental \\
   $\lambda_{a'}$ & $\pi'_a \cap \pi_{D2}$ & anti-fundamental$\times$anti-fundamental \\
   $\bar{\lambda}_{a'}$ & $\pi_{D2} \cap \pi'_a$ & fundamental$\times$fundamental \\
  \hline
 \end{tabular}
 \caption{Charged instanton zero modes\label{chargedzeromodes}}
 \end{center}
\end{table}

\section{Global abelian symmetries}
Important selection rules determining which instantons can generate which terms in the low-energy effective action come from global abelian symmetries \cite{Blumenhagen:2006xt}. These symmetries are remnants of gauge symmetries whose associated gauge bosons have become massive due to the Green-Schwarz mechanism discussed in section \ref{D6branemodelsCY}. \eqref{gaugetransformationcomplexstructuremoduli} implies that the exponential instanton factor \eqref{exponentialinstantonfactor} transforms under gauge transformations of these massive $U(1)$ symmetries as
\begin{eqnarray}
 \exp(-S_{D2}) \rightarrow \exp(-S_{D2}) \exp\left(-4\pi i \sum_{i=0}^{h_{21}} \sum_a N_a m_{D2}^i n_a^i \alpha^a\right) .
 \label{transformationexponentialfactor}
\end{eqnarray}
Gauge invariance implies that the instanton-induced superpotential will have the form
\begin{eqnarray}
 W = \prod_i \Phi_i \exp(-S_{D2}) ,
 \label{superpotentialD2induced}
\end{eqnarray}
where the product of charged matter fields $\prod_i \Phi_i$ has to transform under the $U(1)$ symmetries such as to cancel
the transformation of the exponential factor. The superpotential \eqref{superpotentialD2induced} does however lead to correlators
violating $U(1)$ charge conservation. It is in this sense that the instanton breaks the massive $U(1)$ symmetries.

These abelian symmetries are the diagonal $U(1)$ subgroups of $U(N_a)$ gauge groups.
From table \ref{chargedzeromodes} one can read off the total charge of the D2-instanton under a (possibly massive) $U(1)$ factor. The charge is
\begin{eqnarray}
 && N_a [ (\pi_a \cap \pi_{D2} - \pi_{D2} \cap \pi_a) - (\pi'_a \cap \pi_{D2} - \pi_{D2} \cap \pi'_a) ] \nonumber \\
 &=& N_a ( \pi_a \circ \pi_{D2} - \pi'_a \circ \pi_{D2} )
 = - 2 \sum_{i=0}^{h_{21}} N_a m_{D2}^i n_a^i ,
 \label{instantonu1charge}
\end{eqnarray}
i.e. it is given by the topological intersection numbers of the cycles the instanton and the brane stack $a$ (and its orientifold image) wrap. The charge \eqref{instantonu1charge} also determines the transformation property \eqref{transformationexponentialfactor} of the exponential instanton factor.

In conclusion, the charges of the exponential under the massive $U(1)$ factors are encoded in the instanton zero modes. Integration over all zero modes, which will be discussed shortly, ensures that the instanton-generated superpotential is invariant under the global abelian symmetries.

\section{Corrections to the superpotential}
\label{correctionssuperpotential}
As was mentioned earlier on, the idea \cite{Blumenhagen:2006xt} in setting up a D-instanton calculus is to use the fact that D-instantons are D-branes and as such described by an open string theory, or, in other words, boundary CFT. In order to determine a spacetime correlator in a D-instanton background, one has to compute several BCFT amplitudes and put them together appropriately. For concreteness, consider the spacetime correlator
\begin{eqnarray}
 \langle \prod_{i=1}^N \Phi_i(x_i) \rangle ,
 \label{instantoncorrelator}
\end{eqnarray}
where $\Phi_i(x_i)$ are charged matter fields. The question is now how to compute this correlator in the instanton background. As in every instanton computation, one has to integrate over all zero modes. The fermionic zero modes are of special importance, as they have to be soaked up in order that a non-vanishing result comes out. The idea is to absorb them by inserting each of their vertex operators precisely once on the boundary of  a worldsheet in a CFT amplitude. The bosonic zero modes can be inserted an arbitrary number of times on worldsheet boundaries. When computing the correlator \eqref{instantoncorrelator}, one also has to include the vertex operators for the fields $\Phi_i(x_i)$ in CFT correlators. Clearly, the amplitude has to be connected from the spacetime point of view, so only worldsheets with at least (a part of) one boundary on the instanton are allowed. In the spirit of the path integral as a sum over histories, the correlator \eqref{instantoncorrelator} is given by the sum over all possibilities of distributing the vertex operators for the fermionic zero modes and those for the fields $\Phi_i(x_i)$ on boundaries of any number of worldsheets. Each summand is a product of the CFT correlators for these worldsheets multiplied by any number of CFT amplitudes involving any number of bosonic zero mode vertex operators. The resulting expression will depend on the bosonic zero modes and has to be integrated over them.

The procedure just described implies that determining a correlator exactly is terribly complicated. When computing corrections to holomorphic quantities, i.e. the superpotential and the gauge kinetic function, the string coupling dependence of the individual CFT amplitudes puts strong constraints on which amplitudes can contribute \cite{Blumenhagen:2006xt}. In section \ref{holomorphygaugekineticfunction} the interplay of holomorphy and shift symmetries was used to formulate non-renormalisation theorems for the superpotential and the gauge kinetic function. The same ideas can be employed to determine which individual CFT amplitudes are relevant when computing instanton corrections to the superpotential and the gauge kinetic function.

As explained in section \ref{holomorphygaugekineticfunction}, there are shift symmetries associated with the chiral superfields in whose definition the dilaton, and therefore the string coupling $g_s$, enters. Due to these shift symmetries, the string coupling can appear in the superpotential only in the exponential instanton factor \eqref{exponentialinstantonfactor}, which was argued \cite{Polchinski:1994fq} to arise in the CFT description as the combination of an arbitrary number of vacuum disc diagrams. In order to see how this puts constraints on which CFT amplitudes are relevant when computing corrections to holomorphic quantities, the first thing to notice is that an amplitude on a surface of Euler characteristic $\chi$ depends on the string coupling $g_s$ as $g_s^{-\chi}$. This means that amplitudes of vanishing Euler characteristic, i.e. annulus and Moebius strip diagrams, can contribute. Furthermore, it has been argued \cite{Blumenhagen:2006xt} that the charged zero modes should carry a factor $g_s^{1/2}$ in their vertex operators so that a disc diagram with two of them inserted on the boundary can also contribute, because the string coupling dependence cancels. A positive power of the string coupling in these vertex operators is in agreement with the fact that the instanton should decouple at small string coupling. Furthermore, due to boundary combinatorics, there has to be an even number of charged zero modes on each disc, so two is the minimal number and if such discs are to contribute, one has to include the factor $g_s^{1/2}$. This conjecture is supported by analysing D-instantons that reproduce gauge instantons and comparing with the ADHM \cite{Atiyah:1978ri} construction \cite{Billo:2002hm}. Another argument will be given at the end of the following section.

Being interested in the superpotential, one focuses on a special case of the correlator \eqref{instantoncorrelator}, namely one where two of the charged matter fields are fermions, $\Phi_i(x_i)=\psi_i(x_i)$, $i\in\{1,2\}$, and the other fields are bosons $\Phi_i(x_i)=\phi_i(x_i)$, $i\in\{3,...,N\}$. Concentrating on the (most relevant) case in which there are only charged fermionic zero modes in addition to the universal four bosonic ($x^\mu$) and two fermionic ($\theta^\alpha$) ones, the formula for that part of the correlator under consideration which is relevant for the superpotential involves the vacuum disc and Moebius strip diagrams with the boundary on the instanton, $D^{vac}_{D2}$ and ${M'}^{vac}_{D2}$, and the annulus diagrams ${A'}^{vac}_{D2,D6_a}$ with one boundary on the instanton and the other one on a stack, labelled $a$, of space-filling D-branes. In addition one needs to compute disc diagrams $D(\lambda,\lambda,\Phi_i)$ with two charged zero mode vertex operators $V_{charged\ zero\ mode}$ and an arbitrary positive number of charged matter field vertex operators $V_{\Phi_i}$ inserted. On two discs one also has to insert the neutral zero mode vertex operators $V_{\theta^\alpha}$. As the zero modes are taken care of explicitly by inserting their vertex operators in disc diagrams, they are not to be included in the one-loop diagrams \cite{Blumenhagen:2006xt}. This is indicated by the prime in ${M'}^{vac}_{D2}$ and ${A'}^{vac}_{D2,D6_a}$. The formula for the (relevant part of the) correlator is then given by \cite{Blumenhagen:2006xt}
\begin{eqnarray}
 \langle \psi_1(x_1) \psi_2(x_2) \prod_{i=3}^N \phi_i(x_i) \rangle &=& \sum_{conf}
 \left( \prod D(\lambda,\lambda,\phi_i) \right) \left( \prod_{\alpha=1}^2 D(\theta^\alpha,\lambda,\lambda,\psi_\alpha,\phi_i) \right)
 \nonumber \\ &&
 \exp \left( D^{vac}_{D2} + {M'}^{vac}_{D2} + \sum_a {A'}^{vac}_{D2,D6_a} \right) .
 \label{formulainstantoncorrelator}
\end{eqnarray}
One has to sum over all possible configurations of distributing two charged fer\-mi\-onic zero modes and one or more matter fields $\Phi_i$ on a number of discs. This number clearly is one half of the number of charged fermionic zero modes. It is in principle also possible to attach charged matter fields to annulus diagrams, but this possibility will for simplicity not be considered here.

It is interesting to compare \eqref{formulainstantoncorrelator} to the formula \cite{Witten:1999eg} (adapted to the case of D2-instantons)
\begin{eqnarray}
 W = \exp \left(- V_{D2} + 2 \pi i \int_{\pi_{D2}} C_3 \right) \frac{\rm{Pfaff}'(\mathcal{D}_F)}{\sqrt{\det'(\mathcal{D}_B)}}
 \label{superpotentialphysicalgaugeapproach}
\end{eqnarray}
for the D-instanton induced superpotential derived via a physical gauge approach. The exponential factor in \eqref{superpotentialphysicalgaugeapproach} is clearly equal to \eqref{exponentialinstantonfactor} and, as was already mentioned, is represented by $\exp(D^{vac}_{D2})$ in the CFT approach. The second factor in \eqref{superpotentialphysicalgaugeapproach} is a quotient of a one-loop Pfaffian and the square root of a one-loop determinant and captures the one-loop fluctuations of massive modes around the instanton solution. In the CFT approach, these one-loop fluctuations are encoded in the annulus and Moebius strip diagrams, so the factor $\exp(M^{vac}_{D2} + \sum_a A^{vac}_{D2,D6_a})$ in \eqref{formulainstantoncorrelator} corresponds to the Pfaffian/determinant factor in \eqref{superpotentialphysicalgaugeapproach}. Note that also here, the zero modes are not included in the one-loop contribution, supporting the aforementioned idea that they should not appear in the annulus and Moebius strip diagrams. When computing the latter, it turns out \cite{Akerblom:2006hx,Blumenhagen:2007ip,Billo:2007sw,Billo:2007py} that they can only reproduce the absolute value of the Pfaffian/determinant factor, but not the phase. This is presumably related to the ill-definedness of the one-loop CFT amplitude in the Ramond sector with a $(-1)^F$ insertion \cite{Polchinski:1987tu}, where $F$ is the world-sheet fermion number. The reason for this ill-definedness is a bosonic superghost zero mode.

As the partition functions described in section \ref{abstractCFTs}, the annulus and Moebius strip diagrams appearing in \eqref{formulainstantoncorrelator} are a product of three individual CFT amplitudes. The three CFTs are again the ghost/superghost CFT, the CFT of four free bosons and fermions describing the propagation of a superstring in flat four-dimensional space and the internal CFT. The amplitudes in the ghost and free boson/fermion CFTs are universal for the annulus and Moebius strip diagrams. So one can write the full amplitudes in terms of the amplitude $A^{int}_{D2,a}\genfrac[]{0pt}{}{\alpha}{\beta}$ in the internal CFT as
\begin{eqnarray}
 A^{vac}_{D2,D6_a} &=& \int_0^\infty dl \sum_{\alpha,\beta} (-1)^{\alpha+\beta}
        \frac{\vartheta\genfrac[]{0pt}{}{\alpha}{\beta+1/2}(0)^2}{\vartheta\genfrac[]{0pt}{}{1/2}{0}(0)^2}
        \frac{\eta^3}{\vartheta\genfrac[]{0pt}{}{\alpha}{\beta}(0)} A^{int}_{D2,a}\genfrac[]{0pt}{}{\alpha}{\beta}
        \label{annulusD2a1} \\ &=&
        \int_0^\infty dl \sum_{\alpha,\beta} (-1)^{\alpha+\beta} \frac{\vartheta''\genfrac[]{0pt}{}{\alpha}{\beta}(0)}{\eta^3}
        A^{int}_{D2,a}\genfrac[]{0pt}{}{\alpha}{\beta} . \label{annulusD2a2}
\end{eqnarray}
In deriving the second line from the first, theta function identities as well as the fact that the instanton and the brane are mutually BPS have been used. The latter implies
\begin{eqnarray}
\sum_{\alpha,\beta} (-1)^{\alpha+\beta} \vartheta\genfrac[]{0pt}{}{\alpha}{\beta} A^{int}_{D2,a}\genfrac[]{0pt}{}{\alpha}{\beta}=0 .
\end{eqnarray}
Analogously, the Moebius strip diagrams can be written
\begin{eqnarray}
 M^{vac}_{D2} &=& \int_0^\infty dl \sum_{\alpha,\beta} (-1)^{\alpha+\beta}
        \frac{\vartheta\genfrac[]{0pt}{}{\alpha}{\beta+1/2}(0)^2}{\vartheta\genfrac[]{0pt}{}{1/2}{0}(0)^2}
        \frac{\eta^3}{\vartheta\genfrac[]{0pt}{}{\alpha}{\beta}(0)} M^{int}_{D2}\genfrac[]{0pt}{}{\alpha}{\beta}
        \label{moebiusD21} \\ &=&
        \int_0^\infty dl \sum_{\alpha,\beta} (-1)^{\alpha+\beta} \frac{\vartheta''\genfrac[]{0pt}{}{\alpha}{\beta}(0)}{\eta^3}
        M^{int}_{D2}\genfrac[]{0pt}{}{\alpha}{\beta} \label{moebiusD22}
\end{eqnarray}
with $M^{int}_{D2}\genfrac[]{0pt}{}{\alpha}{\beta}$ the amplitude in the internal CFT. Note that the above expressions still contain the effects of the massless modes which have to be removed before the former can be inserted in \eqref{formulainstantoncorrelator}. The aforementioned ill-definedness of certain terms in the one-loop diagrams can be seen in \eqref{annulusD2a1} and \eqref{moebiusD21}. In the summand with the odd spin structure $(\alpha,\beta)=(1/2,1/2)$ one formally divides by $\vartheta\genfrac[]{0pt}{}{1/2}{1/2}=0$. The procedure that will be adopted in the following is not to include this summand. The phase of the superpotential can later be determined by requiring holomorphy and analytically continuing the resulting expressions.

\section{Holomorphy of the superpotential}
\label{holomorphysuperpotential}
There is an interesting relation between the exponent in \eqref{formulainstantoncorrelator} and the gauge coupling on a fictitious D6-brane that would wrap the cycle the instanton wraps or that, in more abstract CFT terms, would be described by the same boundary state in the internal CFT. By comparing \eqref{exponentialinstantonfactor} and \eqref{gaugekineticfunction} one sees that the complexified tree level gauge coupling on this fictitious D6-brane is proportional to the instanton action $S_{D2}=-D^{vac}_{D2}$. If one interprets the annulus and Moebius strip diagrams in the exponent of \eqref{formulainstantoncorrelator} as the one-loop correction to the instanton action, i.e. 
\begin{eqnarray}
 4\pi^2 S_{D2}^{1-loop}=-{M'}^{vac}_{D2} - \sum_a {A'}^{vac}_{D2,D6_a} ,
 \label{relationmoebiusannulusoneloopaction}
\end{eqnarray}
then \eqref{annulusD2a2}, \eqref{moebiusD22}, \eqref{generalformulathresholdcorrections} and \eqref{generalformulathresholdcorrections2} imply that the latter is equal to the one-loop correction to the gauge coupling on the fictitious D6-brane \cite{Abel:2006yk,Akerblom:2006hx}.

This equality directly leads to the following observation: The quantity $S_{D2}^{1-loop}$ just defined appears in a correlation function \eqref{formulainstantoncorrelator} that is supposed to be encoded in a superpotential in the low energy effective theory. It was just argued to be equal to one-loop corrections to gauge coupling constants, which, as was observed in chapter \ref{gaugecouplingoneloop}, depend non-holomorphically on the moduli. So a priori it does not seem possible to encode the correlator \eqref{formulainstantoncorrelator}, in which the non-holomorphic quantity $S_{D2}^{1-loop}$ appears, in a superpotential, which by definition is holomorphic. The solution to this puzzle \cite{Akerblom:2007uc,Akerblom:2007nh} turns out to be that the non-holomorphic terms cancel against other non-holomorphic terms in \eqref{formulainstantoncorrelator}. The way this cancellation occurs is quite similar to the relation between the one-loop gauge threshold corrections and the holomorphic gauge kinetic function discussed in section \ref{holomorphygaugekineticfunction}.

The crucial formula that allows one to disentangle the holomorphic and non-holomorphic terms in the gauge threshold corrections is \eqref{physicalgaugecoupling}. The equality of the latter and the one-loop corrections to the instanton action suggests writing down a similar formula for $S_{D2}^{1-loop}$. \eqref{physicalgaugecoupling} is to be understood recursively, so the one-loop gauge threshold corrections can be written
\begin{eqnarray}
 && 16\pi^2 \left(g_a^{1-loop}\right)^{-2}(\mu^2) = 16\pi^2 {\rm Im}(f_a^{1-loop}) + b_a \ln\frac{\Lambda^2}{\mu^2} + c_a K^{tree}
             \nonumber \\ &&
             + 2 T_a(adj) \ln  \left(g_a^{tree}\right)^{-2}
             - 2 \sum_r T_a(r) \ln \det K^{ab,tree}_r(\mu^2) .
 \label{physicalgaugecoupling1loop}
\end{eqnarray}
Some of the terms in \eqref{physicalgaugecoupling1loop} need to be reinterpreted in the case of the instanton action. $f_a^{1-loop}$ will be replaced by a quantity which one could call the holomorphic one-loop correction to the instanton action and which can appear in the superpotential. The beta function coefficient $b_a$ as well as the constant $c_a$ depend on how many chiral superfields are charged under the gauge group whose associated gauge coupling one computes corrections to and which representations they transform in. They therefore have to be replaced by quantities that are determined by the number and type of instanton zero modes. The quantity $\left(g_a^{tree}\right)^{-2}$ in \eqref{physicalgaugecoupling1loop} clearly corresponds to the tree-level instanton action. Finally, instead of the charged matter field Kaehler metric $K^{ab,tree}_r$ something like a Kaehler metric for instanton zero modes has to appear in the formula for $S_{D2}^{1-loop}$. Of course, it is not clear a priori what this could be. As one is interested in the moduli dependence of this quantity and this dependence comes solely from the internal CFT, in which the instanton enters just as any ordinary space-filling D-brane, one can define this instanton zero mode Kaehler metric $K^{D2}_\lambda$ to be the Kaehler metric of fictitious charged matter that would arise from open strings between the space-filling D-branes that actually exist in the model and a fictitious space-filling D-brane that would be identical to the instanton in the internal CFT.

It was mentioned in section \ref{gaugecouplingoneloop} that there are in general divergences in the string theory expressions \eqref{generalformulathresholdcorrections} and \eqref{generalformulathresholdcorrections2} due to massless open string modes. As explained, the divergence should be replaced by a term that encodes the running of the gauge coupling, i.e. $b_a \ln\Lambda^2/\mu^2$ on the RHS of \eqref{physicalgaugecoupling1loop}. The same divergence also appears in \eqref{annulusD2a2} and \eqref{moebiusD22}, but in this case it is due to instanton zero modes. It was argued above that these zero modes should not be included when computing the one-loop diagrams in the exponent in \eqref{formulainstantoncorrelator}. A term corresponding to the one with prefactor $b_a$ will therefore not appear in the formula for $S_{D2}^{1-loop}$.

Focusing as before on the case of an instanton with only charged zero modes, which transform in the fundamental representation of the gauge group on the instanton worldvolume, in addition to the universal four bosonic and two fermionic ones, the formula relating the one-loop correction of the instanton action to its holomorphic part is 
\begin{eqnarray}
 16\pi^2 S_{D2}^{1-loop} &=& 16\pi^2 {\rm Im}(S_{D2}^{holo,1-loop}) + c_{D2} K^{tree}
             \nonumber \\ &&
             - 4 \ln {\rm Re}(S_{D2}^{tree})
             - 2 \ln \det K^{D2,tree}_\lambda \ \ ,
 \label{holomorphiconeloopinstantonaction}
\end{eqnarray}
where $c_{D2}=N_f-x$ with $N_f$ the number of fermionic zero modes and $x$ a number that can be determined by explicitly computing the Moebius strip diagram.

To proceed further, one notices that a physical, on-shell correlation function
\begin{eqnarray}
 \langle \prod_i \chi_i(x_i) \rangle
\end{eqnarray}
of a certain number of charged matter fields $\chi_i(x_i)$ computed in a supergravity theory is not just given by the prefactor of the product of these fields in the superpotential, even if the coupling in principle arises from the latter. The reason is that the fields in a supergravity action are usually not canonically normalised. The correlation function depends on the Kaehler potential $K$, the Kaehler metrics $K_{\chi_i}$ of the charged matter fields and the superpotential coefficient $W_{\prod_i \chi_i}$, which encodes the coupling of the fields $\chi_i$, as follows \cite{Cremades:2003qj,Cvetic:2003ch,Abel:2003vv,Cremades:2004wa}:
\begin{eqnarray}
 \langle \prod_i \chi_i(x_i) \rangle = 
 \frac{\exp(K/2) W_{\prod_i \chi_i}}{\sqrt{\prod_i K_{\chi_i}}}
 \label{supergravitycorrelator}
\end{eqnarray}
The correlation functions one computes in string theory using CFT are physical quantities, so that in order to extract the superpotential from such a correlator one must know the Kaehler potential and Kaehler metrics and use \eqref{supergravitycorrelator}. Note that this is true for any correlation function, not just those discussed in this chapter, which arise in an instanton background.

Not only the full spacetime correlator on the LHS of \eqref{formulainstantoncorrelator} is of the form \eqref{supergravitycorrelator}, but also the individual CFT disc amplitudes on the RHS, which are responsible for the absorption of zero modes. This is because they are CFT correlators that in the internal CFT are identical to correlators which involve just charged matter fields and which are encoded in the tree-level superpotential (and, in this context, the Kaehler potential and Kaehler metric) of the low energy effective action. They therefore have the same moduli dependence and are to be disentangled according to \eqref{supergravitycorrelator} into holomorphic superpotential parts and non-holomorphic Kaehler potential/Kaehler metric parts.

So the disc correlators $D(\lambda,\lambda,\phi_i)$ in \eqref{formulainstantoncorrelator} can be written in the form \eqref{supergravitycorrelator} using the aforementioned Kaehler metric for charged zero modes as
\begin{eqnarray}
 D(\lambda,\lambda,\phi_i) &=& \frac{\exp(K/2) W_{\lambda\lambda\prod_i \phi_i}}{\sqrt{K_\lambda K_\lambda \prod_i K_{\phi_i}}}
 \label{discentangled1}
\end{eqnarray}
with $W_{\lambda\lambda\prod_i \phi_i}$ a holomorphic function of the moduli.

In order to determine the form of $D(\theta^\alpha,\lambda,\lambda,\psi_\alpha,\phi_i)$ one has to note that the zero modes $\theta^\alpha$ correspond to gauginos in terms of fields on ordinary D-branes. In contradistinction to charged matter fields, which are rescaled by a square root of their Kaehler metric, gauginos have to be rescaled by the square root of the imaginary part of the gauge kinetic function in order to render them canonically normalised. As the gauge kinetic function corresponds to the instanton action, $\sqrt{{\rm Re}(S_{D2}^{tree})}$ should appear in a formula for $D(\theta^\alpha,\lambda,\lambda,\psi_\alpha,\phi_i)$ that is analogous to \eqref{supergravitycorrelator} or \eqref{discentangled1}. Other than that, this disc diagram does not really correspond to a superpotential coupling due to the appearance of a gaugino-like mode, but is related by supersymmetry to the disc diagram in \eqref{discentangled1} so one expects that
\begin{eqnarray}
 D(\theta^\alpha,\lambda,\lambda,\psi_\alpha,\phi_i) =
 \frac{\exp(K/2) W_{\theta^\alpha\lambda\lambda\psi_\alpha\prod_i \phi_i}}
      {\sqrt{{\rm Re}(S_{D2}^{tree})K_\lambda K_\lambda K_{\psi_\alpha} \prod_i K_{\phi_i}}} ,
 \label{discentangled2}
\end{eqnarray}
where $W_{\theta^\alpha\lambda\lambda\psi_\alpha\prod_i \phi_i}$ is holomorphic.

One can now put everything together, i.e. insert \eqref{relationmoebiusannulusoneloopaction}, \eqref{holomorphiconeloopinstantonaction}, \eqref{discentangled1} and \eqref{discentangled2} into \eqref{formulainstantoncorrelator}. Using $\prod_{i=1}^{N_f} \sqrt{K_{\lambda^i}}=\sqrt{\det(K_\lambda)}$ and reinstating the imaginary part of the holomorphic one-loop correction to the instanton action one finds
\begin{eqnarray}
 && \langle \psi_1(x_1) \psi_2(x_2) \prod_{i=3}^N \phi_i(x_i) \rangle \nonumber \\ &=& \sum_{conf}
 \frac{\exp(xK/4) \left(\prod W_{\lambda\lambda\prod_i \phi_i}\right)
                  \left(\prod_{\alpha=1}^2 W_{\theta^\alpha\lambda\lambda\psi_\alpha\prod_i \phi_i}\right)}
      {\sqrt{\prod_i K_{\phi_i} \prod_{\alpha=1}^2 K_{\psi_\alpha}}} \times \nonumber \\ &&
 \exp\left( - S_{D2} - 4 \pi^2 S_{D2}^{holo,1-loop} \right) ,
\end{eqnarray}
which is of the form \eqref{supergravitycorrelator} if $x=2$. This is thus a consistency condition.

In summary, although the various CFT amplitudes that appear in \eqref{formulainstantoncorrelator} depend non-holomorphically on the moduli, the non-holomorphic terms partly cancel, partly rearrange so as to yield a result that is in agreement with the holomorphy of the superpotential. Note that this only happens if precisely two fermionic zero modes are absorbed with one disc diagram.
\chapter{Applying the D-instanton calculus: The ADS superpotential}
\label{ADSsuperpotential}

The instanton calculus described in the previous chapter cannot be derived from first principles. Quite a bit of guesswork went into setting it up, so it is crucial to test it, e.g. by using it to rederive known results. Therefore, it will now be applied to a certain D-instanton, one which realises a gauge instanton, in a D-brane realisation of an SQCD theory with gauge group $SU(N_c)$ and $N_c-1$ flavours. It can be shown that that gauge instanton generates the so-called ADS-superpotential \cite{Affleck:1983mk,Cordes:1985um}.

In field theory, it can be computed in a semiclassical approximation to the path integral. The fields are expanded around a particular instanton solution, i.e. a topologically non-trivial field configuration solving the Euclidean Yang-Mills equations, and the path integral is evaluated in a one-loop approximation. As string theory can be approximated by field theory at low energies and the ADS superpotential has implications for the low-energy physics, one expects that it can also be obtained from string theory. The first thing to observe when trying to show this is that a (zero size) gauge instanton is realised in string theory by a D(p-4)-brane inside a stack of Dp-branes \cite{Witten:1995im,Douglas:1995bn,Douglas:1996uz,Dorey:2000ww,Dorey:2002ik}. The position of the D(p-4)-brane coincides with the core of the gauge instanton in the field theory on the Dp-branes. The effect of such a D-brane instanton cannot be computed as straightforwardly as that of a gauge instanton, as there is no such thing as a (spacetime) path integral for string theory. But the D-instanton calculus presented in the previous chapter is applicable to D-brane instantons realising gauge theory instantons and can thus be used to rederive the ADS-superpotential \cite{Akerblom:2006hx,Argurio:2007vq,Bianchi:2007wy,Bianchi:2007fx}.
\section{Engineering SQCD}
The first step in reproducing the ADS superpotential by a D-instanton computation is to engineer the SQCD gauge theory in a D-brane setup \cite{Katz:1996fh,Katz:1996th}. This shall here be done in the framework of intersecting D6-brane models on Calabi-Yau manifolds, which were discussed in section \ref{D6branemodelsCY}. One considers a stack $c$ of $N_c$ D6-branes wrapping a special Lagrangian three-cycle in some Calabi-Yau manifold. This setup leads to a $U(N_c)$ gauge theory, but the diagonal $U(1)$ factor of this gauge group decouples at low energies, such that one ends up with the required group $SU(N_c)$. In order to decouple the gravitational modes, one needs to take a limit in which the volume of the Calabi-Yau manifold becomes infinitely large. One must ensure that, when taking this limit, the volume $V_c$ of the three-cycle wrapped by the stack $c$ remains finite such that the gauge coupling does not vanish. Furthermore, reproducing SQCD requires taking the field theory limit, i.e. $\alpha'\rightarrow0$. In addition, one has to make sure that the matter content of the theory is that of SQCD. This means that there must not be any chiral multiplets in the adjoint representation of the gauge group which implies that the three-cycle wrapped by the stack $c$ must be rigid. As was already mentioned, the ADS superpotential is generated by a gauge instanton iff there are $N_c-1$ flavours, i.e. $N_c-1$ chiral multiplets $\Phi_{cf}$, $c\in\{1,...,N_c\}$, $f\in\{1,...,N_c-1\}$, transforming in the fundamental representation of the gauge group and $N_c-1$ chiral multiplets $\widetilde{\Phi}_{ca}$, $c\in\{1,...,N_c\}$, $a\in\{1,...,N_c-1\}$, transforming in the antifundamental representation of the gauge group. Such fields arise in intersecting D-brane models from strings stretching between the stack $c$ and a different stack of branes. This means that one has to introduce two further stacks of branes, denoted $f$ and $a$ in the following and consisting of $N_c-1$ branes each. The intersections of $f$ and $a$ with $c$ have to be oriented differently such that the strings between $f$ and $c$ transform in the fundamental representation and those between $a$ and $c$ in the antifundamental representation of $SU(N_c)$. There are also gauge theories on the stacks $a$ and $f$. Reproducing just SQCD means that they have to decouple. To achieve this one has to take the infinite volume limit of the Calabi-Yau manifold in such a way that the volumes $V_a$ and $V_f$ of the cycles wrapped by $a$ and $f$ tend to infinity.

Summarising, an SQCD theory with gauge group $SU(N_c)$ and $N_c-1$ flavours can be obtained from an intersecting D6-brane model on a Calabi-Yau manifold with three stacks of branes, $c$, $f$ and $a$, by taking the limit
\begin{eqnarray}
 \alpha' &\rightarrow& 0 \nonumber \\
 V_{CY} &\rightarrow& \infty \nonumber \\
 V_f &\rightarrow& \infty \nonumber \\
 V_a &\rightarrow& \infty \nonumber \\
 V_c && finite \ .
 \label{fieldtheorylimit}
\end{eqnarray}
The quiver diagram for the SQCD theory with the three stacks of branes and the charged matter fields is given in figure \ref{SQCDquiver}.
\section{The relevant instanton}
As was already mentioned in the previous chapter, a D2-instanton wrapping the same three-cycle as a space-filling D6-brane is the string theory realisation of a gauge theory instanton. This means that the D-instanton that is expected to reproduce the ADS superpotential is the D2-instanton wrapping the three-cycle that is also wrapped by the stack $c$. According to the analysis of the previous chapter this instanton will have the following zero modes:

There are eight neutral zero modes, four bosonic ones related to broken translational invariance and four fermionic ones related to broken supersymmetries.

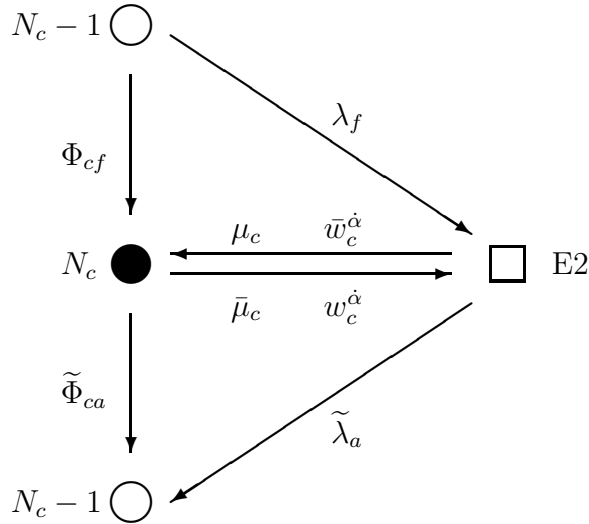
\begin{figure}[ht]
  \begin{center}
    \setlength{\unitlength}{0.75pt}
    \begin{picture}(270,280)(-135,-135)
      \thicklines\put(-100,0){\circle*{20}}
      \thicklines\put(-100,120){\circle{20}}
      \thicklines\put(-100,-120){\circle{20}}
      \thicklines\put(80,-8){\framebox(16,16){}}

      \thicklines\put(-80,-5){\vector(1,0){140}}
      \thicklines\put(60,5){\vector(-1,0){140}}
      \thicklines\put(-100,95){\vector(0,-1){70}}
      \thicklines\put(-100,-25){\vector(0,-1){70}}
      \thicklines\put(-80,115){\vector(3,-2){150}}
      \thicklines\put(70,-20){\vector(-3,-2){150}}

      \put(-135,-5){$N_c$}
      \put(-160,-125){$N_c-1$}
      \put(-160,115){$N_c-1$}
      \put(110,-5){$\mbox{E2}$}

      \put(-50,-26){$\bar{\mu}_c \qquad w^{\dot{\alpha}}_c $}
      \put(-50,13){$\mu_c \qquad \bar{w}^{\dot{\alpha}}_c $}

      \put(-135,-70){$\widetilde{\Phi}_{ca}$}
      \put(-135,50){$\Phi_{cf}$}

      \put(0,-90){$\widetilde{\lambda}_a$}
      \put(0,70){$\lambda_f$}
    \end{picture}
    \caption{SQCD quiver including instanton\label{SQCDquiver}}
  \end{center}
\end{figure}

In addition, there are $4N_c$ bosonic and $2N_c$ fermionic zero modes from strings stretching between the instanton and the stack $c$. As in the previous section they will be denoted $w^{\dot{\alpha}}$, $\bar{w}^{\dot{\alpha}}$, $\mu$ and $\bar{\mu}$, but carry an extra label $c$ which runs from $1$ to $N_c$. Finally there are $N_c-1$ fermionic zero modes $\lambda_f$, $f\in\{1,...,N_c-1\}$ and $N_c-1$ fermionic zero modes $\widetilde{\lambda}_a$, $a\in\{1,...,N_c-1\}$. These zero modes are strings with one end on the instanton and one on the stack f, or a, respectively. The instanton and its charged zero modes are shown in the SQCD quiver depicted in figure \ref{SQCDquiver}.

The formula \eqref{formulainstantoncorrelator} for a spacetime correlator in a D-instanton background needs to be slightly extended for the case at hand as there are additional zero modes, namely the neutral zero modes $\bar{\tau}^{\dot{\alpha}}$ as well as the bosonic charged zero modes $w^{\dot{\alpha}}$ and $\bar{w}^{\dot{\alpha}}$. Also, it is convenient to rewrite it as an explicit integral over all zero modes. This requires that instead of computing a sum over products of disc diagrams one exponentiates the sum of all possible disc diagrams. The Grassmann integration over the fermionic zero modes ensures that each one of them effectively appears precisely once. The integral to be evaluated is
\begin{eqnarray}
 && \int d^4 x d^2\theta d^2\bar{\tau} \prod_{c=1}^{N_c} d^4 w_c d^2\mu_c \prod_{i=1}^{N_c-1} d\lambda_i d\widetilde{\lambda}_i \times \nonumber \\
 && \exp \left( D_{D2}^{vac} + {A'}_{D2}^{vac} + {M'}_{D2}^{vac} + \mathcal{L}_{D2-c} + \mathcal{L}_{D2-c-f} + \mathcal{L}_{D2-c-a} \right) .
 \label{integralADSsuperpotential}
\end{eqnarray}
Some comments on \eqref{integralADSsuperpotential} are in order. $\mathcal{L}_{D2-c}$ denotes the sum of all disc diagrams with boundary segments on the instanton and the brane stack $c$, $\mathcal{L}_{D2-c-f}$ and $\mathcal{L}_{D2-c-a}$ stand for those with boundary segments on the instanton, brane stack $c$ and brane stack $f$ or $a$. Note that disc diagrams with boundaries only on the instanton and brane stack $f$ or $a$ are not possible due to combinatorics. As was explained in the previous chapter, the massless modes are not to be included when computing ${A'}_{D2}^{vac}$ and ${M'}_{D2}^{vac}$. In the present case one has to take the limit \eqref{fieldtheorylimit} which means that massive modes are not be taken into account, too. Therefore one must set ${A'}_{D2}^{vac}={M'}_{D2}^{vac}=0$ in \eqref{integralADSsuperpotential}. The limit \eqref{fieldtheorylimit} also implies that only a subset of all possible disc diagrams contributes to the exponential in \eqref{integralADSsuperpotential}.

It was argued in the previous section that the vacuum disc diagram $D_{D2}^{vac}$ is equal to (minus) the instanton action $S_{D2}$. The latter is equal to the volume of the three-cycle wrapped by the instanton and the brane stack $c$ and therefore to (the inverse square of) the gauge coupling $g^{-2}_{SU(N_c)}$ of the $SU(N_c)$ gauge theory on brane stack $c$. The factor $\exp \left( D_{D2}^{vac} \right)$ thus reproduces the factor $\exp \left( -g^{-2}_{SU(N_c)} \right)$ that appears in the gauge instanton computation.

The first step in evaluating \eqref{integralADSsuperpotential} is to compute $\mathcal{L}_{D2-c}$, $\mathcal{L}_{D2-c-f}$ and $\mathcal{L}_{D2-c-a}$. They consist of disc diagrams involving the charged matter fields $\Phi$, $\widetilde{\Phi}$ and the instanton zero modes. The vertex operators for the charged matter fields in the (-1)-ghost picture take the form \eqref{chiralsuperfieldvertexoperators}
\begin{eqnarray}
 V_{\Phi_{cf}} &=& e^{-\phi(z)} e^{ik_\mu X^\mu(z)} \mathcal{O}^{1/2}_{cf,+1}(z)
 \qquad
 V_{\Phi^*_{cf}} = e^{-\phi(z)} e^{ik_\mu X^\mu(z)} \mathcal{O}^{1/2}_{cf,-1}(z)
 \\
 V_{\widetilde{\Phi}_{ca}} &=& e^{-\phi(z)} e^{ik_\mu X^\mu(z)} \mathcal{O}^{1/2}_{ca,-1}(z)
 \qquad
 V_{\widetilde{\Phi}^*_{ca}} = e^{-\phi(z)} e^{ik_\mu X^\mu(z)} \mathcal{O}^{1/2}_{ca,+1}(z) .
\end{eqnarray}
The worldsheet-$U(1)$-charge of the vertex operators for $\Phi$ and $\widetilde{\Phi}$ is different as these fields transform in conjugate representations of the gauge group. The vertex operators in the (0)-ghost picture read:
\begin{eqnarray}
 V_{\Phi_{cf}} &=& e^{ik_\mu X^\mu(z)} \left( i k_\mu \psi^\mu(z) \mathcal{O}^{1/2}_{cf,+1}(z) + \mathcal{O}^{1}_{cf}(z) \right)
 \\
 V_{\Phi^*_{cf}} &=& e^{ik_\mu X^\mu(z)} \left( i k_\mu \psi^\mu(z) \mathcal{O}^{1/2}_{cf,-1}(z) + \mathcal{O'}^{1}_{cf}(z) \right)
 \\
 V_{\widetilde{\Phi}_{ca}} &=& e^{ik_\mu X^\mu(z)} \left( i k_\mu \psi^\mu(z) \mathcal{O}^{1/2}_{ca,-1}(z) + \mathcal{O}^{1}_{ca}(z) \right)
 \\
 V_{\widetilde{\Phi}^*_{ca}} &=& e^{ik_\mu X^\mu(z)} \left( i k_\mu \psi^\mu(z) \mathcal{O}^{1/2}_{ca,+1}(z) + \mathcal{O'}^{1}_{ca}(z) \right)
\end{eqnarray}
The vertex operators for the neutral zero modes $x$, $\theta$, $\bar{\tau}$ were given in \eqref{vertexoperatorsx} and \eqref{vertexoperatorsthetatau}, those for the charged zero modes $w^{\dot{\alpha}}$ and $\mu$ in \eqref{vertexoperatorsw} and \eqref{vertexoperatorsmu}. The vertex operators for the zero modes $\lambda_f$ and $\widetilde{\lambda}_a$ are
\begin{eqnarray}
 V_{\lambda_f} &=& e^{-\phi(z)/2} \prod_{\mu=0}^3 T_{X^\mu}(z) \mathcal{O}^{3/8}_{cf,-1/2}(z) \\
 V_{\widetilde{\lambda}_a} &=& e^{-\phi(z)/2} \prod_{\mu=0}^3 T_{X^\mu}(z) \mathcal{O}^{3/8}_{ca,+1/2}(z) .
\end{eqnarray}
As the instanton and the brane stack $c$ are described by the same boundary state in the internal CFT, the operators $\mathcal{O}^{3/8}_{cf,-1/2}(z)$ and $\mathcal{O}^{3/8}_{ca,+1/2}(z)$ also appear in the vertex operators of the chiral fermions that are the superpartners of $\Phi$ and $\widetilde{\Phi}$, hence the label $c$.
\section{Computing the disc diagrams}
The relevant disc diagrams will now be computed. From the quiver diagram \ref{SQCDquiver} one can read off which correlators of instanton zero modes and/or charged matter fields are allowed by boundary combinatorics. Further important selection rules come from $U(1)$-worldsheet charge conservation.

One starts by considering the disc diagrams involving only instanton zero modes. These are the diagrams relevant for $\mathcal{L}_{D2-c}$. It turns out \cite{Billo:2002hm} that they reproduce the ADHM constraints \cite{Khoze:1998gy,Dorey:2002ik,Bianchi:2007ft}, which appear in the gauge instanton computation. In order for this to happen, one has to rescale the zero modes in a particular way \cite{Billo:2002hm} and take the limit \eqref{fieldtheorylimit}. Finally, $\mathcal{L}_{D2-c}$ becomes
\begin{eqnarray}
 \mathcal{L}_{D2-c} = \bar{\tau}_{\dot{\alpha}} \left( w^{\dot{\alpha}}_c \bar{\mu}_c + \bar{w}^{\dot{\alpha}}_c \mu_c \right)
 + i D^i \bar{w}^{\dot{\alpha}}_c w^{\dot{\beta}}_c (\tau^i)_{\dot{\alpha}\dot{\beta}} ,
 \label{lagrangianD2c}
\end{eqnarray}
where the $\tau^i$ are the Pauli matrices, $D^i$ are auxiliary fields that need to be integrated over and summation over repeated indices is understood. Exponentiating $\mathcal{L}_{D2-c}$ and integrating over $D^i$ and $\bar{\tau}_{\dot{\alpha}}$ yields
\begin{eqnarray}
 \int d^3D d^2\bar{\tau} \exp\left( \mathcal{L}_{D2-c} \right)
 = \prod_{i=1}^3 \delta\left( \bar{w}^{\dot{\alpha}}_c w^{\dot{\beta}}_c (\tau^i)_{\dot{\alpha}\dot{\beta}} \right)
   \prod_{\dot{\alpha}=1}^2 \delta \left( w^{\dot{\alpha}}_c \bar{\mu}_c + \bar{w}^{\dot{\alpha}}_c \mu_c \right),
\end{eqnarray}
i.e. delta-functions incorporating the bosonic and fermionic ADHM constraints. The bosonic ones can be interpreted as the D- and F-term constraints ensuring supersymmetry on the worldvolume of the instanton. 

Next, the disc diagrams involving charged matter fields and charged fermionic instanton zero modes will be determined \cite{Cvetic:2007ku}. One such diagram is the three-point function
\begin{eqnarray}
 \langle \Phi^*_{cf} \lambda_f \mu_c \rangle
\end{eqnarray}
which can be computed using the correlators ($z_{ij}=z_i-z_j$)
\begin{eqnarray}
 \langle \prod_{\mu=0}^3 T_{X^\mu}(z_1) T_{X^\mu}(z_2) \rangle &=& z_{12}^{-1/2} \\
 \langle e^{-\phi(z_1)} e^{-\phi(z_2)/2} e^{-\phi(z_3)/2} \rangle
         &=& z_{12}^{-1/2} z_{13}^{-1/2} z_{23}^{-1/4} \\
 \langle \mathcal{O}^{1/2}_{cf,-1}(z_1) \mathcal{O}^{3/8}_{cf,-1/2}(z_2) \mathcal{O}^{3/8}_{3/2}(z_3) \rangle
         &=& z_{12}^{-1/2} z_{13}^{-1/2} z_{23}^{-1/4} .
\end{eqnarray}
Taking also the diagram
\begin{eqnarray}
 \langle \widetilde{\Phi}^*_{ca} \bar{\mu}_c \widetilde{\lambda}_a \rangle ,
\end{eqnarray}
which can be computed similarly, into account one obtains
\begin{eqnarray}
 \mathcal{L}_{D2-c-f} + \mathcal{L}_{D2-c-a} = \Phi^*_{cf} \lambda_f \mu_c + \widetilde{\Phi}^*_{ca} \bar{\mu}_c \widetilde{\lambda}_a + ... \ \ .
\end{eqnarray}
Finally, also the disc diagrams with insertions of charged bosonic zero modes and matter fields have to be computed. One such diagram is the four-point amplitude
\begin{eqnarray}
 \langle \bar{w}^{\dot{\beta}}_c \Phi^*_{cf} \Phi_{c'f} w^{\dot{\alpha}}_{c'} \rangle .
\end{eqnarray}
If one chooses the vertex operators for the charged matter fields in the (0)-ghost picture and those for the zero modes in the (-1)-ghost picture, one needs the following correlators \cite{Kostelecky:1986xg}:
\begin{eqnarray}
 && \langle \mathcal{O}^{1/2}_{cf,-1}(z_2) \mathcal{O}^{1/2}_{cf,+1}(z_3) \rangle = z_{23}^{-1}
 \\
 && \langle \mathcal{O'}^{1}_{cf}(z_2) \mathcal{O}^{1}_{cf}(z_3) \rangle = z_{23}^{-2}
 \\
 && \langle e^{-\phi(z_1)} e^{-\phi(z_4)} \rangle = z_{14}^{-1}
 \\
 && \langle S^{\dot{\beta}}(z_1) S^{\dot{\alpha}}(z_4) \rangle = -\epsilon^{\dot{\beta}\dot{\alpha}} z_{14}^{-1/2}
 \\
 && - 2 \langle S^{\dot{\beta}}(z_1) \psi^\mu(z_2) \psi^\nu(z_3) S^{\dot{\alpha}}(z_4) \rangle = 
 (\bar{\sigma}^{\mu\nu})^{\dot{\beta}\dot{\alpha}} z_{14}^{1/2} z_{13}^{-1/2} z_{24}^{-1/2} z_{12}^{-1/2} z_{34}^{-1/2}
 \nonumber \\
 && + \delta^{\mu\nu} \epsilon^{\dot{\beta}\dot{\alpha}}
 (z_{12}z_{34}+z_{13}z_{24})
 z_{23}^{-1} z_{14}^{-1/2} z_{13}^{-1/2} z_{24}^{-1/2} z_{12}^{-1/2} z_{34}^{-1/2}
 \\
 && \langle \left( \Pi_{\mu=0}^3 T_{X^\mu}(z_1) \right)
 e^{ik^{\Phi^*}_\mu X^\mu(z_2)} e^{ik^{\Phi}_\mu X^\mu(z_3)}
 \left( \Pi_{\mu=0}^3 T_{X^\mu}(z_4) \right) \rangle
\end{eqnarray}
The last correlator can be determined from the two point function \cite{Ginsparg:1988ui}
\begin{eqnarray}
 \langle X^\mu(z_2) X^\nu(z_3) \rangle_T = -\delta^{\mu\nu} \ln \frac{z_2-z_3}{(\sqrt{z_2}+\sqrt{z_3})^2}
\end{eqnarray}
in the $\mathbb{Z}_2$-twisted sector of the free boson and the result is:
\begin{eqnarray}
 && \langle \left( \Pi_{\mu=0}^3 T_{X^\mu}(z_1) \right)
 e^{ik^{\Phi^*}_\mu X^\mu(z_2)} e^{ik^{\Phi}_\mu X^\mu(z_3)}
 \left( \Pi_{\mu=0}^3 T_{X^\mu}(z_4) \right) \rangle \\
 &=& z_1^{-1/2} \left(\frac{z_2-z_3}{(\sqrt{z_2}+\sqrt{z_3})^2}\right)^{-(k^\Phi)^2} \delta(k^\Phi+k^{\Phi^*})
\end{eqnarray}
Putting everything together, letting $z_1\rightarrow\infty$, $z_2=1$, $z_4=0$ and substituting $x^2=z_3$ one finds
\begin{eqnarray}
 \langle \bar{w}^{\dot{\beta}}_c \Phi^*_{cf} \Phi_{c'f} w^{\dot{\alpha}}_{c'} \rangle
 = \epsilon^{\dot{\beta}\dot{\alpha}} \int_0^1 dx \frac{(1+x)^{(k^\Phi)^2-2}}{(1-x)^{(k^\Phi)^2+2}} ( 2x + (k^\Phi)^2 (x^2+1) ) .
\end{eqnarray}
The integral converges for $(k^\Phi)^2<-1$ and is defined by analytic continuation for other values of $k^\Phi$ \cite{Polchinski:1998rq}. This means that
\begin{eqnarray}
 \langle \bar{w}^{\dot{\beta}}_c \Phi^*_{cf} \Phi_{c'f} w^{\dot{\alpha}}_{c'} \rangle = -\frac{\epsilon^{\dot{\beta}\dot{\alpha}}}{2} .
\end{eqnarray}
The four-point amplitude $\langle \bar{w}^{\dot{\beta}}_c \widetilde{\Phi}_{ca} \widetilde{\Phi}^*_{c'a} w^{\dot{\alpha}}_{c'} \rangle$ can be computed analogously and one concludes that
\begin{eqnarray}
 \mathcal{L}_{D2-c-f} + \mathcal{L}_{D2-c-a} =
 - \bar{w}^{\dot{\alpha}}_c \Phi^*_{cf} \Phi_{c'f} w_{\dot{\alpha}c'}
 - \bar{w}^{\dot{\alpha}}_c \widetilde{\Phi}_{ca} \widetilde{\Phi}^*_{c'a} w_{\dot{\alpha}c'} + ... \ \ .
\end{eqnarray}
There can in principle be non-vanishing disc diagrams with more insertions of matter fields but they vanish in the limit \eqref{fieldtheorylimit}. Furthermore, as argued in the previous chapter, precisely two fermionic zero modes should appear on each disc diagram. The same is true for bosonic zero modes. Therefore one must use
\begin{eqnarray}
 \mathcal{L}_{D2-c-f} + \mathcal{L}_{D2-c-a} &=& \Phi^*_{cf} \lambda_f \mu_c + \widetilde{\Phi}^*_{ca} \widetilde{\lambda}_a \bar{\mu}_c
 \nonumber \\ &&
 - \bar{w}^{\dot{\alpha}}_c \Phi^*_{cf} \Phi_{c'f} w_{\dot{\alpha}c'}
 - \bar{w}^{\dot{\alpha}}_c \widetilde{\Phi}_{ca} \widetilde{\Phi}^*_{c'a} w_{\dot{\alpha}c'}
 \label{lagrangianD2cfa}
\end{eqnarray}
in \eqref{integralADSsuperpotential}.
\section{Zero mode integration}
The integral to be evaluated becomes
\begin{eqnarray}
 && \int d^4 x d^2\theta d^2\bar{\tau} d^3D \prod_{c=1}^{N_c} d^4 w_c d^2\mu_c \prod_{i=1}^{N_c-1} d\lambda_i d\widetilde{\lambda}_i
 \times \nonumber \\
 && \exp \left( D_{D2}^{vac} + \mathcal{L}_{D2-c} + \mathcal{L}_{D2-c-f} + \mathcal{L}_{D2-c-a} \right)
 \label{integralADSsuperpotential2}
\end{eqnarray}
with $\mathcal{L}_{D2-c}$ given in \eqref{lagrangianD2c} and $\mathcal{L}_{D2-c-f} + \mathcal{L}_{D2-c-a}$ in \eqref{lagrangianD2cfa}. The charged matter fields appearing in these expressions should be thought of as their vacuum expectation values. 

The first step in evaluating \eqref{integralADSsuperpotential2} is to perform the fermionic integral. One finds:
\begin{eqnarray}
 && \int d^2\bar{\tau}\prod_{c=1}^{N_c} d^2\mu_c \prod_{i=1}^{N_c-1} d\lambda_i d\widetilde{\lambda}_i
 \exp \left( \bar{\tau}_{\dot{\alpha}} \left( w^{\dot{\alpha}}_c \bar{\mu}_c + \bar{w}^{\dot{\alpha}}_c \mu_c \right)
 + \Phi^*_{cf} \lambda_f \mu_c + \widetilde{\Phi}^*_{ca} \bar{\mu}_c \widetilde{\lambda}_a \right) \nonumber \\
 && = \sum_{c,c'=1}^{N_c} (-1)^{c+c'}
 \bar{w}^{\dot{\alpha}}_{c} w_{\dot{\alpha}c'} \det\left[ \left( \Phi^*_{df} \widetilde{\Phi}^*_{d'f} \right)'_{cc'} \right]
\end{eqnarray}
In this expression $\Phi^*_{df} \widetilde{\Phi}^*_{d'f}$ is an $N_c\times N_c$ matrix and $( \Phi^*_{df} \widetilde{\Phi}^*_{d'f} )'_{cc'}$ an $(N_c-1)\times (N_c-1)$ matrix obtained from $\Phi^*_{df} \widetilde{\Phi}^*_{d'f}$ by deleting the $c$'th row and $c'$'th column. Note here that the fermionic integral leads to the fact that a superpotential is only generated if there are precisely $N_c-1$ flavours. It otherwise gives zero.

The next step is to evaluate the integral over the bosonic zero modes. After regularising it by adding the term $\epsilon\bar{w}^{\dot{\alpha}}_c w_{\dot{\alpha}c}$ in the exponent it becomes
\begin{eqnarray}
 && \int \prod_{c=1}^{N_c} d^4w_c \bar{w}^{\dot{\alpha}}_{c} w_{\dot{\alpha}c'}
 \exp \left(
            i D^i \bar{w}^{\dot{\alpha}}_c w^{\dot{\beta}}_c (\tau^i)_{\dot{\alpha}\dot{\beta}}
            - \bar{w}^{\dot{\alpha}}_c w_{\dot{\alpha}c'} M_{cc'}
      \right)
 \\ &=& \frac{ M'_{cd} \left( (M^2+D^2)^{-1} \right)'_{dc'} }{ \det(M^2+D^2) } ,
\end{eqnarray}
where $M_{cc'}=\Phi^*_{cf} \Phi_{c'f} + \widetilde{\Phi}_{ca} \widetilde{\Phi}^*_{c'a} + \epsilon \delta_{cc'}$ and $D^2=\sum_i (D^i)^2$. The final integral over $D^i$ can be performed after using the D-flatness condition $\Phi_{cf}=\widetilde{\Phi}^*_{cf}$ \cite{Terning:2006bq}. This is justified as the instanton calculus is only valid for BPS-instantons in supersymmetric configurations. Letting $\epsilon\rightarrow0$, one recovers the ADS superpotential
\begin{eqnarray}
 \int d^4 x d^2\theta \frac{\exp\left(-g_{SU(N_c)}^{-2}\right)}{\det_{fa} \Phi_{cf} \widetilde{\Phi}_{ca}} .
\end{eqnarray}
Note that no non-holomorphic terms stemming from non-canonical Kaehler potentials have to be dealt with as the field theory limit had been taken. In conclusion, the ADS superpotential can be obtained using the D-instanton calculus described in the previous chapter. However, the limit \eqref{fieldtheorylimit} had to be taken which means that there will be corrections to the field theory result if an ADS-like superpotential is generated by a D-instanton in a full globally consistent string theory model.

Interestingly, it turns out that in string theory, a D-instanton wrapping a cycle that is wrapped by only one brane (i.e. $N=1$ in the setup considered above) can generate a superpotential \cite{Aganagic:2007py,GarciaEtxebarria:2007zv,Petersson:2007sc}, although there is no gauge theory interpretation of this D-instanton, as a $U(1)$ gauge theory does not admit topologically non-trivial solutions.
\section{Other gauge groups}
In theories with gauge groups other than $SU(N_c)$, gauge instantons generate superpotentials, too. Examples are $USp(2N_c)$ and $O(N_c)$. It shall be sketched in the following that these superpotentials can also be recovered in a D-instanton computation.

In order to engineer a theory with unitary symplectic or orthogonal gauge group one considers an orientifold of a Calabi-Yau compactification and wraps a stack $c$ of $N_c$ D6-branes on a three-cycle of the internal manifold which is invariant under the antiholomorphic involution that is part of the orientifold projection. The precise form of the latter determines whether the gauge group on the worldvolume of the brane is unitary symplectic or orthogonal. Absence of chiral multiplets transforming in the adjoint representation again requires the three-cycle to be rigid. The relevant D-instanton is once more one that wraps the same cycle as the stack $c$ of space-filling D-branes.

If the gauge group on the D-branes is $USp(2N_c)$, a superpotential is generated if there are $2N_c$ flavours $\Phi_{cf}$, $c,f\in\{1,...,2N_c\}$ transforming in the fundamental representation of $USp(2N_c)$. So one introduces a second stack $f$ of $2N_c$ branes which intersects the first one such that the required chiral multiplets arise as strings stretching between the two stacks of branes. The gauge theory on the second stack of branes must again decouple, so the volume of the cycle it wraps must become infinitely large when taking a limit like \eqref{fieldtheorylimit}.

The gauge group on the instanton in this case is $O(1)$, so, as discussed in the previous chapter, there are four bosonic and two fermionic neutral zero modes. The fermionic zero modes $\bar{\tau}^{\dot{\alpha}}$ as well as some of the charged zero modes are removed by the orientifold projection. It turns out that $2N_c$ fermionic zero modes $\mu_c$ and $4N_c$ bosonic zero modes $w^{\dot{\alpha}}_c$, $c\in\{1,...,2N_c\}$, survive in the sector of open strings between stack $c$ and the instanton. Furthermore, there are $2N_c$ fermionic zero modes $\lambda_f$, $f\in\{1,...,2N_c\}$ from strings stretching between the instanton and brane stack $f$. As the zero modes $\bar{\tau}^{\dot{\alpha}}$ are projected out, there are no fermionic ADHM constraints \cite{Hollowood:1999ev}. There are also no bosonic ADHM constraints \cite{Hollowood:1999ev}. This can be understood from the fact that the gauge theory on the instanton worldvolume is trivial which implies that no D- or F-term constraints arise. The integral to be evaluated is
\begin{eqnarray}
 && \int d^4x d^2\theta \prod_{c=1}^{2N_c} d^2w_c d\mu_c d\lambda_c
 \exp \left( D_{D2}^{vac} + \mu_c \Phi^*_{cf} \lambda_f - w^{\dot{\alpha}}_c \Phi_{cf} \Phi^*_{c'f} w_{\dot{\alpha}c'} \right)
 \nonumber \\
 &=& \int d^4x d^2\theta \frac{\exp\left( - g_{USp(2N_c)}^{-2} \right) }{\det_{cf} \Phi_{cf}} ,
\end{eqnarray}
and one recovers the form of the superpotential found in field theory \cite{Intriligator:1995ne}.

In the case of an $O(N_c)$ gauge theory on the worldvolume of the branes, $N_c-3$ flavours $\Phi_{cf}$, $c\in\{1,...,N_c\}$, $f\in\{1,...,N_c-3\}$, are required for a superpotential to be generated by an instanton. Thus one introduces a second stack $f$ of $N_c-3$ branes. In this case, the gauge group on the instanton worldvolume is $USp(2)$. The zero modes $x^\mu$ and $\theta^\alpha$ transform in the antisymmetric, one-dimensional representation of this group, the zero modes $\bar{\tau}^{\dot{\alpha}}_i$, $i\in\{1,2,3\}$ in the symmetric, three-dimensional representation. This implies that the $3\times 2=6$ fermionic ADHM constraints are reproduced in string theory. There are also $3\times 3=9$ bosonic ADHM constraints, which can again be understood as the D- and F-term supersymmetry conditions on the instanton worldvolume. The charged zero modes transform in the fundamental, two-dimensional representation of $USp(2)$. There are $4N_c$ bosonic zero modes $w^{\dot{\alpha}}_{ac}$ and $2N_c$ fermionic zero modes $\mu_{ac}$, $a\in\{1,2\}$, $c\in\{1,...,N_c\}$, from strings stretching between the instanton and brane stack $c$ as well as $2(N_c-3)$ fermionic zero modes $\lambda_{af}$, $a\in\{1,2\}$, $f\in\{1,...,N_c-3\}$, from those stretching between the instanton and stack $f$.

The fermionic integral can be evaluated and can be seen to lead to the requirement of having $N_c-3$ flavours. The bosonic integral is rather difficult due to the large number of ADHM constraints. Eventually one should recover the superpotential
\begin{eqnarray}
 \int d^4x d^2\theta \frac{\exp\left( - g_{O(N_c)}^{-2} \right) }{\det_{ff'} \Phi_{cf}\Phi_{cf'} }
\end{eqnarray}
found in field theory \cite{Intriligator:1995id}.

\chapter{D-instanton corrections to the gauge kinetic function}
\label{Dinstantoncorrectionsgaugekineticfunction}
The last two chapters have mainly been concerned with D-instanton corrections to the superpotential. Other quantities in the low energy effective action do however receive such corrections, too. This chapter deals with instanton-induced contributions to the gauge kinetic function \cite{Akerblom:2007uc,Blumenhagen:2008ji}, which, just as the superpotential, is holomorphic. Its dependence on the modulus chiral superfields is therefore quite restricted which allows setting up an instanton calculus that can actually be used to perform explicit calculations.

\section{General considerations}
Many of the considerations made in the context of the superpotential carry over to the case of the gauge kinetic function. This is true in particular for those concerning charges under global abelian symmetries and those concerning the dependence on the string coupling. Furthermore, it is clear that the instanton zero modes are once more of crucial importance. So in order to determine which instantons correct the gauge kinetic function one starts by thinking about what the zero mode structure of such instantons should look like. Only corrections to the gauge kinetic function not depending on charged matter fields will be considered here, as corrections that do depend on these fields are normally uninteresting in the most common case where the charged matter fields have vanishing vacuum expectation values. This implies that the relevant instantons will not have charged zero modes because, as was argued in chapter \ref{Dinstantons}, these are absorbed by disc diagrams with charged matter fields inserted. The required structure of neutral zero modes is most easily determined by mapping heterotic worldsheet instantons to the D2-instantons discussed here. The map consists of S- and T-dualities. Of course, one needs to consider heterotic worldsheet instantons which correct the gauge kinetic function \cite{Beasley:2005iu}.

Before doing this, it is worth pausing for a moment to discuss in some detail what zero modes of heterotic worldsheet instantons the different kinds of D2-instanton zero modes mentioned in chapter \ref{Dinstantons} correspond to. Worldsheet instantons of the heterotic string with gauge group $SO(32)$ are mapped to D1-instantons of the type I string under S-duality. The latter become D2-instantons in type IIA orientifolds under mirror symmetry or T-duality. In compactifications on smooth spaces, all D1-instantons in the type I string have an orthogonal gauge group on their worldvolume. As the gauge group on a D-brane is preserved under the mirror map, one immediately sees that, focusing on the easiest and most relevant case of a single instanton, only D2-instantons with gauge group $O(1)$ can be mapped directly to heterotic worldsheet instantons.

The distinction between neutral and charged zero modes in the D2-instanton case has a clear analogue in the heterotic string: Neutral fermionic zero modes, arising as strings with both ends on the D-instanton, correspond to zero modes of the fermions of the right moving superstring, and neutral bosonic zero modes to zero modes of the ten free bosons which are the embedding coordinates of the string in spacetime. Charged fermionic zero modes, which are strings with one end on the instanton and the other one on one of the space-filling D-branes, are mapped to zero modes of the left-moving fermions of the heterotic string that are responsible for the gauge degrees of freedom. Given this distinction, one can ask the question what charged bosonic zero modes correspond to. In order to see what happens, one recalls that such modes arise iff the instanton wraps a cycle that is also wrapped by a space-filling D-brane and therefore represents a gauge instanton. Gauge instantons in the heterotic string are not worldsheet instantons. A zero-size gauge instanton in the heterotic $SO(32)$ theory is a heterotic five-brane \cite{Witten:1995gx}. Therefore, charged bosonic zero modes do not have an analogue in terms of heterotic worldsheet instanton zero modes.

Summarising the discussion in chapter \ref{Dinstantons}, the neutral zero mode structure of a D2-instanton with gauge group $O(1)$ is given as follows: There are always the universal four bosonic and two fermionic zero modes which are Goldstone modes associated with broken translation symmetries and supersymmetries. If the three-cycle $\pi_{D2}$ wrapped by the instanton is not rigid, there will in addition be $2(b_1(\pi_{D2})-x)$ fermionic as well as $2x$ bosonic and $2x$ fermionic zero modes, where the value of $x$ depends on how the anti-holomorphic involution, which is part of the orientifold projection, acts on the three-cycle $\pi_{D2}$. The structure of bosonic and right-moving fermionic zero modes of heterotic worldsheet instantons is rather similar \cite{Beasley:2005iu}. In addition to the universal zero modes, there are $2p$ bosonic and $2p$ fermionic as well as $2g$ fermionic ones, where $p$ is the number of holomorphic sections of the normal bundle of the worldsheet in the compactification manifold and $g$ is the genus of the worldsheet. So when mapping D2-instantons to heterotic world-sheet instantons, $p$ corresponds to $x$ and $g$ corresponds to $(b_1(\pi_{D2})-x$).

Corrections to the gauge kinetic function in the heterotic string come from worldsheet instantons which wrap an isolated, i.e. $p=0$ in the notation used above, curve of genus $g=1$ \cite{Beasley:2005iu}. In view of the above discussion, this means that a D2-instanton can contribute to the gauge kinetic function if it wraps a three-cycle whose first Betti number is one \cite{Akerblom:2007uc}, or, in other words, whose moduli space of deformations is one dimensional. Locally, one can introduce a coordinate $y$ on this moduli space on which the antiholomorphic involution acts as $y\rightarrow y$ or $y \rightarrow -y$. When the three-cycle is wrapped by a D-brane, the moduli space is complexified due to a Wilson line into a one complex dimensional open string moduli space. Before orientifolding, the instanton has two bosonic and four fermionic neutral zero modes $y^i$, $\mu^\alpha$ and $\bar{\mu}^{\dot{\alpha}}$, $i,\alpha,\dot{\alpha}\in\{1,2\}$, with vertex operators \eqref{instantonmodulivertexoperators}, in addition to the universal ones. Some of these zero modes are removed by the orientifold projection. If the anti-holomorphic involution maps $y$ to $y$, it acts on it as on the coordinates in the four-dimensional external space and, just as the zero modes $x^\mu$ and $\theta^\alpha$, the zero modes $y^i$ and $\bar{\mu}^{\dot{\alpha}}$ survive. If $y$ is mapped to $-y$, the zero modes $\mu^\alpha$ survive. Note that the spacetime chirality of the surviving fermionic zero modes that come with $y^i$ is different from that of those coming with $x^\mu$. The reason is that the orientifold projection acts differently on the relevant operators of the internal CFT. According to the above discussion about the relation of heterotic worldsheet instantons and D2-instantons, the relevant D2-instantons are those with one pair of zero modes $\mu^\alpha$.

Summarising, D2-instantons wrapping a cycle with one deformation which is anti-invariant under the anti-holomorphic involution can correct the gauge kinetic function \cite{Akerblom:2007uc}. Furthermore, in order to ensure the absence of charged zero modes, the cycle should not intersect any cycle wrapped by space-filling branes.

According to the previous discussion, D2-instantons that can correct the gauge kinetic function have four fermionic zero modes. The latter have to be absorbed by inserting their vertex operators in CFT correlators. Furthermore, two gauge boson vertex operators have to be inserted because one is interested in the gauge kinetic function. As the latter is holomorphic, only tree and one-loop diagrams can contribute, in analogy to what happens in the case of the superpotential. A correlator of neutral instanton zero modes and gauge bosons is needed, so the relevant diagram must have boundary segments both on the instanton and on the space-filling D-brane whose gauge kinetic function one is interested in. If the correlator were a disc diagram, vertex operators changing the boundary from the instanton to the space-filling D-brane would be needed. Such vertex operators would be associated to charged zero modes, which are absent. Note here that massive charged instanton modes, which always exist, cannot be inserted as, having no four-dimensional momentum, they cannot be on-shell and there is thus no vertex operator of the appropriate conformal weight for them. In conclusion, the diagram needed is an annulus diagram \cite{Akerblom:2007uc}
\begin{eqnarray}
 \langle \prod_{\alpha=1}^2 V_{\theta^\alpha} \prod_{\alpha=1}^2 V_{\mu^\alpha} | \prod_{i=1}^2 V_{gauge\ boson} \rangle
 = \annft{D2}{D6}
 \label{annuluszeromodesgaugebosons}
\end{eqnarray}
with one boundary on the instanton and the fermionic zero modes inserted, and the other boundary on the D-brane with the gauge boson vertex operators inserted.

There is another argument one can give to support the claim that it should be an annulus diagram on which the gauge bosons are inserted and via which the zero modes are absorbed. As the relevant diagram in the heterotic string appears at genus one, i.e. at one loop in the topological expansion, it is reasonable to assume that the relevant diagram in the open string case should also be a one-loop diagram, i.e. an annulus.

Just as the superpotential, the gauge kinetic function has to respect the global abelian symmetries. The absence of charged zero modes implies that the instanton action must be invariant by itself or, in other words, complex structure moduli whose real parts shift under $U(1)$-symmetries must not appear in it.

At this point, one can write down a formula for the correlator of two gauge bosons in the instanton background, from which one can extract the corrections to the gauge kinetic function. In analogy to formula \eqref{formulainstantoncorrelator} for contributions to the superpotential, it involves various vacuum diagrams. In addition, the zero mode absorption diagram \eqref{annuluszeromodesgaugebosons} appears:
\begin{eqnarray}
 \langle F F \rangle = \annft{D2}{D6}
 \exp \left( D^{vac}_{D2} + {M'}^{vac}_{D2} + \sum_a {A'}^{vac}_{D2,D6_a} \right)
 \label{twogaugebosoninstantoncorrelator}
\end{eqnarray}

Similar to what happens in the case of the superpotential, non-holomorphic terms in the different amplitudes in \eqref{twogaugebosoninstantoncorrelator} rearrange so as to give a result that is in agreement with the holomorphy of the gauge kinetic function. Using that in this case there are $N_f=2$ fermionic zero modes (in addition to the universal ones), the $\mu^\alpha$'s, and the previous result $x=2$ derived in section \ref{holomorphysuperpotential}, \eqref{holomorphiconeloopinstantonaction} becomes
\begin{eqnarray}
 16\pi^2 S_{D2}^{1-loop} &=& 16\pi^2 {\rm Im}(S_{D2}^{holo,1-loop})
             \nonumber \\ &&
             - 4 \ln {\rm Re}(S_{D2}^{tree})
             - 2 \ln \det K^{D2,tree}_\mu .
 \label{holomorphiconeloopinstantonactiongkf}
\end{eqnarray}
What is also needed is a formula similar to \eqref{supergravitycorrelator} that allows one to disentangle holomorphic and non-holomorphic terms of the diagram in \eqref{annuluszeromodesgaugebosons}. Given that the $\theta^\alpha$'s are gaugino-like modes, \eqref{annuluszeromodesgaugebosons} is a gauge-kinetic-function- rather than superpotential-like term. Therefore, the factor $\exp(K/2)$ of \eqref{supergravitycorrelator} will not appear and the required formula is
\begin{eqnarray}
 \langle \prod_{\alpha=1}^2 V_{\theta^\alpha} \prod_{\alpha=1}^2 V_{\mu^\alpha} | \prod_{i=1}^2 V_{gauge\ boson} \rangle
 = \frac{\widetilde{f}}{{\rm Re}(S_{D2}^{tree})K_\mu^{tree}} ,
 \label{annuluszeromodesgaugebosonsdisentangled}
\end{eqnarray}
with $\widetilde{f}$ holomorphic. The two non-holomorphic terms in the second line of \eqref{holomorphiconeloopinstantonactiongkf} cancel against the denominator of \eqref{annuluszeromodesgaugebosonsdisentangled}, which accounts for the non-canonical normalisation of fields in a supergravity Lagrangian.

The final formula for the D2-instanton induced correction to the gauge kinetic function is:
\begin{eqnarray}
 f^{np} = \widetilde{f} \times \exp\left( - S_{D2} - 4 \pi^2 S_{D2}^{holo,1-loop} \right)
 \label{formulaholomorphicgkfnp}
\end{eqnarray}
\section{Computation in a concrete model}
\label{Dinstantongkfconcrete}
The next step is to apply the general formulas of the last section to a concrete model \cite{Blumenhagen:2008ji}. They were derived for type IIA orientifolds with intersecting D6-branes, but very similar formulas hold for type I compactifications. The model that will be considered here is the type I orbifold model discussed in section \ref{examplestypeImodel}. The advantage of this model for the present purpose is that an S-dual heterotic model is known for which the gauge threshold corrections including all worldsheet instanton contributions have been determined in section \ref{gaugethresholdstypeImodel}. So after computing the D-instanton corrections to the gauge kinetic function one can compare the results and thereby test the D-instanton calculus.
\subsection{The relevant D-instantons}
\begin{figure}[ht]
  \vspace{10pt}
  \centering
  \hspace{10pt}
  \includegraphics[width=0.7\textwidth]{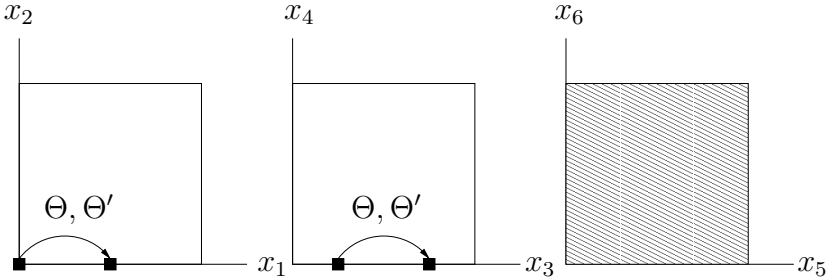}
  \begin{picture}(100,1)
    \put(-285,20){$\Theta,\Theta'$}
    \put(-170,20){$\Theta,\Theta'$}
    \put(-94,95){$x_{6}$}
    \put(-195,95){$x_{4}$}
    \put(-300,95){$x_{2}$}
    \put(-3,0){$x_{5}$}
    \put(-105,0){$x_{3}$}
    \put(-205,0){$x_{1}$}
  \end{picture}
  \vspace{-10pt}
  \caption{Position of a single $O(1)$ instanton: This $E1$ wraps the third
       two-torus and is localised on the fixed points $(x_1,x_2)=(0,0),(1/2,0)$,
      $(x_3,x_4)=(1/4,0),(3/4,0)$ of $\Theta''$.}
  \label{positionE1instanton}
\end{figure}

Recall that the compactification space of the type I model of section \ref{examplestypeImodel} is a shift orbifold of a six-torus with orbifold group $\mathbb{Z}_2\times\mathbb{Z}_2$. Due to the orbifolding, it splits into a product of three two-tori. The relevant D-instantons are D1-instantons wrapping one of the three two-tori, which without loss of generality can be taken to be the third torus. In order to get the result for instantons wrapping another torus one only has to exchange some indices in the following formulas. A D1-instanton wrapping the third torus can either be a bulk brane or it can be charged under the twisted sector associated with the orbifold group element $\Theta''$ . Before orientifolding, a bulk brane has six bosonic open string moduli (not counting the moduli associated with the position of the instanton in the four-dimensional space), four of which describe the position of the instanton on the first two two-tori. The other two are Wilson lines along the two fundamental one-cycles of the third two-torus. A fractionally charged brane only has the two Wilson line moduli. Its position is fixed. According to the discussion in the previous section, it is therefore clear that the relevant instantons are the fractionally charged branes. As they are charged under only one of the three twisted sectors, they are given by a doublet of two branes, which, due to the shift in the orbifold action, are at different points on the first two two-tori. These two points are mapped to each other by the orbifold group elements $\Theta$ and $\Theta'$. Figure \ref{positionE1instanton} illustrates this.

The neutral zero mode structure of D2-instantons wrapping non-rigid cycles was discussed in section \ref{zeromodes}. Analogously, one finds that there are four fermionic zero modes that come with the Wilson line moduli of the D1-instantons under consideration. The orientifold projection removes two of the fermionic zero modes and truncates the moduli space of Wilson lines to a discrete space that will be parameterised by $\beta$ and $\gamma$. Usually, the discrete Wilson line moduli space that survives the orientifold projection is given by $\beta$, $\gamma\in\{0,1/2\}$. In this case, due to the shift included in the orbifold action, it looks a little bit different. The winding numbers in the twisted (closed string) sector associated with $\Theta''$ are of the form (half integer,integer). To see what this implies for the Wilson lines on branes wrapping this torus, one considers the winding mode part
\begin{eqnarray}
 \sum_{w_1\in\mathbb{Z}+\frac{1}{2}} \sum_{w_2\in\mathbb{Z}} \exp\left( 2\pi i (\beta w_1 + \gamma w_2) \right) |w_1,w_2\rangle
 \label{windingpartboundarystate}
\end{eqnarray}
of the twisted sector boundary state describing a brane with Wilson lines $\beta$ and $\gamma$. There are two things to notice from \eqref{windingpartboundarystate}. First, a state with $\beta=1/2$ is not invariant under the orientifold projection and therefore not allowed. To see this one has to take into account that the worldsheet parity operator (which is the full orientifold projection operator in this case) maps $|w_1,w_2\rangle$ to $|-w_1,-w_2\rangle$. Second, a state with $\beta=1$ is not equivalent to a state with $\beta=0$. In conclusion, the Wilson line moduli space in this case is $\beta$, $2\gamma\in\{0,1\}$ \cite{Blumenhagen:2008ji}.

It was already mentioned that the position on the first two two-tori of the D-instantons under discussion is fixed. There is however a discrete moduli space of possible locations. In terms of the coordinates used in figure \ref{positionE1instanton} it is given by $x_2,x_4\in\{0,1/2\}$. The contribution of the instantons to the gauge kinetic function does not depend on their position, so in order to account for this moduli space, one just has to multiply the final expression by four.

The annulus and Moebius strip partition functions of the D1-instanton with (discrete) Wilson lines parameterised by $\beta$ and $\gamma$ are \cite{Camara:2007dy}
\begin{eqnarray}
 A_{D1-D1} &=& \frac{1}{8} \int_0^\infty \frac{dt}{t} \Bigg[
               \frac{\vartheta_3^4-\vartheta_4^4-\vartheta_2^4-\vartheta_1^4}{\eta^{12}} \Lambda^M_3 [0,0,0] \times
               \nonumber \\ &&
               \Big( \Lambda^W_1[0] \Lambda^W_2[0] + \Lambda^W_1[1/2] \Lambda^W_2[1/2] \Big)
               \nonumber \\ &&
               + 4 \frac{\vartheta_3^2\vartheta_4^2-\vartheta_4^2\vartheta_3^2
                   +\vartheta_2^2\vartheta_1^2+\vartheta_1^2\vartheta_2^2}{\eta^6\vartheta_2^2}
               \Lambda^M_3[1/2,0,0] \Bigg] \label{annulusD1D1} \\
 M_{D1} &=& -\frac{1}{8} \int_0^\infty \frac{dt}{t} \Bigg[
            4 \frac{\vartheta_3^2\vartheta_4^2-\vartheta_4^2\vartheta_3^2
              +\vartheta_2^2\vartheta_1^2+\vartheta_1^2\vartheta_2^2}{\eta^6\vartheta_2^2}
            \Lambda^M_3 [1/2,0,0] \times
            \nonumber \\ &&
            \Lambda^W_1[0] \Lambda^W_2[0]
            + 16 \frac{\vartheta_4^4-\vartheta_3^4-\vartheta_1^4-\vartheta_2^4}{\vartheta_2^4}
            \Lambda^M_3[0,0,0] \Bigg] \label{moebiusD1} \\
 A_{D1-D9} &=& \frac{2\times32}{8} \int_0^\infty \frac{dt}{t} \Bigg[
               \frac{\vartheta_2^4-\vartheta_1^4-\vartheta_3^4-\vartheta_4^4}{\vartheta_4^4} \Lambda^M_3 [0,\beta,\gamma]
               \nonumber \\ &&
               + \frac{\vartheta_2^2\vartheta_1^2-\vartheta_1^2\vartheta_2^2
                 + \vartheta_3^2\vartheta_4^2-\vartheta_4^2\vartheta_3^2}{\vartheta_3^2\vartheta_4^2}
               \Lambda^M_3[1/2,\beta,\gamma] \label{annulusD1D9} \Bigg]
\end{eqnarray}
with the momentum and winding sums given in \eqref{typeImomentumsum} and \eqref{typeIwindingsum}. If $\beta=\gamma=0$, there are massless open strings with one end on the D1-instanton and one end on the D9-branes, i.e. charged zero modes. Integration over them makes the amplitude vanish as they cannot be absorbed via disc diagrams due to the lack of charged matter fields. For other values of $\beta$ and $\gamma$, there are no charged zero modes, so there are three different instantons which correct the gauge kinetic function.
\subsection{The one instanton contribution}
\label{oneinstantoncontribution}
The leading instanton corrections will now be computed explicitly. According to \eqref{twogaugebosoninstantoncorrelator} applied to this type I model one has to compute the diagram $\annsft{D1}{D9}$ and one has to evaluate the expressions \eqref{moebiusD1} and \eqref{annulusD1D9} for the Moebius strip and annulus diagrams. Note that \eqref{annulusD1D1} vanishes. As was explained in section \ref{correctionssuperpotential}, the Ramond sector amplitude with a $(-1)^F$ insertion is not to be included when evaluating \eqref{moebiusD1} and \eqref{annulusD1D9}. Furthermore, the contribution of the open string zero modes should be removed from \eqref{moebiusD1}. This is in principle also true for \eqref{annulusD1D9}, but there are no such modes. For the two diagrams one finds:
\begin{eqnarray}
 M'_{D1} &=& 4 {\rm Im}(T^{(3)}) \int_0^\infty dl - 4 {\rm Im} ( i \ln \eta(U^{(3)}) )
 \nonumber \\ &&
 - \ln\left( {\rm Im}(T^{(3)}) {\rm Im}(U^{(3)}) \right)
 \label{moebiusD1evaluated} \\
 A_{D1-D9} &=& - 4 {\rm Im}(T^{(3)}) \int_0^\infty dl + 16 {\rm Im} ( i \ln \frac{\vartheta\genfrac[]{0pt}{}{(1-\beta)/2}{1/2-\gamma}}{\eta}(2U^{(3)}) )
 \label{annulusD1D9evaluated}
\end{eqnarray}
The next step is to evaluate the diagram $\annsft{D1}{D9}$, where the four insertions on the D1 boundary are the Goldstinos $\theta^\alpha$ and the Wilson line modulinos $\mu^\alpha$ and the two insertions on the D9 boundary are gauge bosons. This six point diagram on an annulus is quite hard to compute in CFT. The strategy that will be adopted here \cite{Blumenhagen:2008ji} is to relate $\annsft{D1}{D9}$ to a diagram that can be computed more easily. The first step in doing so is to replace the boundary on the D9-brane with the two gauge boson vertex operators inserted by a boundary on a fictitious D5-brane instanton that is identical to the D9-brane in the internal space, but pointlike in the four-dimensional external space. By comparing \eqref{annulusD2a2} with \eqref{generalformulathresholdcorrections} one finds that this replacement would be allowed if the other boundary was on a space-filling brane with no insertions, i.e. $\annsbt{D9'}{D9}=\annsbb{D9'}{D5}$. It was argued \cite{Billo:2007sw,Billo:2007py} that such a relation should hold more generally, e.g. $\annsft{D1}{D9}=\annsfb{D1}{D5}$. The next step is to use ``T-duality'' along the non-compact directions to replace the boundaries on the instantons by boundaries on space-filling D-branes. Of course, T-duality only makes sense for compact directions, but there should nevertheless exist a relation between the CFT amplitudes, namely $\qquad\annsfb{D1(\theta,\theta,\mu,\mu)\qquad}{D5}=\qquad\annsfb{D5(\lambda,\lambda,\omega,\omega)\qquad}{D9}$
. The four insertions on the boundary of the D1-instanton are Goldstinos and Wilson line modulinos, so the insertions on the boundary of the D5-brane must be gauginos $\lambda$ and fermions $\omega$ of Wilson line chiral supermultiplets. The correlator $\annsfb{D5}{D9}$ would come from the term $L_1=\frac{\partial^2f}{\partial w^2}\omega \omega \lambda \lambda$ contained in the supersymmetric Lagrangian $L=\int d^2\theta f(w+\theta\omega+...) W^2$, where $W=-i\lambda-(\sigma^{\mu\nu}\theta)F_{\mu\nu}+...$ is the gauge field strength superfield. $L$ also contains the term $L_2=\frac{\partial^2f}{\partial w^2} w w F^{\mu\nu}F_{\mu\nu}$. Given that, due to supersymmetry, the prefactor $\frac{\partial^2f}{\partial w^2}$ in $L_1$ and $L_2$ is equal, one concludes that one can replace the gauginos $\lambda$ and Wilson line modulinos $\mu$ in $\annsfb{D5}{D9}$ by gauge boson field strengths $F^{\mu\nu}$ and Wilson line moduli $w$, i.e. $\hspace*{25pt}\annsfb{D5(\lambda,\lambda,\omega,\omega)\qquad}{D9}=\qquad\annsfb{D5(F,F,w,w)\qquad\ }{D9}$. The final step is to use the fact that instead of computing the four point function$\qquad\quad\annsfb{D5(F,F,w,w)\qquad\ }{D9}$ one can compute the two point function $\qquad\annstb{D5(F,F)\qquad}{D9}$ as a function of the Wilson line $w$ and take the second derivative with respect to $w$. In conclusion, one finds the relation \cite{Blumenhagen:2008ji}
\begin{eqnarray}
 \annft{D1(\theta,\theta,\mu\,\mu)\qquad}{D9} = \frac{\partial^2}{\partial w^2} \qquad \anntb{D5(F,F)\qquad}{D9} .
 \label{relationD14D92D52D9}
\end{eqnarray}
The diagram $\qquad\annstb{D5(F,F)\qquad}{D9}$ is a diagram that also appears when computing gauge threshold corrections and can therefore be evaluated using the techniques of chapter \ref{gaugecouplingoneloop}. One finds \cite{Blumenhagen:2008ji}
\begin{eqnarray}
 \anntb{D5(F,F)\qquad}{D9} \varpropto {\rm Im}( i \ln \frac{\vartheta\genfrac[]{0pt}{}{(1-\beta)/2}{1/2-\gamma}(w,2U^{(3)})}{\eta(2U^{(3)})} ).
\end{eqnarray}
Before using this result in \eqref{relationD14D92D52D9}, recall from section \ref{correctionssuperpotential} that the instantonic one-loop vacuum diagrams are partially ill-defined. The procedure proposed there was to take only the well-defined parts into account when computing them and to promote the resulting imaginary parts of holomorphic functions to the full holomorphic functions. A similar procedure should be applied to $\qquad\annsft{D1(\theta,\theta,\mu\,\mu)\qquad}{D9}$. So finally one obtains \cite{Blumenhagen:2008ji}
\begin{eqnarray}
 \annft{D1(\theta,\theta,\mu\,\mu)\qquad}{D9} \varpropto
  \frac{\vartheta''\genfrac[]{0pt}{}{(1-\beta)/2}{1/2-\gamma}}{\vartheta\genfrac[]{0pt}{}{(1-\beta)/2}{1/2-\gamma}} (2U^{(3)}) .
  \label{zeromodeabsorptiondiagramexplicit}
\end{eqnarray}
This expression is a holomorphic function of the moduli and not of the form \eqref{annuluszeromodesgaugebosonsdisentangled}. In fact, this was in some sense to be expected. When relating $\qquad\annsft{D1(\theta,\theta,\mu\,\mu)\qquad}{D9}$ to $\qquad\annstb{D5(F,F)\qquad}{D9}$ the equality of the prefactors of two terms in a supersymmetric Lagrangian was used. It holds for the Lagrangian in the Wilsonian sense, that is a Lagrangian where massless modes have not been integrated out. As the non-holomorphic terms in \eqref{annuluszeromodesgaugebosonsdisentangled} come from massless modes, their effect on the two terms can be different. To correct for this, one recalls that it was shown in the last section that these non-holomorphic terms cancel those appearing in the vacuum Moebius strip diagram. So, what one has to do, is to neglect the term $\ln\left( {\rm Im}(T^{(3)}) {\rm Im}(U^{(3)})\right)$ in \eqref{moebiusD1evaluated}.

Using \eqref{moebiusD1evaluated}, \eqref{annulusD1D9evaluated} and \eqref{zeromodeabsorptiondiagramexplicit} in \eqref{twogaugebosoninstantoncorrelator} and taking into account that the instanton action $S_{D1}=-D_{D1}^{vac}$ is proportional to the complexified worldvolume of the instanton, and therefore to $T^{(3)}$, one finally finds that
\begin{eqnarray}
 \langle F F \rangle =
 \exp \left( \pi i T^{(3)} \right)
 \frac{\vartheta''\genfrac[]{0pt}{}{(1-\beta)/2}{1/2-\gamma}}{\vartheta\genfrac[]{0pt}{}{(1-\beta)/2}{1/2-\gamma}}
 \frac{\vartheta^{16}\genfrac[]{0pt}{}{(1-\beta)/2}{1/2-\gamma}}{\eta^{16}}(2U^{(3)})
 \frac{1}{\eta^4(U^{(3)})} .
\end{eqnarray}
Note that the divergences in \eqref{moebiusD1evaluated} and \eqref{annulusD1D9evaluated} cancel as expected and that the imaginary parts of holomorphic functions in these formulas have been promoted to full holomorphic functions.

Using $\eta^2(U)=\vartheta_4(2U)\eta(2U)$, summing over the three possibilities for the Wilson lines $\beta$ and $\gamma$ and taking the instantons wrapping the first and the second torus into account as well, the one D-instanton correction to the gauge kinetic function becomes \cite{Blumenhagen:2008ji}
\begin{eqnarray}
 \delta^{1-inst} f_I = \sum_{i=1}^3 \exp \left( \pi i T^{(i)} \right) \sum_{a=2}^4
 \frac{\vartheta''_a}{\vartheta_a}
 \frac{\vartheta_a^{16}}{\eta^{16}}
 \frac{1}{\vartheta_4^2\eta^2} (2U^{(i)}) .
 \label{typeI1inst}
\end{eqnarray}
This result precisely matches the leading worldsheet instanton correction to the gauge coupling in the dual heterotic model, i.e. the summand with $p=2k=1$, $j=0$ in \eqref{gaugethresholdsheterotic}. This agreement should be taken as evidence that the D-instanton calculus for corrections to the gauge kinetic function and in particular \eqref{relationD14D92D52D9} are correct.

It is worth pointing out that the sum over the spin structures of the left-moving fermions in \eqref{gaugethresholdsheterotic} corresponds to the sum over the instantons differing in their discrete Wilson lines in \eqref{typeI1inst} \cite{Polchinski:1995df}.
\subsection{Multiply wrapped instantons}
\label{multiplywrappedinstantons}
It remains to be shown that also the higher order terms in \eqref{gaugethresholdsheterotic} can be reproduced by a D-instanton calculation in the type I model. It was argued in similar cases \cite{Bachas:1997xn,Gava:1998sv} that such terms come from corrections induced by multiply wrapped D-instantons or, equivalently, bound states of several instantons \cite{Bachas:1997mc,Kiritsis:1997hf,Bianchi:2007rb}. In the latter description, only certain twisted sectors of the symmetric product orbifold CFT describing the bound state in the infrared contribute and these twisted sectors are essentially equivalent to multiply wrapped instantons. The analysis is easiest in terms of these multiply wrapped instantons so this is the point of view adopted here.

\begin{figure}[ht]
  \centering
  \hspace{10pt}
  \includegraphics[width=0.7\textwidth]{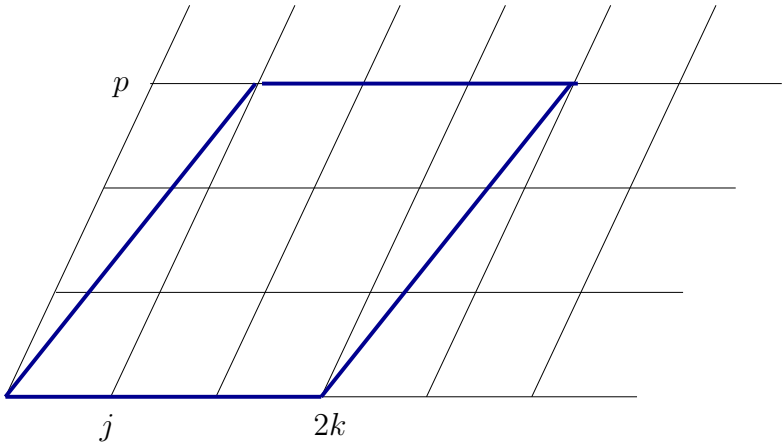}
  \begin{picture}(100,1)
    \put(-255,117){$p$}
    \put(-260,-13){$j$}
    \put(-180,-13){$2k$}
  \end{picture}
  \vspace{5pt}
  \caption{Multiply wrapped D1-instanton.}
  \label{multiplewrapping}
\end{figure}

The objects to be considered are thus D1-instantons whose worldvolume is a multiple cover of one of the three two-tori which constitute the compactification space \cite{Blumenhagen:2008ji}. In other words, the lattice which defines the worldvolume of one of these instantons is a sublattice of the lattice that defines the spacetime two-torus. This is illustrated in figure \ref{multiplewrapping}.

The way the instanton multiply wraps the spacetime torus can be encoded in three numbers that correspond to $p$, $k$ and $j$ in the heterotic worldsheet instanton sum in \eqref{gaugethresholdsheterotic} and that will therefore be denoted by the same letters. From figure \ref{multiplewrapping} one can infer that the instanton worldvolume should be describable by effective Kaehler and complex structure moduli $T^{eff}$ and $U^{eff}$ that are functions of the two-torus moduli ${\rm Im}(T)$ and $U$ as well as $p$, $k$ and $j$. A Comparison of the higher order terms in \eqref{gaugethresholdsheterotic} with the expression \eqref{typeI1inst} suggests $T^{eff}=pk{\rm Im}(T)$ and $U^{eff}=(j+pU)/2k$. By analysing which $p$, $k$ and $j$ yield inequivalent wrappings, one finds that their range is just what is needed to reproduce all the terms in \eqref{gaugethresholdsheterotic}.

Using $T^{eff}=pk{\rm Im}(T)$ and $U^{eff}=(j+pU)/2k$ in \eqref{typeI1inst} to find the contributions of the multiply wrapped instantons does not correctly reproduce the functional form of the heterotic result. What one has to do is the following \cite{Blumenhagen:2008ji}. First, one notes that \eqref{latticesumasymmetric} suggests that the effective complex structure modulus is encoded in
\begin{eqnarray}
 (1,U)A=(A_{11} + A_{21} U, A_{12} + A_{22} U) ,
\end{eqnarray}
where the matrix $A$ will be (a modularly transformed version) of the form \eqref{matrixrepresentativesnondegenerate}. It turns out that $U^{eff}=(A_{12}+A_{22}U)/2(A_{11}+A_{21})$ and that one has to modularly transform $A$ such that $A_{11}$ is half-integer and $A_{12}$ integer before one can apply \eqref{typeI1inst} to the multiply wrapped branes whose worldvolume is described by $T^{eff}$ and $U^{eff}$.

So if $k$ is half-integer and $j$ is integer, one can directly use \eqref{typeI1inst} to reproduce the corresponding terms in \eqref{gaugethresholdsheterotic}. If both $k$ and $j$ are half integer, one finds after a modular T$^{-1}$ transformation $U^{eff}=(j-k+pU)/k$. Inserting this into \eqref{typeI1inst} and performing a modular T transformation yields the required result. Finally, if $k$ is integer and $j$ half-integer one has to perform a modular S transformation which results in $U^{eff}=-k/2(j+pU)$. In analogy to the previous case, one reproduces the corresponding terms in \eqref{gaugethresholdsheterotic}.

To summarise, the higher order terms in the heterotic worldsheet instanton sum in \eqref{gaugethresholdsheterotic} are reproduced in the type I description by multiply wrapped D1-brane instantons. So all the holomorphic parts of the heterotic expression \eqref{gaugethresholdsheterotic} can also be computed in the S-dual type I model.
\chapter{Poly-instantons}
In this chapter, arguments for a new class of instanton corrections \cite{Blumenhagen:2008ji} in four-dimensional string compactifications will be presented. The observation of an equality of certain terms in gauge kinetic functions and instanton amplitudes will lead to the conjecture that D-instanton actions can receive corrections from other D-instantons. These corrections will be reinterpreted in terms of new contributions to the superpotential and the gauge kinetic function arising through multiple instantons. As these configurations of multiple instantons are different from multi-instantons they will be called poly-instantons \cite{Blumenhagen:2008ji}.

After outlining arguments for poly-instantons, some poly-instanton corrections will be computed for the type I orbifold model described in section \ref{examplestypeImodel}. The general discussion will therefore focus on type I compactifications, but the ideas and results carry over to other orientifold compactifications of type II string theories.
\section{Arguments for poly-instantons}
Consider a supersymmetry preserving compactification of the type I string on some Calabi-Yau manifold or on an orbifold. The model will contain D9-branes with orthogonal and D5-branes with unitary symplectic gauge groups \cite{Gimon:1996rq}. The instantons of interest here are D1-brane instantons with an orthogonal gauge group on their worldvolume.

Recall from section \ref{holomorphysuperpotential} that there is a relation between the action of a D2-instanton and the gauge coupling/gauge kinetic function on a fictitious D6-brane. (A brane which does not exist in the model(s) under consideration, but which is useful to establish relations between certain expressions will be referred to as a fictitious brane in the following.) In analogy, there is a relation between the action $S_{D1}$ of a D1-instanton and the gauge coupling $g_{D5}$ on a fictitious D5-brane that wraps the same cycle in the internal space as the instanton or that is described by the same boundary state.

More precisely, the tree-level instanton action and the tree-level gauge coupling/gauge kinetic function are both given by dimensionally reducing the Dirac-Born-Infeld and Chern-Simons actions on the same cycle, so one finds:
\begin{eqnarray}
 \left(g_{D5}^{tree}\right)^{-2} &=& {\rm Im}(f_{D5}^{tree}) = {\rm Im} (S_{D1}^{tree}) \\
 f_{D5}^{tree} &=& S_{D1}^{tree} \label{gaugekineticfunctioninstantonactionequalitytree}
\end{eqnarray}
In exact analogy to the case of D2-instantons and D6-branes discussed in section \ref{holomorphysuperpotential}, the equality of certain CFT amplitudes implies that the one-loop gauge threshold corrections $\left(g_{D5}^{1-loop}\right)^{-2}$ and the one-loop correction to the instanton action $S_{D1}^{1-loop}$, defined in analogy to \eqref{relationmoebiusannulusoneloopaction}, are equal.
\begin{eqnarray}
 \left(g_{D5}^{1-loop}\right)^{-2} = S_{D1}^{1-loop}
 \label{gaugecouplinginstantonactionequality1loop}
\end{eqnarray}
The threshold corrections depend non-ho\-lo\-mor\-phi\-cally on the moduli, but one can use \eqref{physicalgaugecoupling} to determine the one-loop correction to the holomorphic gauge kinetic function. \eqref{gaugecouplinginstantonactionequality1loop} implies that $S_{D1}^{1-loop}$ is also non-holomorphic. It was argued in section \ref{holomorphysuperpotential} that a precisely analogous formula, \eqref{holomorphiconeloopinstantonaction}, can be used to disentangle holomorphic and non-holomorphic pieces in $S_{D1}^{1-loop}$ and to thereby define the holomorphic quantity $S_{D1}^{holo,1-loop}$. Consequently,
\begin{eqnarray}
 {\rm Im}(f_{D5}^{1-loop}) = {\rm Im}(S_{D1}^{holo,1-loop}) .
\end{eqnarray}
It is plausible to assume
\begin{eqnarray}
 f_{D5}^{1-loop} = S_{D1}^{holo,1-loop} .
 \label{gaugekineticfunctioninstantonactionequality1loop}
\end{eqnarray}
The type I version of the non-renormalisation theorem of section \ref{holomorphygaugekineticfunction} implies that the gauge kinetic function $f_{D5}$ does not receive perturbative corrections beyond one loop. On the other hand, the holomorphic part $S_{D1}^{holo}$ of the instanton action is to be considered as the exponent that appears in the instanton induced superpotential and gauge kinetic function. It was argued in chapters \ref{Dinstantons} and \ref{Dinstantoncorrectionsgaugekineticfunction} that in order to compute them no (perturbative) amplitudes beyond one-loop in the open string coupling have to be taken into account. This implies that also $S_{D1}^{holo}$ does not receive higher order perturbative corrections. Keeping this in mind and denoting the full perturbative gauge kinetic function by $f_{D5}^{pert}$ and the full perturbative holomorphic part of the instanton action by $S_{D1}^{holo,pert}$, \eqref{gaugekineticfunctioninstantonactionequalitytree} and \eqref{gaugekineticfunctioninstantonactionequality1loop} imply \cite{Blumenhagen:2008ji}
\begin{eqnarray}
 f_{D5}^{tree} + f_{D5}^{1-loop} = f_{D5}^{pert} = S_{D1}^{holo,pert} = S_{D1}^{tree} + S_{D1}^{holo,1-loop} .
 \label{gaugekineticfunctioninstantonactionequalitypert}
\end{eqnarray}
It was shown in chapter \ref{Dinstantoncorrectionsgaugekineticfunction} that $f_{D5}$ can receive corrections $f_{D5}^{D1'}$ from D1'-instantons. The prime on D1' indicates that these instantons are different from the D1-instantons whose action is equal to $f_{D5}$. If the equality \eqref{gaugekineticfunctioninstantonactionequalitypert} between the holomorphic part of the D1-instanton action and the gauge kinetic function on the fictitious D5-brane, which holds to all orders in perturbation theory, is true exactly, also the instanton action must receive corrections $S_{D1}^{D1'}$ from D1'-instantons.
\begin{eqnarray}
 S_{D1}^{D1'} \overset{?}{\neq} 0 \qquad S_{D1}^{holo,D1'} \overset{?}{=} f_{D5}^{D1'}
 \label{equalitynpgkfD1action}
\end{eqnarray}

It shall now be discussed what instanton corrections to instanton actions imply. One starts by considering a formula for a D1-instanton correction to some quantity. Obviously, the instanton action will appear in such an expression. Given \eqref{twogaugebosoninstantoncorrelator} and \eqref{formulaholomorphicgkfnp} one can schematically write the D1-instanton correction to the gauge kinetic function on some stack of branes labelled $a$ as
\begin{eqnarray}
 f_a^{D1} = \anntf{a}{D1} \exp \left( - S_{D1}^{holo,pert} \right) .
 \label{D1correctiongkf}
\end{eqnarray}
This formula together with the conjectured equality \eqref{equalitynpgkfD1action} between the D1'-in\-stan\-ton corrections to the D1-instanton action and the gauge kinetic function on the fictitious D5-brane imply
\begin{eqnarray}
 S_{D1}^{holo,D1'} = \annbf{D1}{D1'} \exp \left( - S_{D1'}^{holo,pert} \right) .
\end{eqnarray}
The next step is to add $S_{D1}^{holo,D1'}$ to $S_{D1}^{holo,pert}$ in \eqref{D1correctiongkf} and to expand the exponential.
\begin{eqnarray}
 f_a^{D1} &=& \anntf{a}{D1} \exp \left( - S_{D1}^{holo,pert} - S_{D1}^{holo,D1'} \right) \nonumber \\
          &=& \anntf{a}{D1} \exp \left( - S_{D1}^{holo,pert} -  \annbf{D1}{D1'} e^{- S_{D1'}^{holo,pert}} \right) \nonumber \\
          &=& \anntf{a}{D1} \exp \left( - S_{D1}^{holo,pert} \right) \nonumber \\ &&
              - \anntf{a}{D1} \annbf{D1}{D1'} \exp \left( - S_{D1}^{holo,pert} - S_{D1'}^{holo,pert} \right) + ...
 \label{polyinstantonexpansion}
\end{eqnarray}
The last line in \eqref{polyinstantonexpansion} should be interpreted as a two-instanton correction to the gauge kinetic function \cite{Blumenhagen:2008ji}. Generalising this observation means that instanton corrections to instanton actions can be rephrased as corrections from multiple instantons to more physical quantities such as the gauge kinetic function or the superpotential. It is important to stress that these corrections come from the interplay of distinct instantons and are therefore different from the usual multi-instantons which appear in gauge or string theory. In string theory multi-instantons are multiply wrapped worldsheets or stacks of D-instantons. So, in order to distinguish this new type of multiple instantons they will be referred to as poly-instantons \cite{Blumenhagen:2008ji}.

It is clear that, although the equality \eqref{gaugekineticfunctioninstantonactionequalitypert} of the perturbative part of the D1-instanton action and that of the gauge kinetic function on the fictitious D5-brane makes it plausible that this equality holds exactly \eqref{equalitynpgkfD1action}, this does not have to be the case. In the following another argument \cite{Blumenhagen:2008ji} in favour of D-instanton corrections to D-instanton actions, or, equivalently, poly-instantons, will be given.

Consider a string compactification with an $SU(N)$ factor in the low energy gauge group, which arises from a stack of $N$ D-branes. Assume that the only massless states charged under this $SU(N)$ group are the vector supermultiplets transforming in the adjoint representation, $N-1$ chiral multiplets transforming in the fundamental representation and $N-1$ chiral multiplets in the antifundamental representation. Assume furthermore that the gauge coupling associated to the $SU(N)$ gauge group factor receives non-perturbative D-instanton corrections such that the full gauge coupling can be written
\begin{eqnarray}
 \frac{1}{g_{full}^2}=\frac{1}{g_{tree}^2}+\frac{1}{g_{1-loop}^2}+\frac{1}{g_{np}^2} .
\end{eqnarray}
If one is now interested in the low energy physics of this string compactification, one can use an effective field theory. In the case described, the latter will contain an SQCD sector, more precisely a supersymmetric $SU(N)$ gauge theory with $N-1$ flavours. A gauge instanton in this theory will generate an ADS superpotential \cite{Affleck:1983mk}
\begin{eqnarray}
 W_{ADS} = \frac{1}{\det \Phi \widetilde{\Phi}}\exp\left(-\frac{1}{g_{full}^2}\right)
         = \frac{1}{\det \Phi \widetilde{\Phi}}\exp\left(-\frac{1}{g_{tree}^2}-\frac{1}{g_{1-loop}^2}-\frac{1}{g_{np}^2}\right) .
 \nonumber \\ \label{ADSsuperpotentialgaugetheory}
\end{eqnarray}
This superpotential must also be derivable in the full string theory. There it is generated, as was shown in section \ref{ADSsuperpotential}, by a D-brane instanton and takes the form
\begin{eqnarray}
 W_{ADS}^{string\ theory} = \frac{1}{\det \Phi \widetilde{\Phi}} \exp\left(-S_{D-inst.}\right)
 \label{ADSsuperpotentialstringtheory}
\end{eqnarray}
It is clear that if \eqref{ADSsuperpotentialstringtheory} is to reproduce \eqref{ADSsuperpotentialgaugetheory}, the D-instanton action $S_{D-inst.}$ must receive instanton corrections, as $1/g_{np}^2$ is instanton induced.

\section{Computing poly-instanton corrections}
In order to determine which poly-instantons can contribute to which quantities one needs to carefully analyse their zero modes. This shall now be done for the example of the poly-two-instanton which yields the term in the last line of \eqref{polyinstantonexpansion} \cite{Blumenhagen:2008ji}. The instanton D1 corrects the gauge kinetic function $f_a$, so it has four bosonic zero modes $x^\mu$ and four fermionic ones $\theta^\alpha$ and $\mu^\alpha$. The instanton D1' corrects the D1-instanton action or the gauge kinetic function on the fictitious D5-brane. This means that it also has four bosonic zero modes ${x'}^\mu$ and four fermionic ones ${\theta'}^\alpha$ and ${\mu'}^\alpha$ and that there are no charged zero modes in the D1-D5 sector. The latter fact implies that there are no zero modes from strings stretching between the D1- and the D1'-instanton. In total, there are thus eight fermionic zero modes in this poly-two-instanton sector. They are absorbed via the diagrams $\annstf{a}{D1}$ and $\annsbf{D1}{D1'}$. In order to see how the eight bosonic zero modes are absorbed, one notes that the linear combination $(x+x')^\mu/2$ describes the position of the center of mass of the two-instanton configuration and the integral over it is just the integral over the four-dimensional space that is not performed explicitly when computing corrections to the gauge kinetic function. But the orthogonal linear combination $(x-x')^\mu$ needs to be integrated over. It will be shown later on in this section how this is done.

When performing the expansion in \eqref{polyinstantonexpansion} to higher orders, the zero mode absorption diagram $\annsbf{D1}{D1'}$ will appear several times, say $n$ times. This means that the instanton D1' shows up $n$ times in the corresponding poly-instanton sector. This implies on the one hand that a combinatorial factor $1/n!$ has to be included in the amplitude and on the other hand that there will be more zero modes at those loci in instanton moduli space where two or more instantons of type D1' are at the same point in the four-dimensional space. The strategy that will be pursued here is to evaluate such poly-instanton amplitudes at a point in moduli space where these zero modes are absent. If no singularities appear when performing the integral over moduli space, the result should be trustworthy.

It is clear that in order to find the full correction coming from a fixed poly-instanton, one has to sum over all possibilities of distributing the fermionic zero modes amongst annuli ending on two of the instantons that are part of the poly-instanton. One has to ensure that the combination of all diagrams in each summand is connected from the spacetime perspective.

The next thing to discuss is how to compute \cite{Blumenhagen:2008ji} the zero mode absorption diagrams $\annsbf{D1}{D1'}$. This four fermion amplitude on an annulus cannot be straightforwardly computed using CFT techniques, but in close analogy to what was done for the six-point amplitude $\annsft{D1}{D9}$ in section \ref{oneinstantoncontribution}, it can be related to other diagrams that are computable \cite{Blumenhagen:2008ji}. Applying T-duality, employing supersymmetry to replace the four fermions by four bosons and using the fact that the four point diagram $\quad\annsbf{D5\quad}{\qquad D5'(F,F,w,w)}\qquad$ is equal to the second derivative of the two point diagram $\annsbt{D5}{\qquad D5'(F,F)}\qquad$ with respect to $w$ one finds a relation of the form
\begin{eqnarray}
 \annbf{D1}{\qquad D1'(\theta,\theta,\mu,\mu)} \qquad \sim \frac{\partial^2}{(\partial w)^2} \annbt{D5}{\qquad D5'(F,F)} .
\end{eqnarray}
There are two things to note before such an identification can be made. First, as was already mentioned, in a poly-two-instanton sector one has to integrate over the relative position $(x-x')^\mu$ of the two instantons. As the open strings between the instantons D1 and D1' receive a mass contribution proportional to $(x-x')^2$, the annulus amplitude $\annsbf{D1}{D1'}$ contains a factor $\exp(-\pi t (x-x')^2)$, where $t$ is the modular parameter of the annulus, so one effectively has the integral
\begin{eqnarray}
 \int d^4(x-x') \exp(-\pi t (x-x')^2) = t^{-2} .
 \label{positionintegration}
\end{eqnarray}
Second, the open strings between two space-filling D5-branes have momenta along the four non-compact directions in contrast to the open strings between two D1-instantons. Integration over these momenta gives a factor $t^2$ in the annulus amplitude. When applying ``T-duality'' to relate $\annsbf{D1}{D1'}$ to $\annsbf{D5}{D5'}$ one should take this factor into account. It cancels the factor \eqref{positionintegration}, such that one can equate \cite{Blumenhagen:2008ji}
\begin{eqnarray}
 \int d^4(x-x') \annbf{D1}{\qquad D1'(\theta,\theta,\mu,\mu)} \qquad = \frac{\partial^2}{(\partial w)^2} \annbt{D5}{\qquad D5'(F,F)} .
 \label{relation4zeromodeabsorptionGT}
\end{eqnarray}

A look at \eqref{polyinstantonexpansion} shows that, in order to determine poly-instanton corrections, one needs to compute the amplitudes $\annstf{a}{D1}$, $\annsbf{D1}{D1'}$ and, in order to determine $S_{D1}^{holo,pert}$, $\annsbb{b}{D1}$ and $\annscb{}{D1}$. $\annstf{a}{D1}$ and $\annsbf{D1}{D1'}$ are related to gauge threshold correction diagrams via \eqref{relationD14D92D52D9} and \eqref{relation4zeromodeabsorptionGT}. This means that all CFT amplitudes one has to compute are gauge threshold correction diagrams and instantonic vacuum diagrams. Once this is done, determining poly-instanton corrections becomes a combinatorial exercise.
\section{Computation in a concrete model}
The next step is to apply \cite{Blumenhagen:2008ji} the general formulas of the previous sections. As poly-instanton corrections are closely related to D-instanton corrections to gauge kinetic functions, it is natural to consider the type I orbifold of section \ref{examplestypeImodel}, for which the D-instanton corrections to the gauge kinetic function were computed in section \ref{Dinstantongkfconcrete}.

The building blocks for the poly-instantons are the three single instantons described in section \ref{Dinstantongkfconcrete}, which correct the gauge kinetic function and which differ in their discrete Wilson lines. To shorten the notation, they shall be denoted D1$_i$, $i\in\{2,3,4\}$ in the following. D1$_2$ is characterised by $\beta=0$, $\gamma=1/2$, D1$_3$ by $\beta=1$, $\gamma=1/2$ and D1$_4$ by $\beta=1$, $\gamma=0$. The labelling should be clear from \eqref{vacuumannulusD1iD9}. As in section \ref{Dinstantongkfconcrete}, the explicit computation will be performed for D1-instantons wrapping the third torus.

To start with, the necessary one-loop amplitudes will be determined. Suppressing the divergences due to tadpoles that cancel in the final results and promoting the computed expressions to holomorphic functions as discussed in chapters \ref{Dinstantons} and \ref{Dinstantoncorrectionsgaugekineticfunction}, the relevant vacuum diagrams are \eqref{moebiusD1evaluated}, \eqref{annulusD1D9evaluated}
\begin{eqnarray}
 \annbb{D9}{D1_i} &=& - 16 \ln \frac{\vartheta_i}{\eta}(2U^{(3)}) \label{vacuumannulusD1iD9} \\
 \anncb{O9}{D1_i} &=& 4 \ln \eta (U^{(3)}) \label{vacuummoebiusD1i} .
\end{eqnarray}
The six-point diagrams read \eqref{zeromodeabsorptiondiagramexplicit}
\begin{eqnarray}
 \anntf{D9}{D1_i} = \frac{\vartheta''_i}{\vartheta_i}(2U^{(3)}) .
 \label{sixpointannulus}
\end{eqnarray}
Finally one has to compute the annulus diagrams $\raisebox{4pt}[10pt][6pt]{\annsbf{\ \ D1_i^{\phantom{I^I}}}{\ \ D1_j^{\phantom{I^I}}}}$ with the boundaries on two different instantons and four fermionic zero modes inserted on one of the boundaries. This can be done by determining $\raisebox{0pt}[0pt][0pt]{\ \ \annsbt{D5_i^{\phantom{I^I}}}{\ \ D5_j^{\phantom{I^I}}}}$ and using \eqref{relation4zeromodeabsorptionGT}. One finds \cite{Blumenhagen:2008ji}
\begin{eqnarray}
 \annbt{D5_3}{D5_4} = \annbt{D5_4}{D5_3} &=& \ln \frac{\vartheta_2}{\vartheta_3}(w,2U^{(3)}) \\
 \annbt{D5_2}{D5_4} = \annbt{D5_4}{D5_2} &=& \ln \frac{\vartheta_3}{\vartheta_2}(w,2U^{(3)}) ,
\end{eqnarray}
which gives
\begin{eqnarray}
 \int d^4(x_3-x_4) \annbf{D1_3}{D1_4} = \int d^4(x_3-x_4) \annbf{D1_4}{D1_3} =
      \frac{\vartheta''_2}{\vartheta_2} - \frac{\vartheta''_3}{\vartheta_3} &=& -\pi^2 \vartheta_4^4(2U^{(3)}) \nonumber \\
 \int d^4(x_2-x_4) \annbf{D1_2}{D1_4} = \int d^4(x_2-x_4) \annbf{D1_4}{D1_2} =
      \frac{\vartheta''_3}{\vartheta_3} - \frac{\vartheta''_2}{\vartheta_2} &=& \pi^2 \vartheta_4^4(2U^{(3)}) , \nonumber \\
 \label{fourpointannulus}
\end{eqnarray}
where $x_i$ is the position of the instanton D1$_i$ in the external space.

Note that, due to extra charged zero modes, D1$_2$ would not correct the gauge kinetic function on D5$_3$ and D1$_3$ would not correct that of D5$_2$. D1$_2$ and D1$_3$ are therefore expected not to mutually correct their instanton actions. Indeed, there are extra zero modes from strings between D1$_2$ and D1$_3$. This leads to a divergence in $\raisebox{2pt}[10pt][6pt]{\annsfb{D1_2^{\phantom{I^I}}}{\ D1_3^{\phantom{I^I}}}=\annsfb{D1_3^{\phantom{I^I}}}{\ D1_2^{\phantom{I^I}}}}$. These diagrams can therefore not appear in poly-instanton amplitudes \cite{Blumenhagen:2008ji}.

Having collected all the relevant one-loop CFT amplitudes, one can now compute poly-instanton corrections. One starts with sectors consisting of two instantons. Out of the three possibilities D1$_2$-D1$_3$, D1$_2$-D1$_4$ and D1$_3$-D1$_4$, D1$_2$-D1$_3$ does not contribute due to the aforementioned zero modes. Taking all combinatorial possibilities into account, the amplitude in the D1$_3$-D1$_4$ sector yields the following correction to the gauge kinetic function \cite{Blumenhagen:2008ji}:
\begin{eqnarray}
 \int d^4(x_3-x_4) \left( \anntf{D9}{D1_3} \annbf{D1_3}{D1_4} + \anntf{D9}{D1_4} \annbf{D1_4}{D1_3} \right)
      \times \nonumber \\
      \exp \left( 2\pi i T^{(3)} - \annbb{D9}{D1_3} - \anncb{O9}{D1_3} - \annbb{D9}{D1_4} - \anncb{O9}{D1_4} \right) ,
 \label{twopolyinstantonD13D14NE}
\end{eqnarray}
which, using \eqref{vacuumannulusD1iD9}, \eqref{vacuummoebiusD1i}, \eqref{sixpointannulus} and \eqref{fourpointannulus} can be evaluated to
\begin{eqnarray}
 && - \pi^2 e^{2\pi i T^{(3)}} \frac{\vartheta_4^4(2U^{(3)})}{\eta^8(U^{(3)})} \frac{\vartheta_3^{16}\vartheta_4^{16}}{\eta^{32}} (2U^{(3)})
 \left( \frac{\vartheta''_3}{\vartheta_3} + \frac{\vartheta''_4}{\vartheta_4} \right) (2U^{(3)})
 \\
 &=& - 4 \pi^2 \frac{e^{2\pi i T^{(3)}}}{\eta^2\vartheta_4^2(4U^{(3)})} \frac{\vartheta_4^{16}}{\eta^{16}} (4U^{(3)})
     \frac{\vartheta''_4}{\vartheta_4} (4U^{(3)}) ,
 \label{twopolyinstantonD13D14}
\end{eqnarray}
where in the last step some theta/eta-function identities have been used. Completely analogously one finds for the poly-instanton D1$_2$-D1$_4$ \cite{Blumenhagen:2008ji}
\begin{eqnarray}
 && \int d^4(x_2-x_4) \left( \anntf{D9}{D1_2} \annbf{D1_2}{D1_4} + \anntf{D9}{D1_4} \annbf{D1_4}{D1_2} \right)
      \times \nonumber \\ &&
      \exp \left( 2\pi i T^{(3)} - \annbb{D9}{D1_2} - \anncb{O9}{D1_2} - \annbb{D9}{D1_4} - \anncb{O9}{D1_4} \right)
 \label{twopolyinstantonD12D14NE} \\
 &=& \pi^2 e^{2\pi i T^{(3)}} \frac{\vartheta_4^4(2U^{(3)})}{\eta^8(U^{(3)})} \frac{\vartheta_2^{16}\vartheta_4^{16}}{\eta^{32}} (2U^{(3)})
 \left( \frac{\vartheta''_2}{\vartheta_2} + \frac{\vartheta''_4}{\vartheta_4} \right) (2U^{(3)})
 \\
 &=& - 4 \pi^2 \frac{e^{2\pi i T^{(3)}}}{\eta^2\vartheta_2^2(\frac{1}{2}+U^{(3)})} \frac{\vartheta_2^{16}}{\eta^{16}} (\frac{1}{2}+U^{(3)})
     \frac{\vartheta''_2}{\vartheta_2} (\frac{1}{2}+U^{(3)}) .
 \label{twopolyinstantonD12D14}
\end{eqnarray}
By comparing these results with the expression for the contributions of worldsheet instantons to the gauge threshold corrections in the dual heterotic  model, one finds that \eqref{twopolyinstantonD13D14} is equal to the summand with $p=2$, $k=1/2$, $j=0$ and $a=4$ in \eqref{gaugethresholdsheterotic}, and \eqref{twopolyinstantonD12D14} to that with $p=1$, $k=1$, $j=1/2$ and $a=2$. This also means, given that, as shown in section \ref{multiplywrappedinstantons}, all terms in \eqref{gaugethresholdsheterotic} with $kp>1/2$ are reproduced in the type I description by multiply wrapped instantons, that \eqref{twopolyinstantonD13D14} and \eqref{twopolyinstantonD12D14} are equal to corrections to the gauge kinetic function that come from these multiply wrapped instantons.

This equality can have two reasons. Either, poly-instantons are just a complicated way of reproducing contributions from multiply wrapped instantons and are already included in the heterotic result. Or, this equality is just a coincidence having its reason in the fact that a D-brane doubly wrapped along some cycle has the same partition function as two singly wrapped branes with relative Wilson line $1/2$ along this cycle. One can show that the poly-instanton sectors D1$_3$-D1$_4$ and D1$_2$-D1$_4$ are related to the doubly wrapped instantons D1$_4$ and D1$_2$, whose contribution they reproduce, in precisely this way. In conclusion, this means that at the poly-two-instanton order one cannot conclusively see whether poly-instantons give new contributions.

In order to do so, one has to compute a poly-three-instanton amplitude. There are various possibilities what the three instantons can be. In order to see that poly-instantons really give new contributions, it suffices to consider the poly-instanton consisting of D1$_2$, D1$_3$ and D1$_4$. Taking all possibilities to absorb the zero modes on different annuli into account, the amplitude reads:
\begin{eqnarray}
 && \int d^4(x_2-x_3) d^4(x_2+x_3-2x_4) \Bigg( \anntf{D9}{D1_2} \annbf{D1_2}{D1_4} \annbf{D1_4}{D1_3}
 \nonumber \\ &&
 + \anntf{D9}{D1_3} \annbf{D1_3}{D1_4} \annbf{D1_4}{D1_2}
 + \anntf{D9}{D1_4} \annbf{D1_4}{D1_2} \annbf{D1_4}{D1_3} \Bigg)
 \times \exp \Bigg( 3 \pi i T^{(3)}
 \nonumber \\ &&
 - \annbb{D1_2}{D9} - \annbc{D1_2}{O9} - \annbb{D1_3}{D9} - \annbc{D1_3}{O9}
      - \annbb{D1_4}{D9} - \annbc{D1_4}{O9} \Bigg)
 \label{threepolyinstantonD12D13D14NE}
\end{eqnarray}
Using the one-loop amplitudes given in \eqref{vacuumannulusD1iD9}, \eqref{vacuummoebiusD1i}, \eqref{sixpointannulus} and \eqref{fourpointannulus}, one finds the following expression for this poly-three-instanton amplitude:
\begin{eqnarray}
 -\pi^4 \exp\left( 3\pi i T^{(3)} \right) \frac{\vartheta_4^8(2U^{(3)})}{\eta^{12}(U^{(3)})}
 \frac{\vartheta_2^{16}\vartheta_3^{16}\vartheta_4^{16}}{\eta^{48}}(2U^{(3)})
 \sum_{a=2}^4 \frac{\vartheta''_a}{\vartheta_a}(2U^{(3)})
 \label{polythreeinstantonfinal}
\end{eqnarray}
The expression \eqref{gaugethresholdsheterotic} for the gauge threshold corrections in the heterotic model does not contain such a term. This means that poly-instantons give new corrections which are not visible in the standard approach to gauge threshold corrections in heterotic string compactifications.

The poly-two-instanton corrections were shown to be equal to corrections from doubly wrapped instantons. It was argued that this could have its reason in the fact that a doubly wrapped D-brane has the same partition function as two singly wrapped branes with relative Wilson line one-half. A poly-three-instanton correction could have been expected to be equal to a contribution from a triply wrapped brane. In order to reproduce the partition function of a triply wrapped brane, one needs singly wrapped branes with relative Wilson lines one-third. Such Wilson lines are not allowed in the model under consideration due to the orientifold projection. It is thus clear why the poly-three-instanton amplitude \eqref{polythreeinstantonfinal} does not reproduce a correction from a triply wrapped brane, but gives a new correction.

The argument in favour of poly-instantons started from the observation that D-instanton actions should receive instanton corrections. For the model under consideration, D1$_2$ and D1$_4$ mutually correct their instanton actions as well as D1$_3$ and D1$_4$. If one includes these corrections in the expression
\begin{eqnarray}
 \delta f = \anntf{D9}{D1_2} e^{-S_2} + \anntf{D9}{D1_3} e^{-S_3} + \anntf{D9}{D1_4} e^{-S_4}
\end{eqnarray}
for the one D-instanton corrections to the gauge kinetic function one finds
\begin{eqnarray}
 \delta f = \anntf{D9}{D1_2} \exp
   \left( - S_2 + \annbf{D1_2}{D1_4} e^{-S_4+\annsbf{D1_4}{D1_2}e^{-S_2+...}+\annsbf{D1_4}{D1_3}e^{-S_3+...}} \right)
 \nonumber \\
 + \anntf{D9}{D1_3} \exp
   \left( - S_3 + \annbf{D1_3}{D1_4} e^{-S_4+\annsbf{D1_4}{D1_3}e^{-S_3+...}+\annsbf{D1_4}{D1_2}e^{-S_2+...}} \right)
  \nonumber \\
 + \anntf{D9}{D1_4} \exp
   \left( - S_4 + \annbf{D1_4}{D1_3} e^{-S_3+\annsbf{D1_3}{D1_4}e^{-S_4+...}}
   \annbf{D1_4}{D1_2} e^{-S_2+\annsbf{D1_2}{D1_4}e^{-S_4+...}} \right) .
 \nonumber \\ \label{instantoncorrectionsinstantonactions}
\end{eqnarray}
By expanding all the exponentials in \eqref{instantoncorrectionsinstantonactions} to sufficiently high order, one reproduces the expressions \eqref{twopolyinstantonD13D14NE}, \eqref{twopolyinstantonD12D14NE} and \eqref{threepolyinstantonD12D13D14NE}.

In order to obtain all D-instanton corrections to the gauge kinetic function, one has to include the corrections from multiply wrapped instantons in \eqref{instantoncorrectionsinstantonactions}, both to the gauge kinetic function itself and to the instanton actions.

In conclusion, poly-instantons give new corrections to holomorphic couplings in four-dimensional string compactifications, which should be computable whenever D-instanton corrections to gauge kinetic functions can be computed. Note that, although only poly-instanton corrections to the gauge kinetic function have been computed explicitly, there can also be such corrections to the superpotential. This should be clear both from the general arguments and the explicit computation.

\section{Poly-instantons and the heterotic string}
After having demonstrated that poly-instantons yield new corrections, one can ask the question what they correspond to in the heterotic string \cite{Blumenhagen:2008ji}. A naive application of the S-duality map from the type I to the heterotic string leads one to believe that poly-instanton corrections should arise in the heterotic string from instanton sectors consisting of several worldsheets which must interact in some way in order for the resulting amplitude not to be factorisable. These interactions cannot be the usual splitting and joining processes of strings, but might come from terms in the effective action of several heterotic worldsheets which are not present if there is just one of them. It is then clear that the usual Polyakov path-integral approach does not take poly-instantons into account because it deals with only one worldsheet.

Heterotic worldsheet instantons that correct instanton actions would have to wrap curves of genus one, just as those that can correct gauge kinetic functions. According to a non-renormalisation theorem, worldsheet instantons correcting the superpotential wrap curves of genus zero. A poly-instanton correcting the superpotential would have to consist of one worldsheet of genus zero and several worldsheets of genus one. The conjecture that genus one worldsheets contribute to the superpotential at first sight seems to contradict the non-renormalisation theorem. However, a genus one worldsheet does not induce a dilaton dependence in addition to that coming from the genus zero worldsheet, so there is no contradiction.

It is also possible that there are no poly-instantons in the heterotic string. If this is true, there are two possibilities how to reconcile the two different results of the type I and heterotic models. Either, S-duality does no longer hold after including poly-instanton corrections, or the S-duality map receives corrections in the sense that the moduli of the two models are mapped to each other in such a way that the exponentials of exponentials, which are characteristic for poly-instantons, disappear when mapping the type I result to the heterotic string.

At present, it is not clear which of the three possibilities sketched is realised.

\vskip 1cm 
 {\noindent  {\Large \bf Acknowledgements}} 
\vskip 0.5cm 
The author would like to thank his collaborators Nikolas Akerblom,
Ralph Blumenhagen, Dieter L\"ust and Erik Plauschinn. Also, he would like
to thank Florian Gmeiner, Sebastian Moster and Timo Weigand for
discussions.
\clearpage
\appendix
\chapter{Gauge threshold corrections for fractionally charged D6-branes}
\label{appendixgaugethresholdcorrectionsfractional}
In this appendix, formulas for the gauge threshold corrections in intersecting D6-brane models on the $\mathbb{Z}_2\times\mathbb{Z}_2$ orbifold with $h_{21}=51$ of the six-torus are displayed. Four cases, those described in section \ref{examplesfractional} when writing down the partition functions, will be distinguished.

{\bf Case 1:}
\begin{eqnarray}
 \left(g_{aa}^{(1)}\right)^{-2} &=& 32 \pi\, N_a^2  \int_0^\infty dl 
      \sum_{i=1}^3 \sigma_{aa}^i
      \frac{(V_a^i)^2}{R_1^{(i)} R_2^{(i)}} \widetilde{L}_{aa}^{(i)}
 \label{gccorr1} \\ &=& 
      32 \pi\, N_a^2 \int_0^\infty dl 
      \sum_{i=1}^3 \sigma_{aa}^i
      \frac{(V_a^i)^2}{R_1^{(i)} R_2^{(i)}}
  \label{appendixAtwtadpole1} \\ &&
      + 32 \pi\, N_a^2 \left( \sigma_{aa} \ln\left[ \frac{M_s^2}{\mu^2}\right] 
      - \sum_{i=1}^3 \sigma_{aa}^i \ln\left[ (V_a^i)^2\right] \right)
  \label{appendixAannurunning1annukaehlermet1} \\ &&
      - 32 \pi\, N_a^2 \sum_{i=1}^3 4\, \sigma_{aa}^i \, {\rm Im}(i\ln \eta(T^{(i)}))
  \label{appendixAholocorr1} \\ &&
      - 32 \pi\,  N_a^2 \sigma_{aa} \ln [4\pi]\, .
  \label{constants1}
\end{eqnarray}
{\bf Case 2:}
\begin{eqnarray}
 \left(g_{ab}^{(2)}\right)^{-2} &=& 32 \pi\, N_a N_b  \int_0^\infty dl 
      \sum_{i=1}^3 \sigma_{ab}^i
      \frac{(V_a^i)^2}
               {R_1^{(i)} R_2^{(i)}} \widetilde{L}_{ab}^{(i)} 
 \label{gccorr2} \\ &=&
      32 \pi\,  N_a N_b \int_0^\infty dl 
      \sum_{i=1}^3 \sigma_{ab}^i
      \frac{(V_a^i)^2}{R_1^{(i)} R_2^{(i)}}
  \label{appendixAtwtadpole2} \\ &&
      + 32 \pi\,  N_a N_b
      \sum_{i=1}^3 \sigma_{ab}^i \Bigg( \ln\left[ \frac{M_s^2}{\mu^2}\right]
      - \ln\left[ (V_a^i)^2\right]
  \label{appendixAannurunning2annukaehlermet2} \\ &&
      - 4 {\rm Im}(i\ln \eta(T^{(i)})) \Bigg)
  \label{appendixAholocorr2} \\ &&
      - 32 \pi N_a N_b \sigma_{ab} \ln [4\pi]
  \label{constants2}
\end{eqnarray}
{\bf Case 3:}
\begin{eqnarray}
 \left(g_{ab}^{(3)}\right)^{-2} &=& 32 \pi\, N_a N_b  \int_0^\infty dl 
      \sum_{i=1}^3 \sigma_{ab}^i
      \frac{(V_a^i)^2}
               {R_1^{(i)} R_2^{(i)}} \widetilde{L}_{ab}^{(i)} 
 \label{gccorr3} \\ &=&
      32 \pi N_a N_b \int_0^\infty dl 
      \sum_{i=1}^3 \sigma_{ab}^i
      \frac{(V_a^i)^2}{R_1^{(i)} R_2^{(i)}}
  \label{appendixAtwtadpole3} \\ &&
      - 64 \pi N_a N_b \sum_{i=1}^3 \sigma_{ab}^i
      Im\left(i\ln\frac{\vartheta
      \genfrac[]{0pt}{}{(1-|\delta_a^i-\delta_b^i|)/2}{(1-|\lambda_a^i-\lambda_b^i|)/2}
      (0,T^{(i)})}{\eta(T^{(i)})}\right)
  \label{holocorr3} 
\end{eqnarray}
{\bf Case 4:}
\begin{eqnarray}
 && \left(g_{ab}^{(4)}\right)^{-2} \nonumber \\ &=& N_a N_b 
 \int_0^\infty dl \Bigg[ 8 \left({\textstyle \prod_{i=1}^3 I_{ab}^i}\right)\, \sum_{i=1}^3
 \frac{\vartheta'_1 (\theta_{ab}^i,2il)}{\vartheta_1(\theta_{ab}^i,2il)}
 + \sum_{i\ne j\ne k} 32 I_{ab}^i\,  \sigma_{ab}^i \nonumber \\ &&
 \left( \frac{\vartheta'_1(\theta_{ab}^i,2il)}{\vartheta_1(\theta_{ab}^i,2il)} 
 +  \frac{\vartheta'_4(\theta_{ab}^j,2il)}{\vartheta_4(\theta_{ab}^j,2il)}+
    \frac{\vartheta'_4(\theta_{ab}^k,2il)}{\vartheta_4(\theta_{ab}^k,2il)} \right)
 \Bigg] \label{gccorr4} \\ &=& 
        N_a N_b \int_0^\infty dl \ 8 \left({\textstyle \prod_{i=1}^3 I_{ab}^i}\right) \sum_{i=1}^3 \pi 
 \cot\left[\pi \theta_{ab}^i\right]
    \label{appendixAuntwtadpole} \\ &&
        + N_a N_b \int_0^\infty dl \sum_{i=1}^3 32\,  I_{ab}^i\,
        \sigma_{ab}^i\, 
        \pi \cot\left[\pi\theta_{ab}^i\right]
    \label{appendixAtwtadpole4} \\ &&
        + 16 \pi\, N_a N_b \Upsilon_{ab} \sum_{i=1}^3 \left( s_{ab}^i \ln\left[ \frac{M_s^2}{\mu^2}\right] +
              \ln \left[ \frac{\Gamma(1-|\theta_{ab}^i|)}
              {\Gamma(|\theta_{ab}^i|)}\right]^{s_{ab}^i} \right)
    \label{appendixAannurunning4annukaehlermet4} \\ &&
        + 64 \pi\,  N_a N_b \ln [2] \sum_i I_{ab}^i\,
        (\theta_a^i-\theta_b^i)\,  \sigma_{ab}^i
    \label{appendixAannuuni4} \\ &&
        + 16 \pi N_a N_b \left( (\ln 2 - \gamma_E) \Upsilon_{ab} \sum_i s_{ab}^i
        + \ln 4 \sum_{i\ne j\ne k} I_{ab}^i \sigma_{ab}^i ( s_{ab}^j + s_{ab}^k )
        \right)
    \label{constants4}
\end{eqnarray}
The abbreviation $s_{ab}^i=sign(\theta_{ab}^i)$ was used.

\clearpage
\nocite{*}
\bibliography{revklein}
\bibliographystyle{utphys}
\end{document}